\documentclass[12pt]{article}

\advance\topmargin by -0.85in
\usepackage{booktabs}
\usepackage{textcomp}
\usepackage{framed}
\usepackage{slashbox}
\usepackage{makecell}
\usepackage{times}
\usepackage{mathptm}
\usepackage{graphicx}
\usepackage{array}
\usepackage{url}
\usepackage{color}
\usepackage{amsmath}
\usepackage[ruled]{algorithm2e}
\usepackage[utf8]{inputenc}

\DeclareUnicodeCharacter{2212}{-}

\begin{document}


\title{RAID Organizations for Improved Reliability and Performance: 
A Not Entirely Unbiased Tutorial}
\author{Alexander Thomasian\footnote{Thomasian \& Associates, Pleasantville, NY, alexthomasian@gmail.com}
}
\date{}
\maketitle

\subsection*{Abstract}
This is a followup to the 1994 tutorial by Berkeley RAID researchers
whose 1988 RAID paper foresaw a revolutionary change 
in storage industry based on advances in magnetic disk technology,
i.e., replacement of large capacity expensive disks with arrays of small capacity inexpensive disks.
NAND flash SSDs which use less power, incur very low latency, provide high bandwidth, 
and are more reliable than HDDs are expected to replace HDDs as their prices drop.  
Replication in the form of mirrored disks and 
erasure coding via parity and Reed-Solomon codes are two methods 
to achieve higher reliability through redundancy in disk arrays.
RAID(4+k), k=1,2,... arrays utilizing k check strips makes them k-disk-failure-tolerant 
with maximum distance separable coding with minimum redundancy.
Clustered RAID, local recovery codes, partial MDS, and multilevel RAID 
are proposals to improve RAID reliability and performance.
We discuss RAID5 performance and reliability analysis in conjunction with HDDs w/o and with latent sector errors - LSEs,
which can be dealt with by intradisk redundancy and disk scrubbing, 
the latter enhanced with machine learning algorithms.
Undetected disk errors causing silent data corruption are propagated by rebuild.
We utilize the M/G/1 queueing model for RAID5 performance evaluation,
present approximations for fork/join response time in degraded mode analysis,
and the vacationing server model for rebuild analysis.
Methods and tools for reliability evaluation with Markov chain modeling and simulation are discussed.
Queueing and reliability analysis are based on probability theory and
stochastic processes so that the two topics can be studied together.
Their application is presented here in the context of RAID arrays in a tutorial manner.

Categories and Subject Descriptors:
B.4.2 [Input/Output and Data Communications]: Input/Output Devices;
B.4.5 [Input/Output and Data Communications]: Reliability, Testing, and Fault-Tolerance;
D.4.2 [Operating Systems]: Storage Management;
E.4 [Data]: Coding and Information Theory.

TABLE OF CONTENTS
\footnote{Sections indicated by asterisks require knowledge of queueing theory
and stochastic processes and can be skipped without loss in continuity.}

\begin{small}
\noindent
\ref{sec:intro}. Introduction.                                                            \newline
\ref{sec:capacity}. Limiting storage utilization to meet performance goals.               \newline
\ref{sec:sched}. Disk scheduling policies.                                                 \newline
\ref{sec:compress}. Data compression, compaction and deduplication.                       \newline
\ref{sec:tail}. Dealing with high tail latency.                                           \newline
\ref{sec:RAID5}. RAID classification and proposal to eliminate large disks.               \newline
\ref{sec:RAID6}. Maximum distance separable RAID(4+K) arrays.                             \newline 
\ref{sec:RDP}. Rotated Diagonal Parity - RDP coded arrays.                                \newline 
\ref{sec:xcode}. X-code 2DFT array with diagonal parities.                                \newline 
\ref{sec:PMDS}. Partial MDS code.                                                         \newline    
\ref{sec:tiger}. Disk adaptive redundancy scheme.                                         \newline
\ref{sec:multiD}. Multidimensional coding for higher reliability.                        \newline 
\ref{sec:gridfiles}. Grid files.                                                         \newline
\ref{sec:RESAR}. REliable Storage at Exabyte Scale - RESAR.                              \newline
\ref{sec:2d}. 2-d and 3-dimensional parity coding.                                       \newline 
\ref{sec:LRC}. Local Recoverable Code - LRC.                                             \newline
\ref{sec:LRCcomparison}. A systematic comparison of existing LRC schemes.                \newline
\ref{sec:widestripe}. Wide stripe erasure codes.                                         \newline
\ref{sec:practical}. Practical design considerations for wide stripe LRCs  .             \newline
\ref{sec:distr}. Reducing rebuild traffic in distributed RAID.                           \newline 
\ref{sec:pyramid}. Pyramid codes.                                                        \newline 
\ref{sec:Xorbas}. Hadoop Distributed File System - HDFS-Xorbas.                          \newline 
\ref{sec:HACDFS}. Hadoop Adaptively-Coded Distributed File System - HACDFS.              \newline 
\ref{sec:cidon}. Copyset Replication for Reduced Data Loss Frequency                     \newline
\ref{sec:DiskReduce}. More Efficient Data Storage: Replication to Erasure Coding.        \newline 
\ref{sec:CRAID}. Clustered RAID5.                                                        \newline
\ref{sec:BIBD}. Balanced Incomplete Block Design - BIBD.                                 \newline
\ref{sec:Thorp}. CRAID implementation based on Thorp shuffle.                            \newline
\ref{sec:NRP}. Nearly Random Permutation - NRP.                                          \newline
\ref{sec:shifted}. Shifted parity group placement.                                       \newline
\ref{sec:misc}. Miscellaneous topics.                                                    \newline
\ref{sec:distrsparing}. Distributed sparing in RAID5.                                    \newline
\ref{sec:PCM}. Permanent Customer Model - PCM for RAID5 rebuild.                         \newline 
\ref{sec:prioritize}. Rebuild processing in Heterogeneous Disk Arrays - HDAs.            \newline
\ref{sec:singlediskrebuild}. Optimal single disk rebuild in EVENODD, RDP, X-code.        \newline 
\ref{sec:RAIDperfeval}. RAID performance evaluation.                                      \newline    
\ref{sec:RAID5perfanal}$^*$. RAID performance analysis with M/G/1 queueing model.         \newline 
\ref{sec:diskservice}$^*$. Components of disk service time w/o and with ZBR.             \newline 
\ref{sec:seekdistance}$^*$. Seek distance distribution without and with ZBR.                  \newline
\ref{sec:normal}$^*$. RAID5 performance analysis in normal mode.                         \newline
\ref{sec:degraded}$^*$. Fork/Join response time analysis for degraded mode.              \newline  
\ref{sec:RAID5rebuild}. Rebuild Processing in RAID5.                                       \newline
\ref{sec:RAID5rebuildanalysis}$^*$. Analysis of rebuild processing with {\t Vacationing Server Model - VSM}. \newline 
\ref{sec:VSM}$^*$. M/G/1 VSM analysis with multiple vacations of two types.              \newline 
\ref{sec:alternative}$^*$. An alternative method to estimate rebuild time.               \newline 
\ref{sec:RAIDrel}. RAID5 reliability analysis.                                           \newline
\ref{sec:7.2}$^*$. RAID5/6 reliability analysis with unrecoverable errors.                 \newline
\ref{sec:DEH}. Disk scrubbing and Intra-Disk Redundancy to deal with Latent Sector Errors - LSEs.    \newline 
\ref{sec:IDRschemes}. Schemes to implement IDR.                                          \newline
\ref{sec:IDRperf}. IDR's effect on disk performance.                                     \newline 
\ref{sec:comparison}. Comparison of disk scrubbing and IDR with IPC.                     \newline
\ref{sec:combined}. Combining scrubbing with SMART and Machine Learning.                 \newline
\ref{sec:UDE}. Undetected Disk Errors - UDE and Silent Data Corruption - SDC.            \newline
\ref{sec:mirhyb}. Mirrored and Hybrid Disk Arrays Description and Reliability Comparison.\newline
\ref{sec:shortcut}. Shortcut Method to Compare the Reliability of Mirrored  Disks and RAID(4+k). \newline
\ref{sec:multilevel}. Reliability analysis of multilevel RAID arrays.                    \newline 
\ref{sec:RAID15}. Mirrored RAID5 - RAID1/5 reliability analysis.                         \newline 
\ref{sec:RAID51}. RAID5 with mirrored disks - RAID5/1.                                   \newline 
\ref{sec:shortcut2}. Shortcut reliability analysis to compare RAID1/5 and RAID5/1.       \newline
\ref{sec:autoRAID}. AutoRAID hierarchical array.                                         \newline
\ref{sec:simul}. Discrete-event simulation for reliability evaluation.                   \newline 
\ref{sec:uber}. Simulation study of a digital archive.                                   \newline
\ref{sec:HRAID}. Simulation of hierarchical RAID reliability.                            \newline 
\ref{sec:Proteus}. Proteus open-source simulator.                                        \newline
\ref{sec:ATT}. CQSIM\_R Tool Developed at AT\&T.                                         \newline
\ref{sec:SIMEDC}. SIMedc simulator for erasure coded data centers.                       \newline
\ref{sec:tools}. Reliability modeling tools.                                             \newline 
\ref{sec:ARIES}. Automated Reliability Interactive Estimation System - ARIES Project at ARIES.  \newline
\ref{sec:SAVE}. System AVailability Estimator - SAVE Project at IBM Research.            \newline
\ref{sec:SHARPE}. Symbolic Hierarchical Automated Reliability and Performance Evaluator - SHARPE.  \newline
\ref{sec:ZRL}. Research at IBM Zurich Research Lab.                                      \newline 
\ref{sec:metrics}. System reliability metrics.                                           \newline 
\ref{sec:placement}. Clustered versus declustered data placements.                       \newline 
\ref{sec:tapelibrary}. Performance analysis of a tape library system at ZRL.             \newline 
\ref{sec:flash}. Flash Solid-State Drives - SSDs.                                        \newline
\ref{sec:WOM} Write-Once Memory - WOM codes to enhance SSD lifetimes.                    \newline
\ref{sec:predictable}. Predictable microsecond level support for flash.                  \newline
\ref{sec:diffRAID}. Differential RAID for SSDs.                                          \newline
\ref{sec:FAWN}. Fast Array of Wimpy Nodes - FAWN.                                        \newline
\ref{sec:RAMCloud}. Distributed DRAM-based storage - RAMCloud.                           \newline
\ref{sec:conclusions}.  Conclusions.                                                     \newline   
Appendix I. Transient analysis of state probabilities in Markov chains.                  \newline
Appendix II. The mathematics behind SHARPE reliability modeling package.                 \newline
Abbreviations.  \newline
Bibliography. 
\end{small}

\section{Introduction}\label{sec:intro}

The tutorial is intended for computer science and engineering students and IT professionals
interested in storage system organization, performance and reliability evaluation.
It is a followup to Chen et al. 1994 \cite{Che+94}, Thomasian and Blaum 2009 \cite{ThBl09},
and an extension of Thomasian 2021 \cite{Thom21}.
The title derives from the fact that this is a rapidly growing field and some topics remain uncovered. 

Reliability analysis of storage systems initially dealt with disk failures at a course granularity,
whole {\it Hard Disk Drives - HDDs} failures. 
More complex analysis was introduced to take into account the effect of disk scrubbing and 
{\it IntraDisk Redundancy - IDR} to deal with {\it Latent Sector Errors - LSEs} in HDDs. 

We discuss the queueing analysis of RAID arrays based on HDDs, 
where performance is more of an issue than NAND SSDs invented at Toshiba in 1989,
which provides a high throughput, low latencies, and higher reliability than HDDs.

Some sections require exposure to probability theory and stochastic processes, 
e.g., Kobayashi et al. 2014 \cite{KoMT14}, 
Trivedi 2002 \cite{Triv02} with heavy reliability modeling coverage,
and elementary queueing theory, e.g., Kleinrock 1975/76 \cite{Klei75,Klei76}, 
Coding theory applicable to RAID design are presented in MacWilliams and Sloane 1977 \cite{MaSl77},
Fujiwara 2006 \cite{Fuji06}, and short tutorial with good references Blaum 2009 \cite{Blau09}.

There are chapters dedicated to storage systems in textbooks on computer architecture, 
e.g.,  Hennessey and Patterson 2019 \cite{HePa19}, 
operating systems, e.g., Arpaci-Dusseau 2018 \cite{ArDu18},
databases, e.g., Ramakrishnan and Gehrke 2002 \cite{RaGe02},
which can be complemented by Thomasian 2021 \cite{Thom21}.
Cache, DRAM, Disk are discussed in Jacob, Ng and Wang 2008\cite{JaNW08}.
A book on NAND flash memories is Micheloni et al. 2010 \cite{MiCM10},
whose reading requires electrical engineering background. 
We have provided url's of web pages on novel developments in the field
and some technical topics covered by wikipedia.

Table of contents lists topics covered in the paper.
We start the discussion with a simple queueing analysis of disks.

\section{Limiting Storage Utilization to Meet Performance Goals}\label{sec:capacity}

Due to exponential growth of data projected to reach 175 zettabytes - ZB=$10^{21}$B) by  2025
and the main role HDDs play in storing data. \newline 
\begin{scriptsize}
\url{https://www.networkworld.com/article/3325397/}
\end{scriptsize}

In spite of multiterabyte capacity disks their storage cannot be fully utilized
due to overload in providing access to randomly places data blocks.
This is especially true for {\it Online Transaction Processing - OLTP} applications
where the transfer time for small 4 or 8KB data blocks tends to be negligible,
but there is the unproductive positioning time, which is the sum of two components:
\begin{description}
\item[\bf Seek time] Moving the {\it Read/Write - R/W} head attached to the disk arm to the target track. 
\item[\bf Rotational latency] The time for the data on the rotating track to reach the R/W head.
\end{description}

HDDs are classified according to their use: {\it Personal Storage - PS} versus {\it Enterprise Storage - ES},
their interface: {\it Advanced Technology Attachment - ATA} 
versus {\it Small Computer System Interface - SCSI} Anderson et al. 2003 \cite{AnDR03}.
{\it Serial ATA - SATA} is replacing disks with parallel ATA.
SATA and SCSI vary according to their bandwidth and access latency.
Less expensive SATA 7200 {\it Rotations Per Minute - RPM} disks are mainly used in data centers. 

Bandwidth is especially important in HDDs accessed by {\it OnLine Transaction Processing - OLTP} workloads,
where accesses to small randomly placed disk blocks with high positioning time, but negligible transfer time.
More expensive ES drives provide faster seek time via more expensive circuitry than PS drives.
Rotational latency is reduced by adopting SCSI 10K or 15K RPM disks with 3 and 2 ms mean latencies,
but also faster seek times.
Such a workload was observed by analyzing I/O traces Ramakrishnan et al. 1992 \cite{RaBK92}. 
Such insights were used in specifying database benchmarks by the {\it Transaction Processing Council - TPC}.  \newline
\begin{scriptsize}
\url{https://www.tpc.org.}
\end{scriptsize}

The use of few tracks allows lower seek times due to short stroking.
This is accomplished by just utilizing higher capacity outer disk cylinders,
which is due to {\it Zoned Bit Recording - ZBR} Jacob et al. 2008 \cite{JaNW08}.  
Zoning maintains almost the same linear recording density across disk tracks,
so that outer tracks have a higher capacity than inner tracks by a factor of 1.7.

That block transfer time is negligible with respect to rotational latency,
which is one half of disk rotation time or 4.17 msec for 7200 RPM disks,
while the number of sectors transferred is less than ).1\% of sectors on a track.

Rotational latency can be shortened by providing two disk arms which are 180$^\circ$ apart,
The same effect can be achieved by storing records twice on opposing sides of each track,
but this has the disadvantage of adversely affecting sequential accesses.
The two-actuator Seagate HDD from Seagate is discussed in Section \ref{sec:flash}.

We model disks as single server M/M/1 queues 
with FCFS scheduling Kleinrock 1975 \cite{Klei75}.
Disk requests arrive according to a Poisson process with rate $\lambda$ 
and have exponentially distributed service times with mean $\bar{x}=1/\mu$, where $\mu$ is the service rate. 
Disk utilization is the fraction of time the disk is busy: $\rho  = \lambda \bar{x}$ 
Since $\rho < 1$ the maximum sustainable arrival rate is $\lambda_{max} \leq 1 / \bar{x}_{disk}$,
e.g., for $\bar{x}=10$ milliseconds (ms) $\lambda_{max}=100$ {\it IO Per Second - IOPS}  

If $\lambda \bar{x} > 1$ then $m$ disks should be provided disks so that $\rho = \lambda \bar{x} / m < 1$.
Due to Little's result Kleinrock 1975 \cite{Klei75} 
$\rho$ is also the mean number of requests in service.
The distribution of number of requests in M/M/m queues given below
can be used to estimate the probability that a request has to wait.           \newline
\begin{scriptsize} 
\url{https://en.wikipedia.org/wiki/M/M/c_queue}                               \newline
\end{scriptsize}
In fact we have a different queueing system where a single requests is sent to all $m=3$ servers
and assuming slightly different maen disk service times ($\rho_i = \lambda \bar{x}_i, i=1,23$) 
the probability that a request does not have to wait is:
$\prod_1^3 (1- \rho^i)$

Replicating data on multiple disks is a costly solution,
which is adopted by companies operating hyperscalars.
It has the advantage of increased reliability and reduced response time
which may be necessitated by {\it Service Level Agreements - SLAs}.
SLA's in {\it Internet Protocol - IP} networks are discussed in Verma 2000 \cite{Verm04}.    \newline
\begin{scriptsize}
\url{https://en.wikipedia.org/wiki/Internet_Protocol}              \newline
\url{https://en.wikipedia.org/wiki/Service-level_agreement}
\end{scriptsize}

Consider $N$ heterogeneous disks indexed in nonicreasing speed order with a large number of small files, 
which can be moved across disks to minimized overall mean disk response time ($\bar{R}$).
The Lagrange multiplier method is applied to response times obtained 
by the M/G/1 queueing model in Section \ref{sec:normal} 
to determine optimal disk arrival rates in Piepmeier 1975 \cite{Piep75}. \newline 
\begin{scriptsize}
\url{https://en.wikipedia.org/wiki/Lagrange_multiplier}                
\end{scriptsize}
The optimization yields $\lambda_i \geq 0, 1 \leq i \leq \leq  N$ 
\vspace{-1mm}
$$\bar{R}= \sum_{i=1}^M \frac{\lambda_i}{\Lambda} R_i, \mbox{  with  }\sum_{i=1}^N \lambda_i = \Lambda
\mbox{ and } \lambda_i = 0,  M < i \leq N.$$ 
Note that only $M<N$ disks have $\lambda_i > 0$ 
and it is assumed they have adequate capacity to hold the data. 

Georgiadis et al. 2004 \cite{GeNT04} considered the min-max policy
to obtain an allocation $\lambda_i, 1 \leq u \leq M$,
which equalizes the mean response times over $M \leq N $ fastest devices.
The allocation results in higher utilization for faster disks.

The mean disk access  time ($R_{disk}$) is an important contributor 
to transaction response times in OLTP.
Given its maximum: $R_{disk}^{max}$ the corresponding arrival rate 
using the M/M/1 queueing system is given as follows:
\vspace{-1mm}
$$R_{disk} = \frac{ \bar{x}_{disk} }{1- \rho} < R_{disk}^{max} \mbox{ which implies }
{\lambda}_{max} < \frac{1}{\bar{x}_{disk}} - \frac{1}{R_{disk}^{max}}.$$
If $R_{disk}^{max} =20$ then ${\lambda}_{target}=50$, which is $\lambda_{max}/2$.  

The response time distribution in M/M/1 queueing system is exponentially distributed: 

\vspace{-2mm}
\begin{eqnarray}\label{eq:Rdistr}
R (t) = 1 - e^{ t / R (\rho) }.  
\end{eqnarray}
The arrival rate ($\lambda_p$) to achieve a certain percentile ($p$) for disk response time is: 
$R_{disk}^{(p)}  =  - R (\rho) \mbox{ln} (1-p)$, hence:  
\vspace{-1mm}
$$\lambda_p = \frac{1}{\bar{x}_{disk}} + \frac{\mbox{ln} (1-p) } {R_{disk}^{(p)} }$$
Note that $\mbox{ln}(1-p) < 0$ for $0 \leq p < 1$.

Given the second moments of disk service time both analyses can be repeated
more accurately using an M/G/1 queueing model Kleinrock 1975 \cite{Klei75}.

\begin{framed}
\subsection{Disk Scheduling Policies}\label{sec:sched}

Disk access time can be  reduced by adopting judicious disk scheduling policies,
such as the {\it Shortest Seek Time First - SSTF} and SCAN by Denning 1967 \cite{Denn67}
and {\it Shortest Access Time First - SATF} by Jacobson and Wilkes 1991 \cite{JaWi91}. 
Disk scheduling policies are compared via simulation in 
Worthington et al. 1994 \cite{WoGP94}, 
Hsu and Smith 2005 \cite{HsSm04},
Jacob et al. 2008 \cite{JaNW08}, 
and Thomasian 2011 \cite{Thom11}.
Reported in the latter is an extension of SATF with lookahead and priorities
and an empirical equation for SATF mean disk service versus queue-length ($q$):
\vspace {-1mm}
$$\bar{x}_{SATF} (q) = \bar{x}_{FCFS} / q^p \mbox{  where  }p \approx 1/5$$  
so that disk service time is halved for $q=32$.
Increase in throughput versus $q$ is given in Table 2 in Anderson et al. 2003 \cite{AnDR03}
and plotted as Figure 6.2 in Hennessey and Patterson 2019 \cite{HePa19}.

Discussed below are some other disk scheduling methods and {\it Complete Fairness Queueing - CFQ} 
which provides a fair allocation of the disk I/O bandwidth for all the processes which request an I/O operation. \newline
\begin{scriptsize}
\url{https://en.wikipedia.org/wiki/I/O_scheduling}                             \newline
\url{https://www.kernel.org/doc/Documentation/block/cfq-iosched.txt}
\end{scriptsize}

\end{framed}

There have been several studies to reorganize disk data to improve disk access times.
This is done dynamically by the {\it Automatic Locality Improving Storage - ALIS} tool 
based on common sequences of disk requests Hsu et al. 2005 \cite{HsSY05}. 
This study reviews earlier articles on this topic, which deal with static data allocation.
This discussion is relevant to LSA where performance can be improved 
by maintaining seek affinity Menon 1995 \cite{Meno95}.

Anticipatory arm placement of idle disks is another scheme 
to reduce disk access time Thomasian and Fu 2006 \cite{ThFu06}.
In the case of single disks with $C$ cylinders the arm is placed at cylinder C/2, 
while in mirrored disks the arms are placed at C/4 and 3C/4. 
The mean seek distance for reads in the two cases is C/2 and C/4,
which is otherwise $C/3$ and $C/5$ as discussed in Section \ref{sec:mirhyb}.
The paper considers disks with hot spots (frequently accessed cylinders) and zoning. 

HDDs due to their lower cost per GB and high capacities are the preferred medium in large data centers 
Brewer et al. 2016 \cite{Brew16,Bre+16}, where the following metrics are identified: 
(1) Higher IOPS, 
(2) Higher capacity,
(3) Lower tail latency, which is discussed in Subsection \ref{sec:tail}. 
(4) Security requirements, and 
(5) {\it Total Cost of Ownership - TCO}.  
Higher bandwidth attainable by NAND Flash SSDs require fewer drives to sustain the same workload.
Reliability, performance, and cost are identified as key cloud storage parameters in Huang 2013 \cite{Huan13}. 

Hyperscalars are data centers with a very large number of disks.
The four largest hyperscalar platforms out of twelve largest are — 
AWS (Amazon Web Services), Google Cloud, Meta, and Microsoft Azure, 
which represent 78\% of capacity in this category and consume 13,177 MegaWatts. \newline
\begin{scriptsize}
\url{https://www.datacenterknowledge.com/manage/2023-these-are-world-s-12-largest-hyperscalers}
\end{scriptsize}   

While SSDs cost more per GB than HDDs they incur lower power consumption, 
but it is the lower TCO that matters.  \newline
\begin{scriptsize}
\url{https://www.snia.org/sites/default/files/SNIA_TCOCALC_Workpaper_Final_0.pdf}  \newline
\end{scriptsize}
Power saving in disks is achievable by different levels of disk powering, 
such as spindown in laptops to save battery power. 
With dropping NAND Flash storage costs NAND Flash disks are expected to replace disks in laptops. \newline
\begin{scriptsize}
\url{https://www.techradar.com/news/ssd-vs-hdd-which-is-best-for-your-needs}
\end{scriptsize}

{\it Massive Array of Idle Disks - MAID} save power by powering down 
the majority of disks in a data center Colarelli and Grunwald 2002 \cite{CoGr02}.
RAID5 requires at least two disks should be powered up for updates,
but {\it NonVolatile RAM - NVRAM}, which is DRAM powered by {\it Uninterruptible Power Supply - UPS}
is another option to preserve updates without disk spin-up.
The MAID proposal led to the Copan startup for archival storage.  \newline
\begin{scriptsize}
\url{http://www.thic.org/pdf/Jul05/copansys.aguha.050720.pdf}
\end{scriptsize}
Lower power consumption is attainable by NAND Flash SSDs as discussed in Section \ref{sec:flash}.
Other schemes to save disk power are discussed in Chapter 7 in Thomasian 2021 \cite{Thom21}. 

{\it Power Usage Effectiveness - PUE},
which is the ratio of total power consumption (including cooling plus power conversion)
to the power used to run IT equipment is a metric used to determine data center energy efficiency.
Variation of PUE over (2008-22) and across different Google sites is as follows. \newline
\begin{scriptsize}
\url{https://www.google.com/about/datacenters/efficiency/}                        
\end{scriptsize}

Seagate has announce a 30 TB {\it Heat Assisted Magnetic Recording - HAMR} HDDs in April 2023. \newline
\begin{scriptsize} 
\url{https://blocksandfiles.com/2023/04/21/seagate-30tb-hamr/}        \newline
\end{scriptsize}
While Flash NAND SSDs tended to smaller in capacity than HDDs they may exceed HHDs, 
as in the case of PureStorage 300 TB flash drives in 2026, which will exceed HDD capacity ten-fold. \newline
\begin{scriptsize}
\url{https://blocksandfiles.com/2023/03/01/300tb-flash-drives-coming-from-pure-storage/} \newline
\end{scriptsize}

\subsection{Data Compression, Compaction and Deduplication}\label{sec:compress}

While memory costs at all levels are dropping more efficient storage utilization 
can be attained by preserving storage capacity by using data deduplication, compaction, and compression.
The first two methods are applicable at secondary storage level and 
the third method has been applied to main memory.
Data compression methods such as Huffman, Lempel-Ziv - LZ
and arithmetic encoding are discussed in Sayood 2017 \cite{Sayo17}.
Compressed data in main memory is decompressed as it is loaded into CPU caches 
and vice-versa as discussed in Section 2.14 in Thomasian 2021 \cite{Thom21}. 

Deduplication is a technique to duplicate data at block, page, file level.
It has been effectively applied to archival and backup storage to reduce storage costs 
Data deduplication is discussed in Section 2.15 in 
Thomasian 2021 \cite{Thom21} and Mohammed and Wang 2021 \cite{MoWa21}.
A classification of deduplication systems according to Paulo and Pereira 2014 \cite{PaPe14} is
(1) granularity, (2) locality, (3) timing, (4) indexing, (5) technique, and (6) scope. \newline
\begin{scriptsize}
\url{https://en.wikipedia.org/wiki/Data_deduplication}               	
\end{scriptsize}

Duplication can be checked quickly by applying has functions such as SHA-2 
and in the case of a match a byte-by-byte comparison can be applied.  \newline
\begin{scriptsize}
\url{https://en.wikipedia.org/wiki/SHA-2}                   
\end{scriptsize}

Data compression is suited to LFS and LSA paradigms, 
since files are not written in place and this allows their size to vary.
Data compaction creates large empty spaces by removing older versions of files
and cleanup of defunct objects based on their {\it Time-to-Live - TTL}.
Storage fragmentation is dealt with by continuous garbage collection and data compaction, 
which is the shifting of the location of stored data to release unused storage for reuse.

A performance study of LFS was reported in McNutt 1994 \cite{McNu94}.
A directory indicates the physical location of each logical object
and the location of the most recent copy for objects written more than once.
A mathematical model of garbage collection shows how collection load
relates to the utilization of storage and the amount of update locality present in the pattern of updates.
A realistic statistical model of updates, based upon trace data analysis is applied.
Alternative policies are examined for determining which areas to target.

Performance of RAID5 and LSA is compared using an approximate analysis in Menon 1995 \cite{Meno95},
which is based on the following relationship used to estimate the rate of garbage collection.
Given {\it Average Segment Occupancy - ASO} and lowest {\it Best Segment Occupancy} we have:
\vspace{-1mm}
$$\mbox{ASO}=(1-\mbox{BSO}) / \mbox{ln}(1- \mbox{BSO}), \hspace{5mm}\mbox{e.g., if ASO-0.6 then BSO=0.324}.$$

There are two types of write hits: 
(1) to dirty blocks in {\it NonVolatile Storage - NVS} caches, (2) to clean pages.
Given a request rate $K$ the fraction of request to clean pages is $C$ 
and the fraction of writes to dirty ages is $D$.

The reading of segments is attributable to garbage collection
and only a fraction BSO of segment writing is attributable to garbage collection and the rest to destage.
Let $KB$ be the read miss rate, $KC$ the miss rate due to write hits in clean pages 
and $KD$ the destage rate due to write misses

The rate at which dirty blocks are created in the NVS caches is $K(C+D)/X$
assuming that $X$ blocks are destaged together. 
It is argued that segments are written and read at rate $(KC+KD/(X (1-BSO) Z)$
where $Z$ is the segment size in tracks.
The effect of data compression and seek affinity is investigated in this study 
and it is observed that LSA outperforms RAID5 in throughput, 
but offers higher response times.

Menon and  Stockmeyer 1998 \cite{MeSt98} proposed a new algorithm 
for choosing segments for garbage collection in LFS and LSA
and compared it against a greedy and cost-benefit algorithms via simulation.
The basic idea is that segments which have been recently filled by writes
should not be considered for garbage collection until they achieve a certain age threshold.
This is because as time progresses data in such segments is less likely to be updated,
making old segments eligible for garbage collection.
The algorithm chooses segments that yield the most free space.
Also given is a method to determine the optimal age-threshold
under certain assumptions about the workload.

\subsection{Dealing with High Tail Latency}\label{sec:tail}

The following discussion is based on Dean and Barroso 2013 \cite{DeBa13}. 
Reasons for high tail latency are:

\begin{description}

\item[\bf Shared resources]  
Sharing resources such as CPU cores, processor caches, memory and network bandwidth.

\item[\bf Daemons]     
Periodically scheduled background daemons use only limited resources on average, 
but generate delays when active.

\item[\bf Global resource sharing] Network switches, shared file systems.

\item[\bf Maintenance] 
This includes data reconstruction in distributed file systems,  
periodic log compaction in storage systems as in the case of LSA.
In the case of SSDs there is 100-fold increase in random read accesses 
for a modest write activity as discussed in Section \ref{sec:flash}. 

\item[\bf Queueing delays] Enhances variability.   

\end{description}

Increased variability may be due to the following:

CPUs may temporary overrun their power limits,    
but throttle to lower temperature but that limits CPU speed.

Powersaving  modes  to save energy in various devices 
incurs additional  latency  when  moving  from  inactive  to  active modes,
Spindown is used in laptops to save battery power,
but a delay is incurred until spinup is completed.

The effect of subsecond response times in a timesharing computer system  
on programmer productivity was observed in Thadhani 1981 \cite{Thad81}. 
There is the double effect that programmers respond faster when response time is subsecond 
and this reduces the total time to accomplish a task. 
The effect of response time on successful conversion in e-commerce is discussed here. \newline
\begin{scriptsize}
\url{https://queue-it.com/blog/ecommerce-website-speed-statistics/}   
\end{scriptsize}

Response time in search engines is reduced by concurrently accessing $n$ replicated copies of a file 
rather than just a single copy on a single device, 
which may be temporarily or permanently unavailable.
Response time is then the minimum of $n > 1$ exponentially distributed responses times.
In the case of homogeneous M/M/1 queues $R_{min} = R / n$.

Distributed {\it Shortest Processing Time First - SPTF} 
is a request distribution protocol for decentralized brick storage system with high-speed interconnects.
D-SPTF dynamically presented in Lumb and Golding 2004 \cite{LuGo04} selects servers to satisfy a request 
while balancing load, exploiting aggregate cache capacity, 
and reduces positioning times for cache misses. 

IBM Intelligent Bricks project is described by Wilcke et al. \cite{WGF+06} is an example of brick storage systems.
Each brick is a standalone RAID(4+k), but bricks collaborate in 2006.
{\it Hierarchical RAID} proposal in Subsection \ref{sec:HRAID} is conceptually similar,
but its reliability analysis led to conflicting conclusions Thomasian 2022 \cite{Thom22}. 

\section{RAID Classification and Proposal to Eliminate Large Disks}\label{sec:RAID5}

Two main categories of RAID redundancy are replication (mirroring) 
and erasure coding (parity in the simplest case).
In the case parity coded RAID3, RAID4, RAID5 described in Chen et al. \cite{Che+94}
$N-1$ surviving disks out of $N$ disks need be accessed 
to recover a lost block and for RAID(4+k) the number is $N-k+1$ disks.

Erasures are failures whose location is known and 
hence a single parity is adequate to recover a lost block 
by {\it eXclusive-ORing - XORing} corresponding surviving blocks.
{\it Redundant Array of Independent Memories - RAIM} Meaney et al. 2012
and {\it Redundant Array of Independent Libraries - RAILs} (tapes) 
Ford et al. 1998 \cite{FoMB98} utilize the RAID5 paradigm.                                \newline
\begin{scriptsize}
\url{https://en.wikipedia.org/wiki/Redundant_array_of_independent_memory}                \newline
\url{https://blocksandfiles.com/2021/10/08/quantums-exabyte-munching-scale-out-modular-tape-library/}  
\end{scriptsize}

Hamming codes Fijiwara 200 \cite{Fuji06} was adopted by Thinking Machines for the Connection Machine 5 - CM5.  \newline
\begin{scriptsize}
\url{https://en.wikipedia.org/wiki/Hamming_code}                               \newline
\url{https://en.wikipedia.org/wiki/Connection_Machine}                         \newline
\end{scriptsize}
With the positions of failed disks known up to two disk failures can be recovered,
as also noted in Gibson 1992 \cite{Gibs92}.

The RAID proposal by Patterson, Gibson, and Katz 1988 \cite{PaGK88}
advocated replacing expensive, large form-factor disks, reliable 
disks used in conjunction with IBM mainframes with arrays of inexpensive small form-factor,
less reliable disks used by PCs - Personal Computers in 1980s.

Large form-factor {\it Direct Access Storage Devices - DASD}, 
using variable block sizes with the {\it Extended Count Key Data - ECKD} format,
which is similar to CKD in layout, but support five additional {\it Channel Command Words -CCWs}. \newline
\begin{scriptsize}
\url{https://en.wikipedia.org/wiki/Count_key_data}  \newline
\end{scriptsize}
ECKD is supported by IBM's {\it Multiple Virtual Storage - MVS} operating system 
and its descendant z/OS operating system family.
In IBM mainframes running MVS the {\it Disk Array Controller - DAC} emulates 
CCWs issued to ECKD format disks on {\it Fixed Block Architecture - FBA} disks 
with 512 B (byte) and 4096 B disk sectors since 2010.

There is reduction in reliability when expensive, large capacity, more reliable disks
are replaced by an array of inexpensive, smaller capacity, less reliable disks.
The added component count with less reliable disks was addressed 
by introducing redundancy so that a RAID5 array with the same capacity 
as IBM's 3390 drives can be made as reliable, 
while providing parallel access and consuming less power Gibson 1992 \cite{Gibs92}. 
Power was not a major consideration at the time, 
but Chapter 7 in Thomasian 2021 \cite{Thom21} is dedicated to this topic.

The {\it Mean Time TO Failure - MTTF} of small disks was estimated crudely in Gibson 1992 \cite{Gibs92}
based on the time to return broken disks to manufacturers. 
The statistical methods in first edition of Lawless 2003 \cite{Lawl03} were followed in this study.
There have been several studies of disk failures Schroeder and Gibson 2007 \cite{ScGi07}, 
Jiang et al. \cite{JHZK08}, Schroeder et al. 1010 \cite{ScDG10},
and NAND flash-based SSDs in Schroeder et al. 2017 \cite{ScML17},
Maneas et al. 2021 \cite{MMES21}.

Striping partitions files into fixed size {\it Stripe Units - SUs} or strips, 
which are placed in round-robin manner on successive rows of disks modulo the number of disks $N$.
A striped RAID with no redundancy is classified as RAID0.
RAID5 uses striping like RAID0 but dedicates one strip to parity.
\vspace{-1mm}
$$P_N = D_1 \oplus D_2 \oplus D_3 \ldots \oplus P_{N-1}.$$
Parity strips in RAID5 are placed in right to left diagonals to balance the parity update workload. 
This is referred to left symmetric organization. 
Various parity placements are considered in Chen et al. 1994 \cite{Che+94},
which also reviews studies to determine optimal strip size. 
Default 128 KB and 256 KB strip sizes are adopted by some operating systems. \newline 
\begin{scriptsize}
\url{https://ioflood.com/blog/2021/02/04/optimal-raid-stripe-size-and-filesystem-readahead-for-raid-10/}
\end{scriptsize}

Fault-tolerance in RAID5 is attained at the cost of extra processing.
Given a modified disk block $d_{new}$, $d_{old}$, 
which is not cached is read from disk to compute $d_{diff} = d_{old} \oplus d_{new}$.
$p_{old}$ if not cached is read to compute $p_{new} = p_{old} \oplus p_{diff}$.
Both $d_{new}$ and $p_{new}$ need to be written to disk eventually.
The fact that a single write requires four disk accesses 
is called the {\it Small Write Penalty - SWP} in Chen et al. 1994  \cite{Che+94}.
In the case of RAID$(4+k), k \geq 1$ 
$2(k+1)$ reads and writes are required to update data and $k$ check blocks.

The purpose of RAID striping was disk load balancing 
and was ascertained by Ganger et al. 1996 \cite{GWHP96}.
Parity striping proposed in Gray et al. 1990 \cite{GrHW90} 
preserves the original data layout by adding parities contiguously. 
preferably at the middle disk cylinders to achieve seek affinity.
This is because in early RAID5 arrays with no NVS support, 
the update of a data block was considered completed only when both the data block 
and the corresponding parity block were written to disk. 

Parity striping has the advantage of obviating accesses to multiple disks 
when large blocks are accessed and the strip size is small.
It has he disadvantage of possibly unbalanced disk loads, known as access skew, 
which can be dealt with via disk load balancing support.
Load balancing can be achieving by {\it Hierarchical Storage Management - HSM} 
which moves data data horizontally across fast disks to balance loads,
but also vertically from faster to slower disks and vice-versa.
HSM or tiered storage is discussed here. \newline
\begin{scriptsize}
\url{https://www.ibm.com/support/pages/operating-system-disk-balancing-support}  \newline
\url{https://en.wikipedia.org/wiki/Hierarchical_storage_management}              \newline
\end{scriptsize}
HP's AuroRAID discussed in Subsection \ref{sec:autoRAID} 
moves data  from RAID1 to RAID5 arrays and vice-versa.

RAID5 is compared with parity striping in Chen and Towsley 1993 \cite{ChTo93}.
The analysis does not take into account the fact that strip size 
should be chosen large enough to satisfy most requests sizes
and these are satisfied by multi-disk {\it Fork/Join - F/J} requests 
resulting in a significant increase in disk loads and this favors parity striping.

In a RAID5 with parity group size $G$ if $n > G/2$ strips are updated then it may be more efficient 
to read the remaining $G - n$ strips to compute the parity 
by XORing them according to the {\it ReConstruct Write - RCW} method Thomasian 2005 \cite{Thom05},
which may be more efficient than a {\it Read Modify Write - RMW} which leads to SWP. 
Linux's RAID5 employs likewise a simple majority rule to determine a strategy for writing,
i.e., if a majority of pages for a stripe are dirty then {\it Parity Computation - PC} is chosen, 
otherwise {\it Parity Increment - PI}  is chosen and is used in degraded mode.

Upon the failure of a single disk in RAID5, say Disk$_1$, 
its blocks can be reconstructed on demand as follows:
\vspace{-1mm}
$$d_1 = d_2 \oplus d_3 \oplus \ldots \oplus p_{N}$$
The RAID5 read load  on surviving disks is doubled when a single disk fails. 

The load increase in processing read and write requests 
as a function of the fraction of read requests ($f_r$) 
is given in Ng and Mattson 1994 \cite{NgMa94}:

\vspace{-2mm}
\begin{eqnarray}\label{eq:degraded}
\mbox{LoadIncr}(f_r) =
\frac{ U_{faulty} }{ U_{faultfree} } = \frac{N}{N-1} + \frac{ (N-2)f_R + (N-8) f_W}{(N-1)(f_R + 4 f_W)},
\end{eqnarray}
e.g., $\mbox{LoadIncr} (0) = 1.333$ and $\mbox{LoadIncr}{1}=2$.              

Rebuild is the systematic reconstruction of strips of the failed disk on a spare disk.
Strip $D_1$ on Disk$_1$ is reconstructed in the DAC buffer by reading and XORing $N-1$ corresponding strips, 
before they are written to a spare disk ${D'}_1$, which replaces Disk$_1$.
\vspace{-1mm}
$${D'}_1 = D_2 \oplus D_3 \oplus \ldots P_{N}.$$

Several schemes to attain higher reliability are proposed by the  Berkeley RAID team 
Hellerstein et al. 1994 \cite{Hel+94}.
2-dimensional arrays are analyzed in Newberg and Wolf 1994 \cite{NeWo94},
but the analysis of Full-2 codes by Lin et al. 2009 \cite{LZW+09} took much longer.
Simulation was used to investigate the effect of varied frequency of repairs 
(replacing broken disks) on the MTTDL \cite{Hel+94}

\begin{framed}
\subsection*{Reliability of 2D squares and 3D cube}

2-D size $n^2$ and 3-D size $n^3$ arrays  
are considered in Basak and Katz 2015 \cite{BaKa15}.
There are $n$ disks in each dimension protected by $n$ parity disks 
with an extra parity disk protecting the parities
so that the number of disks for $n$ dimensions is $n^d + d \times (n+1), n=2,3$.
{\it Continuous Time Markov Chains - CTMCs} are used to estimate MTTDL.
It is obvious that four disk failures constituting a rectangle lead to data loss
and the number of cases leading to data loss is $\binom{n}{2}n$.
Given there are a series $\ell$ of such squares the total number of disks is $N=(n^2 + 2n + 2)\ell$.
The probability of failure with $i=3$ and $i=4$ already failed disks is 
\vspace{-1mm}
$$\alpha_3 =  \frac{ \binom{N}{2}n \ell }{ \binom{N}{4} }, \hspace{5mm}
\alpha_4 = \frac { [ \binom{N}{2} n (n^2 + 2n-2 )+ n^2) ]  \ell }  { \binom{N}{5} }$$

The analysis is extended to $\alpha_i = \mbox{min}(1,(i+2)\alpha_i), \hspace{2mm} i \geq 5$
and 3D cubes with $N_3=4^3$ data and 15 parity disks 
which are compared with 2d square $N_2 = 8^2 =64$ data and 18 parity disks.
The cube RAID provides a superior MTTDL with respect to square RAIDs
and RAID6 arrays with the same number of data disks.

\end{framed}

\section{Maximum Distance Separable RAID(4+K)}\label{sec:RAID6}

RAID(4+k) {\it Maximum Distance Separable - MDS} erasure-coded arrays 
uses the capacity of $k$ disks to tolerate $k$ disk failures,
which is the minimum redundancy known as the Singleton bound Thomasian and Blaum 2009 \cite{ThBl09}.
RAID(4+k) kDFT arrays for $k \geq 1$ can be implemented using the RS code and its variations,
while more efficient computationally efficient parity codes are known for $k=2$ and $k=3$

StorageTek's Iceberg was an early RAID6 product with RS coding,
adopting the {\it Log-Structure Array - LSA} paradigm, 
which is an extension of the {\it Log-structured File System - LFS} scheme 
Rosenblum and Ousterhout 1992 \cite{RoOu92}.
LSA accumulates a stripe's worth of data in the DAC cache, 
before the strips are written out as a full stripe.
This allows check strips to be computed on-the-fly.  
LSA with a single parity was compared to RAID5 in Subsection \ref{sec:compress}.
Conceptually LSA is not suitable for databases/OLTP  
applications with frequent updates to records (relational tuples),
since extra processing is required to update the indices of updated records not written in place.
StorageTek was acquired by Sun Microsystems, itself acquired by Oracle Corp. \newline
\begin{scriptsize}       
\url{https://www.oracle.com/it-infrastructure/}
\end{scriptsize}

{\it Thin Provisioning - TP} provides a method for optimizing utilization 
of available storage was first used in Iceberg, but has since been adopted by others. 
TP relies on on-demand allocation of blocks of data 
versus the traditional method of allocating all the blocks in advance. \newline
\begin{scriptsize}
\url{https://en.wikipedia.org/wiki/Thin_provisioning}
\end{scriptsize}
3PAR, an HPE subsidiary, advocates TP using {\it Dedicate on Write - DoW} 
rather than {\it Dedicate on Allocate - DoA} paradigm, to reduce disk space.          \newline
\url{https://web.archive.org/web/20061118125440/http://www.byteandswitch.com/document.asp?doc_id=35674}.      \newline
DoA paradigm used by IBM's MVS OS where a user had to specify an initial allocation and increments.

Three 2DFTS are EVENODD by Blaum et al. 2002 \cite{Bla+02} 
X-code by Xu and Bruck \cite{XuBr99}, and 
{\it Rotated Diagonal Parity - RDP} by Corbett et al. \cite{Cor+04}. 
EVENODD has been extended to $k=3$ most notably the STAR code in Huang and Xu 2005 \cite{HuXu05},
whose decoding complexity is  lower than comparable codes. 
RDP has also been extended twice to $k=3$ Thomasian and Blaum \cite{ThBl09}, 

RAID7.3 is an MDS 3DFT, whose coding details are not specified.
It is argued that 3DFT is required to deal with additional disk failures
in view of long rebuild times with HDDs (see Table \ref{tab:SSD}).
Given 12 TB drives 540 TB capacity is achieved by 12 RAID6 arrays each with six disks each
or RAID7.3 with data disks and 24 P and Q check disks,
but the same MTTDL is achieved by 45 data and only 3 check disks.     \newline 
\begin{scriptsize}
\url{https://www.hyperscalers.com/jetstor-raidix-1185mbs-storage-nas-san-record}
\end{scriptsize}

While EVENODD, RDP, and X-code are described in Thomasian and Blaum 1992 \cite{ThBl09}, 
but the latter two are described in this paper for the sake of self-completeness,
since they are discussed in Section \ref{sec:gridfiles} and  Section \ref{sec:singlediskrebuild}.

RM2 is a non-MDS 2DFT parity code where 1-out-m rows are dedicated to check blocks Park 1995 \cite{Park95}.
Blocks in the first $m-1$ rows each protected by two parities in the $m^{th}$ row, 
RM2 is described and evaluated in detail in Thomasian et al. 2007 \cite{ThFH07}. 
RM2 was patented at about the same time as EVENODD
it lost its importance given that it is not MDS and more complex rebuild for two failed disks.

Coding is applied to $m$ rows at a time, which are referred here as segments.
$m-1$ rows are dedicated to data blocks and the $m^{th}$ row to parity,
hence the redundancy level is $p=1/M$.
There are $2(m-1)$ data blocks associated with each parity block,
but each data block is protected by two parities as exemplified 
by Figure~\ref{fig:RM2} $N=7$ disks and $M=3$ rows.
Parity strips in the bottom two protect the data strips in top two rows.
$P_0$ protects a PG consisting of four strips:
$( D_{0,6}, D_{0,1}, D_{0,4}, D_{0,3} )$,
with the first two strips in row 1 and the other two strip in rows 2,
and $P_1$ protects $\left( D_{0,1}, D_{1,2}, D_{1,4}, D_{1.5} \right)$, etc.
Nite that RM2 is not MDS since $2/7 < p=1/3$.

\begin{figure}
\begin{center}
\begin{scriptsize}
\begin{tabular}{|c|*{6}{c|}}\hline  
Disk0 &Disk1 &Disk2 &Disk3 &Disk4 &Disk5 &Disk6                      \\ \hline \hline
$D_{2,3}$ & $D_{3,4}$ & $D_{4,5}$ & $D_{5,6}$ & $D_{6,0}$ & $D_{0,1}$ & $D_{1,2}$ \\ \hline
$D_{1,4}$ & $D_{2,5}$ & $D_{3,6}$ & $D_{4,0}$ & $D_{5,1}$ & $D_{6,2}$ & $D_{0,3}$ \\ \hline 
$P_{0}$ & $P_{1}$ & $P_{2}$ & $P_{3}$ & $P_{4}$ & $P_{5}$ & $P_{6}$         \\ \hline
\end{tabular}
\end{scriptsize}
\caption{RM2 data layout with $N=7$ disks and $M=3$ rows per segment,
with redundancy $p=1/M=33.3\%$. \label{fig:RM2}.}
\end{center}
\end{figure}

\subsection{Rotated Diagonal Parity - RDP Coded Arrays}\label{sec:RDP}

RDP is an MDS 2DFT code developed ten years after the EVENODD code 
and was adopted in NetApp's 2DFT array Corbett et al. 2004 \cite{Cor+04}. 
There are $p+1$ blocks in each row where the first $p-1$ blocks hold data,
the $p^{th}$ block holds a horizontal parity and $p+1^{st}$ block holds the diagonal parity.
The controlling parameter $p > 2$ should be a prime number.
When  a block is updated the horizontal parity is updated first and 
is used in updating the corresponding diagonal parity,  
RDP with $p=5$ is shown in Figure~\ref{fig:RDP}.

\begin{figure}[b]
\begin{footnotesize}
\begin{center}
\begin{tabular}{|c|c|c|c|c|c|}   \hline
Disk$_0$       &Disk$_1$       &Disk$_2$       &Disk$_3$       &Disk$_4$       &Disk$_5$    \\ \hline \hline
$d_{0,0}^0$    &$d_{0,1}^1$    &$d_{0,2}^2$    &$d_{0,3}^3$    &$d_{0,4}^4$    &$d_{0,5}$   \\ \hline
$d_{1,0}^1$    &$d_{1,1}^2$    &$d_{1,2}^3$    &$d_{1,3}^4$    &$d_{1,4}^0$    &$d_{1,5}$   \\ \hline
$d_{2,0}^2$    &$d_{2,1}^3$    &$d_{2,2}^4$    &$d_{2,3}^0$    &$d_{2,4}^1$    &$d_{2,5}$   \\ \hline
$d_{3,0}^3$    &$d_{3,1}^4$    &$_{3,2}^0$     &$d_{3,3}^1$    &$d_{3,4}^2$    &$d_{3,5}$   \\ \hline
\end{tabular}
\end{center}
\end{footnotesize}
\caption{\label{fig:RDP}Storage system with RDP code with $p=5$.
The superscripts are the diagonal parity groups on Disk$_5$. The parity of diagonal 4 is not stored.}
\end{figure}

Horizontal parity blocks hold the even parity of the data blocks in that row. 
Diagonal parity blocks hold the even parity of data and row parity blocks in the same diagonal.
Horizontal parities on Disk$_4$ and diagonal parities at Disk$_5$ are computed as follows 
A proof that two disk failures can be dealt with is given in the paper.

\vspace{-2mm}
\begin{align} \nonumber
d^{i,4} &= d^{i,0} \oplus d^{i,1} \oplus d^{i,2} \oplus d^{i,3}, i=0,3. \\ \nonumber
d_{0,5} &= d_{0,0} \oplus d_{3,2} \oplus d_{2,3} \oplus d_{1,4}      \\ \nonumber
d_{1,5} &= d_{0,1} \oplus d_{1,0} \oplus d_{3,3} \oplus d_{2,4}      \\ \nonumber
d_{2,5} &= d_{0,2} \oplus d_{1,1} \oplus d_{2,0} \oplus d_{3,4}      \\ \nonumber
d_{3,5} &= d_{0,3} \oplus d_{1,2} \oplus d_{2,1} \oplus d_{3,0}      \\ \nonumber
\end{align}

According to Theorem 2 in Thomasian and Blaum 2009 \cite{ThBl09} given an array of $n$ disks $2-2/n$ 
is the minimum number of XORs required to protect against two disk failures.
RDP protects $(p-1)^2$ data blocks using $2p^2 - 6 p + 4$ XORs.
Setting $n=p-1$ we have $2n^2 -2n$ XORs, hence RDP requires $2-2/n$ XORs per block.

EVENODD with $n$ data disk each with $n-1$ data blocks requires $(n-1)^2$ XORs 
to compute row parities and $(n-2)n$ XORs to compute diagonal parities .
EVENODD requires a further $n-1$ XORs to add the parity of a distinguished diagonal
to the parity of each of the other $n - 1$ diagonals.
$2n^2 -3n$ XORs are required to encode $n(n-1)$ blocks or $2-1/(n-1)$ XORs per block.    
Hence RDP outperforms EVENODD in the number of XOR operations. 

If the block size in RDP and EVENODD was a strip then a full stripe write 
would require updating the four diagonal parity strips.
This inefficiency is alleviated by selecting the prime number $p=2^n+1$,
which allows defining diagonal parities within a group of $2^n$ stripes.
This allows the block size for RDP to be the usual system block size divided by $2^n$.
If the system's disk block size is 4 KB we can select $p = 17$, so that sixteen 256 B blocks per strip.
This allows efficient processing of full stripe writes.
EVENODD developed a similar scheme, but used $p=2^{10}+1=257$.

It is shown in Blaum and Roth 1999 \cite{BlRo99} that the minimum update complexities 
of MDS codes for an $m \times n$ array with $k < n $ data and $r=n-k$ check columns are:  
\vspace{-1mm}
$$
\mbox{MDS single bit update-cost} = 
\begin{cases}
r=2   & 2 +\frac{1}{m} ( 1 {−} \frac{1}{k})                 \\
r=3   & 3 + \frac{3}{m} (\frac{2}{3} {−} \frac{1}{k}).
\end{cases}
$$

In the generator matrix $G = [I|P]$, the update cost can be viewed as the number of 1's per row in P.    \newline
\begin{scriptsize}
\url{https://en.wikipedia.org/wiki/Generator_matrix}                   \newline
\end{scriptsize}
One is deducted because the number of 1's per row in P instead of in G are counted. \newline
For $r=2$ (resp $r=3$) the equation is given as Proposition 5.2
(resp. 5.5) in the paper (Clarification by Prof. P. P.~C. Lee at CUHK).


\begin{figure}[t]
\begin{footnotesize}
\begin{center}
\begin{tabular}{|c||c|c|c|c|c|c|c|}\hline
    Row\# & $D_0$ & $D_1$ & $D_2$ & $D_3$ & $D_4$ & $D_5$ & $D_6$ \\ \hline\hline
    0     & $B_{0,0}^2$ & $B_{0,1}^3$ & $B_{0,2}^4$ & $B_{0,3}^5$ & $B_{0,4}^6$ & $B_{0,5}^0$ & $B_{0,6}^1$ \\ \hline
    1     & $B_{1,0}^3$ & $B_{1,1}^4$ & $B_{1,2}^5$ & $B_{1,3}^6$ & $B_{1,4}^0$ & $B_{1,5}^1$ & $B_{1,6}^2$ \\ \hline
    2     & $B_{2,0}^4$ & $B_{2,1}^5$ & $B_{2,2}^6$ & $B_{2,3}^0$ & $B_{2,4}^1$ & $B_{2,5}^2$ & $B_{2,6}^3$ \\ \hline
    3     & $B_{3,0}^5$ & $B_{3,1}^6$ & $B_{3,2}^0$ & $B_{3,3}^1$ & $B_{3,4}^2$ & $B_{3,5}^3$ & $B_{3,6}^4$ \\ \hline
    4     & $B_{4,0}^6$ & $B_{4,1}^0$ & $B_{4,2}^1$ & $B_{4,3}^2$ & $B_{4,4}^3$ & $B_{4,5}^4$ & $B_{4,6}^5$ \\ \hline
    5     & $B_{5,0}^0$ & $B_{5,1}^1$ & $B_{5,2}^2$ & $B_{5,3}^3$ & $B_{5,4}^4$ & $B_{5,5}^5$ & $B_{5.6}^6$ \\ \hline
    6     & & & & & & &\\ \hline
\end{tabular}
\hspace{5mm}
\begin{tabular}{|c||c|c|c|c|c|c|c|}\hline
\hline
      Row\# & $D_0$ & $D_1$ & $D_2$ & $D_3$ & $D_4$ & $D_5$ & $D_6$ \\ \hline\hline
    0 & $B_{0,0}^5$ & $B_{0,1}^6$ & $B_{0,2}^0$ & $B_{0,3}^1$ & $B_{0,4}^2$ & $B_{0,5}^3$ & $B_{0,6}^4$ \\ \hline
    1 & $B_{1,0}^4$ & $B_{1,1}^5$ & $B_{1,2}^6$ & $B_{1,3}^0$ & $B_{1,4}^1$ & $B_{1,5}^2$ & $B_{1,6}^3$ \\ \hline
    2 & $B_{2,0}^3$ & $B_{2,1}^4$ & $B_{2,2}^5$ & $B_{2,3}^6$ & $B_{2,4}^0$ & $B_{2,5}^1$ & $B_{2,6}^2$ \\ \hline
    3 & $B_{3,0}^2$ & $B_{3,1}^3$ & $B_{3,2}^4$ & $B_{3,3}^5$ & $B_{3,4}^6$ & $B_{3,5}^0$ & $B_{3,6}^1$ \\ \hline
    4 & $B_{4,0}^1$ & $B_{4,1}^2$ & $B_{4,2}^3$ & $B_{4,3}^4$ & $B_{4,4}^5$ & $B_{4,5}^6$ & $B_{4.6}^0$ \\ \hline
    5   &  & & &  &  &  & \\ \hline
    6 & $B_{6,0}^0$ & $B_{6,1}^1$ & $B_{6,2}^2$ & $B_{6,3}^3$ & $B_{6,4}^4$ & $B_{6,5}^6$ & $B_{6,6}^6$\\ \hline
\end{tabular}
\end{center}
\end{footnotesize}
\caption{ \label{fig:slope--1}
X-code array with $N=7$ disks with parity group for $p(i)=B(5,i), 0 \leq i \leq 6$ with slope=+1 
and parity group for $q(i) = B(6,j), 0 \leq i \leq 6$ at row 6 with slope=-1.}
\end{figure}

\subsection{X-code 2DFT Array with Diagonal Parities}\label{sec:xcode}

Given $N$ disks, which is a prime number, we have segments consisting of $N \times N$ strips.
Rows in a segment are indexed as $0:N-1$, rows $0:N-3$ hold data strips,
row $N-2$ holds P diagonal parities with positive slope, 
and row $N-1$ Q parities with negative slope, thus the redundancy level is $2/N$
and each data block is protected by two parity groups, which are diagonals with $\pm 1$ slopes.
hence we have an MDS array since it is shown in Xu and Bruck \cite{XuBr99}, 
where it is shown that recovery is possible for all two disk failures.

Figures~\ref{fig:slope--1} show the data layouts for $N=7$ columns.
$B_{i,j}$ denotes a data block (strip) at the $i^{th}$ row (stripe) and the $j^{th}$ column (disk).
PG memberships are specified by the number for the corresponding P and Q parities.
The two parities are independent and given that ${\langle{X}\rangle}_N \equiv X \ \mbox{mod} \ N $
then $p(i)$ and $q(i)$ for $0 \le i \le N-1$ are computed as follows:
\vspace{-1mm}
$$p(i) = B_{N-2,i} = \sum_{k=0}^{N-3} B_{k, {\langle{i-k-2}\rangle}_N},  \hspace{5mm}
q(i) = B_{N-1,i} = \sum_{k=0}^{N-3} B_{k, {\langle{i+k+2}\rangle}_N}.$$

The parities to be updated are determined by the {\it Parity Group - PG} to which the block belongs. 
The PG size is $N-1=6$ for both parities and either PG can be accessed to reconstruct a block,
so that the load increase in degraded mode is the same as RAID6.
Data recovery with two disk failures in X-code is investigated in Thomasian and Xu 2011 \cite{ThXu11},
where it is shown that the load increase is affected by the distance of two failed disks.
A balanced load increase can be attained by rotating successive $N \times N$ segments.

\subsection{Partial Maximum Distance Separable Code}\label{sec:PMDS}

{\it Partial MDS - PMDS} relies on local and global parities Blaum et al. 2013 \cite{BlHH13}.
Given an $m \times n$ array of sectors $r$ erasures can be corrected in a horizontal row.
The case of 1DFT $r=1$ with $p_i, 0 \leq i \leq 3$
and $s=2$ with two global parities $g_1$ and $g_2$ is shown in Figure~\ref{fig:PMDS},
Given six check sectors as many failed sectors can be corrected.

\begin{figure}[t]
\begin{center}
\begin{footnotesize}
\begin{tabular}{|c|c|c|c|c|c|c|}\hline
Column 1  & Column 2   & Column 3  &  Column 4  & Column 5   & Column 6  & Column 6  \\ \hline
$d_{0}$   & $d_{1}$    & $d_{2}$   &  $d_{3}$   & $d_{4}$    & $d_{5}$   &$p_0$      \\ \hline
$d_{6}$   & $d_{7}$    & $d_{8}$   &  $d_{9}$   & $d_{10}$   & $d_{11}$  &$p_1$      \\ \hline
$d_{12}$  & $d_{13}$   & $d_{14}$  &  $d_{15}$  & $d_{16}$   & $d_{17}$  &$p_2$      \\ \hline
$d_{18}$  & $d_{19}$   & $d_{20}$  &  $d_{21}$  & $g_1$      & $g_21$    &$p_3$      \\ \hline
\end{tabular}
\end{footnotesize}
\end{center}
\caption{\label{fig:PMDS}Layout of the local and global parities for Partial MDS with $r=1$ and $s=2$,
The four local parities are $p_i, 0 \leq i \leq 3$ and  $g_1$ and $g_2$ are the global parities.}
\end{figure}

\begin{description}
\item[Recoverable case I:] One drive failure: $(d_3, d_9, d_{15}, d_{21})$ 
and two additional sector failures: $d_2$ and $d_{12}$
\item[Recoverable case II:] $r=1$ failures per row $(d_2, d_11, d_{12}, d_{19}$
and two additional failures anywhere: $d_4$ and $d_19$
\item
[Recoverable case III:]
$d_2$ and $d_4$ in row zero and $d_12$ and $d_14$ in row 2.
$d_{11}$ and $d_{19}$ in row 1 and three.
$d_1$ ad $d_{12}$ can be recovered by $g_1$ and $g_21$ 
and then remaining blocks can be recovered using row parities.

\end{description}

RAID6 is an overkill to tolerate the failure of one disk and one sector 
according to Plank et al. 2023, Plank and Blaum 2014 \cite{PlBH13,PlBl14}.
{\it Sector Disk - SD} is an erasure code based on the PMDS concept 
to cope with storage systems really fail.
The decoding process in SD is based on straightforward linear algebra.
A brief summary is as follows.

\begin{description}

\item[PMDS:] Given a stripe defined by the parameters $(n,m,s,r)$ 
a PMDS code tolerates the failure of any $m$ blocks per row, and any additional blocks in the stripe. 
PMDS codes are maximally fault-tolerant codes that are defined with $m \times r$ local parity equations 
and $s$ global parity equations \cite{GHSY12}.  
However, PMDS codes make no distinction for blocks that fail together because they are on the same disk.

\item[SD:] Given a similarly defined stripe as PMDS 
SD code tolerates the failures of any $m$ disks (columns of blocks) plus any additional $s$ sectors in the stripe

\end{description}

For a given code construction, the brute force way to determine whether 
it is PMDS or SD is to enumerate all failure scenarios and test to make sure that decoding is possible. 
Given $n$ disk and $m$ as the number of coding disks
$s$ sectors per stripe are dedicated to coding, and there are $r$ rows per stripe.
A set of $m  \times r + s$ equations are set up each of which sums to zero.
For SD codes the number is: $ \binom{n}{m} \binom{r(n-m)}{s}$

There are many constructions which are valid as SD codes, but not as PMDS codes.
SD codes can recover the failure of a column 4 $(d_3, d_9, d_{15}, d_{21})$ and $d_3$ and $d_12$,
but not not aforementioned case II.


``We provided a direct methodology and constructive algorithms to implement a universal and complete solution
to the recoverability and non-recoverability of lost sectors. 
This method and algorithm meets the User Contract that says that what is theoretically recoverable shall be recovered. 
Our solution can be applied statically or incrementally. 
We demonstrated the power of the direct method by showing how it can recover data in lost sectors 
when these sectors touch more strips in the stripe than the fault tolerance of the erasure code. 
The direct method can be joined with any code-specific recursive algorithm 
to address the problem of efficient reconstruction of partial strip data. 
Alternatively, the incremental method can be reversed when some data is recovered 
to provide a completely generic method to address this same partial strip recovery problem. 
Finally, we provided numerical results that demonstrate significant performance gains 
for this hybrid of direct and recursive methods'' Hafner et al. 2005 \cite{HDRT05}. 

\subsection{Tiger: Disk Adaptive Redundancy Scheme}\label{sec:tiger}

Tiger {\it Disk-Adaptive Redundancy - DAR} scheme dynamically tailors itself 
to observed disk failure rates Kadekodi et al. 2022 \cite{Kad+22}. 
Tiger results in significant space and cost savings with respect to existing DAR schemes, 
which are constrained in that they partition disks into subclusters with homogeneous failure rates.  

The earlier Pacemaker DAR scheme is less desirable 
from the viewpoint of the fraction of viable disks for higher n-of-k's 
as shown in Figure 2a for 7-of-9, 14-of-17, 22-of-25, and 30-of-33.
Pacemaker also reduces intra-stripe diversity 
and is more susceptible to unanticipated changes in a make or model's failure rate and 
only works for clusters committed to DAR.

Tiger avoids constraints by introducing eclectic stripes 
in  which redundancy is  tailored to  diverse disk failure rates. 
Tiger ensures safe per-stripe settings given that device failure rates change over time. 
Evaluation of real-world clusters shows that Tiger provides better space-savings, 
less bursty IO to change redundancy schemes and better robustness than prior DAR schemes.

Appendix A in \cite{KSCM23} derives MTTDL for eclectic stripes using the Poisson binomial distribution,
whose disks have different failure rates.  \newline
\begin{scriptsize}
\url{https://en.wikipedia.org/wiki/Poisson_binomial_distribution}
\end{scriptsize}

\section{Multidimensional Coding for Higher Reliability}\label{sec:multiD}

Several multidimensional coding methods are presented starting 
with Grid files in Subsection \ref{sec:gridfiles}, 
a disk array referred to as RESAR in Subsection \ref{sec:RESAR}, 
and 2D and 3D arrays for archival storage in Subsection \ref{sec:2d}.

\subsection{Grid Files}\label{sec:gridfiles}

Grid files by Li et al. 2009 in \cite{LiSZ09} use coding in two dimensions to protect data strips.
The Horizontal (H) and Vertical (V) {\it Parity Code - PC} or HVPC code 
is the simplest Grid code over $k_1$ rows and $k_2$ columns:
\vspace{-1mm}
$$H_{i, k_2} = D_{i,1} \oplus D_{i,2} \oplus \ldots  D_{i, k_2 }, \hspace{5mm}i = 1, . . . , k_1 $$
\vspace{-1mm}
\[ V_{k_1, j} = D_{1, j} \oplus D_{2, j} \oplus \ldots D_{k_1, j},  \hspace{5mm}j =  1, . . . , k_2 \]
\vspace{-1  mm}
$$P_{k_1,k_2}= 
\bigoplus_{i=1}^{k_1} H_{i,k_2} = 
\bigoplus_{j=1}^{k_2} V_{k_1,j} = 
\bigoplus_{i=1}^{k_1} \bigoplus_{j=1}^{k_2} D_{i,j} $$ 

If the fault-tolerance of row and column  codes are given 
as is $t_r$ and $t_c$ the total number of faults tolerated is:

\vspace{-2mm} 
\begin{eqnarray}\label{eq:FT}
t_{ft} = ( t_r+1 ) ( t_c+1 ) -1.
\end{eqnarray}
Informally, the strips of a rectangle with $(t_r+1)(t_c+1)$ strips cannot be recovered,
but just removing one failed strip allows recovery, hence the minus one in Eq. \ref{eq:FT}.  
An approximation (lower bound) to the reliability of grid files 
can be obtained by considering $k=t_{ft}$ in Eq. \ref{eq:RAID(4+k)}.
In fact the summation can be carried up to the maximum number 
of failures tolerated for cases that do not lead to data loss
with the reliabilities multiplied by the probability that so many failures can be tolerated.
as given in Newberg and Wolf 1994 \cite{NeWo94}.

In addition to {\it Single Parity Code - SPC} or RAID5, 
EVENODD, STAR, and X-code codes are considered in this study.
RDP which is computationally more efficient than EVENODD should be added to this list.
SPC or RAID5 requires one row or column, EVENODD, RDP, X-code require two, and STAR code three. 
X-code requires a prime number of disks and unlike other codes is vertical,
but can be placed horizontally and vice-versa.

In Figure \ref{fig:2xcodes} we consider a $p \times p$ X-code array,
which is protected by a horizontal SPC code with P' parities. 
There is data update, two parity updates for the X-code and three updates for the SPC code, total of five.


\begin{figure}[h]
\begin{center}
\begin{footnotesize}
\begin{tabular}{|c|c|c|c|c||c|c|}\hline
Column\_1      &Column\_2      &Column\_3  &Column\_4         &Column\_5          &Column\_6    \\ \hline \hline
$D_{1,1}$      &$D_{1,2}$      &$\ldots$   &$D_{1,p-2}$       &${P}_{1,p-1}$      &${Q}_{1,p}$  \\ \hline
$D_{2,1}$      & $D_{2,2}$     &$\ldots$   &$D_{2,p-2}$       &${P}_{2,p-1}$      &${Q}_{2,p}$  \\ \hline
$\vdots$       &$\vdots$       &$\ddots$   &$\vdots$          &$\vdots$           &$\vdots$     \\ \hline
$D_{p,1}$      &$D_{p,2}$,     &$\ldots$   &$D_{p,p-2}$       &${P}_{p,p-1}$      &${Q}_{p,p}$  \\ \hline   \hline
${P'}_{p+1,1}$ &${P'}_{p+1,2}$ &$\ldots$   &${P'}_{p+1,p-2}$  &${P'}_{p+1,p-1}$   &${P'}_{p+1,p}$ \\ \hline
\end{tabular}
\end{footnotesize}
\end{center}
\caption{ \label{fig:2xcodes}Data in the first $p$ rows is protected by an X-code with P and Q parities 
in columns $p+1$ and $p+2$. 
(left upper corner) is inside a ($(p+2)\times(p+2)$) X-code with P and Q parities.
The $2 \times p$  space in upper right hand corner with $D'$s is only covered by the larger X-code.
Data block updates affect $P$ and $Q$ parities and the P' parities.}
\end{figure}

Data recovery in Grid files is carried alternating between row-wise and column-wise recovery
and stops when there are no further fault strips (refer to Figure 6 in the paper).

\subsection{Reliable Storage at Exabyte Scale - RESAR}\label{sec:RESAR} 

RESAR has parity groups weaving through a 2-D disk array Schwarz et al. 2016 \cite{SAK+16} 
As shown in Figure \ref{fig:RESAR} has disklets numbered 
such that coordinates can be converted to a single number:
\vspace{-1mm}
$$(i,j) = i \times  (k+2) +j, \hspace{3mm} 0 \leq j \leq k+1 \mbox{ with } k=4, i=2, \ldots k+1$$  
Disklet $39 = 6 \times 6 +3$ is protected by P6 horizontally and D7 diagonally.
If disklets 33, 35,  38, and 40 were placed on the same disk or disks on the same rack, 
then the failure of this disk or rack causes data loss, 
since the four disklets share two reliability stripes. 
Details on achieving appropriate placements are given in the paper.

\begin{figure}[h] 
\centering	
\begin{footnotesize}
\begin{tabular}{|c|c|c|c|c|c|}\hline
    &2  &3 &4 &5 &$P_0$   \\ \hline
$D_0$ &8 &9 &10 &11 &$P_1$  \\ \hline
$D_1$ &14 &15 &16 &17 &$P_2$ \\ \hline
$D_2$ &20 &21 &22 &23 &$P_3$ \\ \hline 
$D_3$ & 26 & 27 &28 &29 &$P_4$ \\ \hline
$D_4$ &32 &33 &34 &35 &$P_5$  \\ \hline
D$_5$ &38 &39 &40 &41  &P$_6$ \\ \hline
D$_6$ &44 &45 &46 &47 &P$_7$ \\ \hline 
D$_7$ &50 &51 &52 &53 &P$_8$ \\ \hline
D$_8$ &56 &57 &&& P$_9$ \\ \hline
D$_9$ &&&&&             \\ \hline
\end{tabular}
\end{footnotesize}
\caption{\label{fig:RESAR}A small bipartite RESAR layout
connecting $P_i$ and $d_i$ parities for $0 \leq i \leq 9$ via intervening disklets.
Each data disk is protected by a P parity in its row 
and a D parity at a lower row as member of diagonal parity group.}
\end{figure}

RESAR-based layout with 16 data disklets per stripe has about 50 times lower probability of suffering data loss
in the presence of a fixed number of failures than a corresponding RAID6 organization.
The simulation to estimate the probability of data loss for 100,000 disks required 10,000 hours.

\subsection{2D- and 3-Dimensional Parity Coding}\label{sec:2d}

2D and 3D parity protection for disk arrays is discussed in Paris and Schwarz 2021 \cite{PaSc21}
A small example of 2D protection is given in Figure~\ref{fig:3D}.

There are the following parity groups: 
\vspace{-1mm}
$$(P_1, D_1, D_3, D_8), (P_2, D_1, D_2, D_5, D_9), (P_3, D_3, D_4, D_7, D_{10}),
(P_4, D_3, D_4, D_6, D_9), (P_5, D_5, D_6, D_7, D_8, D_9).$$

Three disk failures can result in data loss in this case.
Two examples are $(P_1, P_2, D_1)$, and $(D_1,D_5,D_8)$.
In the first case there is data loss since $D_1$ loses both of its parities.
In the second case there are two failures in the three parity groups, 
which can support only one disk failure.

\begin{figure}[h]
\centering
\begin{footnotesize}
\begin{tabular}{|c|c|c|c|c|c|}\hline
       &         &$P_1$    &         &         &       \\       \hline
       &$P_2$    &         &$P_3$    &         &       \\       \hline
$P_4$  &         &         &         &$P_5$    &                \\       \hline \hline
       &         &$D_1$    &         &         &       \\       \hline
       &$D_2$    &         &$D_3$    &         &$P_6$           \\       \hline \hline  
$P_7$  &         &         &         &$P_8$    &                \\       \hline
       &         &$D_4$    &         &         &       \\       \hline
       &$D_5$    &         &$D_6$    &         &$P_9$           \\       \hline \hline 
$\vdots$  &$\vdots$ &$\vdots$ &$\vdots$ &$\vdots$  &$\vdots$      \\       \hline
\end{tabular}
\end{footnotesize}
\caption{\label{fig:3D}3D parity where each data disk is protected 
by three parities one vertically and two diagonally.}  
\end{figure}

A small example with 3-D parities with nine  parity groups is given in Figure \ref{fig:3D}.
$$(P_1, D_1, D_4), (P_1, D_2, D_5), (P_3, D_3, D_6),\hspace{1mm}
(P_4, D_1, D_3), (P_5,D_1, D_2), (P_6, D_2,D_3),  \hspace{1mm}
(P_7, D_4, D_6), (P_7 ,D_4, D_5), (P_9,D_5,D_6),$$ 

As few as four disk failures can result is data loss with 3D coding, 
e.g., $(P_2, P_5, D_2, P_6)$, since $D_2$ has lost the three parities protecting it.

\section{Local Redundancy Code}\label{sec:LRC}

The {\it Local Redundancy Code - LRC} adds a few parity disks 
to provide speedier recovery by reducing the number of disks involved in recovery.
There is also  reduced data transmission cost in distributed storage systems with a large number of disks.
LRC differs from CRAID in Section \ref{sec:CRAID} which achieves faster rebuild by parallelizing disk accesses.

An early example of LRC is the (8,2,2) Pyramid code Huang et al. 2007 \cite{HuCL07},
which has two local parities, one over first four and the second over the next four disks; 
and the two global parities are over all eight disks.
The recovery cost due to local parities is lower as shown in Table~\ref{tab:pyramid}.
also discussed in Section \ref{sec:pyramid} 

\begin{figure}
\begin{center}
\begin{footnotesize}
\begin{tabular}{|c|c|c|c|c|}      \hline
$d_1$   &$d_2$   &$d_3$  &$d_4$   &$c_{1,1}$ \\ \hline
$d_5$   &$d_6$   &$d_7$  &$d_8$   &$c_{1,2}$ \\ \hline
$c_2$   &$c_3$   &     &        &            \\ \hline
\end{tabular}
\end{footnotesize}
\end{center}
\caption{\label{fig:pyramid}An (8,2,2) pyramid code with 8 data and 4 check disks.}
\end{figure}

\begin{table}
\begin{footnotesize}
\begin{center}
\begin{tabular}{|c|c|c|c|c|c|}\hline
\# of failed blocks &                   &1      &2     &3       &4           \\ \hline \hline
MDS code            &recovery (\%)      &100    &100   &100     &0           \\ \hline
(11,8)              &read overhead      &1.64   &2.27  &2.91    &-           \\ \hline \hline
Pyramid code        &recovery (\%)      &100   &100   &100.0    &68.89        \\ \hline
(12,8)              &read overhead      &1.25  &1.74  &2.37     &2.83         \\ \hline
\end{tabular}
\end{center}
\end{footnotesize}
\caption{\label{tab:pyramid}Comparison between MDS code (11,8) and Pyramid code (12,8)}
\end{table}

{\bf Windows Azure Storage - WAS} improves over Pyramid codes
by reading less data from more fragments Huang et al. 2012 \cite{HSX+12}.
With the (12,2,2) code 100\% of 3 and 86\% of 4 disk failures are recoverable.
\vspace{-1mm}
$$\left( X_0, X_1,X_2, X_3, X_4, X_5, X_6, p_X \right),
\left( Y_0, U_1, Y_2, Y_3, Y_4, Y_5, Y_6, p_Y \right), P_0, P_1 [\forall X, \forall Y ]$$

WAS design choices are justified in Table \ref{tab:WAS}:

\begin{table}
\begin{footnotesize}
\begin{center}
\begin{tabular}{|c|c|c|c|}\hline
Scheme         &Storage    &Reconstruction   &Savings       \\
               &overhead   &cost             &              \\ \hline \hline
RS(6+3)        &1.5        &6                &              \\ \hline
RS(12+4)       &1.29       &14               &14\%          \\ \hline
LRC(12,2,2)    &1.29       &7                &14\%          \\ \hline
\end{tabular}
\end{center}
\end{footnotesize}
\caption{\label{tab:WAS}Choice of Windows Azure Storage}
\end{table}

We continue the discussion with a smaller (6,2,2) example
for which the MTTDL is $2.6\times 10^{12}$, 
versus $3.5 \times 10^9$ for replication
and $6.1 \times 10^{11}$ for (6.3) RS. 
i.e., a ten fold increase in MTTDL at the cost of an extra disk.

\vspace{-2mm}
$$ \left( X_0,X_1,X_2,p_X \right), \left( Y_0, Y_1, Y_2,p_Y \right), p_0,p_1 $$

This LRC can tolerate arbitrary 3 failures by using the following sets of coefficients:

\vspace{-2mm}
$$q_{x,0} = \alpha_0 x_0 + \alpha_1 x_1 + \alpha_2 x_2, \hspace{5mm}
q_{x,1} = \alpha_0^2 x_0 + \alpha_1^2 x_1 + \alpha_2^2 x_2 \hspace{5mm} q_{x,2} =  x_0 + x_1 + x_2 $$

\vspace{-2mm}
$$q_{y,0} = \beta_0 y_0 + \beta_1 y_1 + \beta_2 y_2, \hspace{5mm}
q_{y,1} = \beta^2_0 y_0 + \beta_1^2 u_1 + \beta_2^2 u_2 \hspace{5mm} q_{y,2} =  y_0 + y_1 + y_2 $$

\vspace{-2mm}
$$p_0 = q_{x,0}+q_{y,0}, \hspace{5mm} p_1 = q_{x,1}+q_{y,1}, \hspace{5mm} p_x = q_{x,2}, \hspace{5mm} p_y = q_{y,2} $$

The values of $\alpha$s and $\beta$s are chosen such that the LRC can decode all decodable four failures.

For disk failure rates: $\lambda$ and repair rates $\rho$ we have the following transitions
among the states of a Markov reliability model, see e.g., Trivedi 2002 \cite{Triv02}.

\vspace{-2mm}
$${\cal S}_{i} \rightarrow {\cal S}_{i-1}: i \lambda, \hspace{5mm}  10 \leq i \leq 8 \hspace{5mm}
{\cal S}_{i} \leftarrow {\cal S}_{i-1}: \rho_i \hspace{5mm}  9 \leq i \leq 7$$

Let $p_d$ be the fraction of decodable four disk failures

\vspace{-2mm}
$${\cal S}_7  \rightarrow 7 {\cal S}_6: \lambda p_d, \hspace{5mm}
{\cal S}_7 \rightarrow  {\cal S}_{Failed}: 7 \lambda (1- p_d) $$

With $N$ storage nodes with capacity $S$ and network bandwidth $B$
the average repair rate of a single failure is $\rho_9=\epsilon (N-1) B / S C$,
where $\epsilon$ is the fraction of network bandwidth at each node.
This is yet another option for rebuild processing.
It takes 3 fragments to repair the 6 data fragments and 2 local parities.
It takes 6 fragments to repair the 2 global parities.
The {\it Average Repair Cost - ARC} is: $C = (3 \times 8 + 6 \times 2 )/10=3.6$
Enumerating all failure patterns $p_d=0.86$.
The detection and triggering time $T=30$ minutes dominates repair time for more than one fragment failure,
so that $\rho_8=\rho_7=\rho_6=1/T$, $N=300$, $S=16$ TB, $B=1$ GB per second, $\epsilon=0.1$

\subsection{A Systematic Comparison of Existing LRC Schemes}\label{sec:LRCcomparison}

Xorbas, Azure's LRCs, and Optimal-LRCs are compared using the {\it Average Degraded Read Cost - ADRC}, 
and the {\it Normalized Repair Cost - NRC} in Kolosov et al. 2018 \cite{KYL+18}.
The trade-off between these costs and the code's fault tolerance offer different choices.

There are two  types  of $(n,k,r)$ LRCs.
With information-symbol locality, only the data blocks can be repaired in a local fashion by $r$ blocks,
while global parities require $k$ blocks for recovery.
Pyramid and Azure-LRC are examples of so-called data-LRCs.

In codes with all-symbol locality, all the blocks,
including the global parities, can be repaired locally from $r$ surviving blocks.
Xorbas discussed in Subsection \ref{sec:Xorbas} is an example of a Full-LRC code.

The experimental evaluation on a Ceph cluster deployed on Amazon {\it Elastic Compute Cloud - EC2} \newline
\url{https://en.wikipedia.org/wiki/Amazon_Elastic_Compute_Cloud}           \newline
demonstrates the different effects of realistic network and
storage bottlenecks on the benefit from various LRC approaches.
Ceph is an open-source software-defined storage platform 
that implements object storage on a single distributed computer cluster and 
provides 3-in-1 interfaces for object-, block-, and file-level storage Maltzahn et al. 2010 \cite{MMK+10}.
The study shows the benefits of full-LRCs and data-LRCs depends
on the underlying storage devices, network topology, and foreground application load.


The normalized repair cost metric can reliably identify
the LRC approach that would achieve the lowest repair cost in each setup.
The {\it Average Repair Cost - ARC} is based on the assumption
that the probability of repair is the same for all blocks:
$$ARC = \sum_{i=1}^n \mbox{cost}(b_i) / n.$$
In the case of (10,6,3) Azure code is $ARC = (8 \times 3) +(2 \times 6) /10 = 3.6$.

The {\it Normalized Repair Cost - NRC} which amortizes the cost 
of repairing parity blocks over data blocks is given as:
\vspace{-1mm}
$$NRC = ARC \times\frac{n}{k}, \hspace{5mm} NRC = [ (8 \times 3) + (2 \times 6)]/6=6.$$

{\it Degraded Read Cost -DRC} is the full-node repair cost and
which for (10,6,3) Azure LRC is $6 \times 3 / 6 =3$.

Xorbas (16,10,5) has RS-based four global parities
which can be recovered from local parities via an implied parity block as shown in Figure 2 in the paper.
\vspace{-1mm}
$$( X_1, X_2, X_3, X_4, X_5, P_{1:5}, (X_6, X_7, X_8, X_9, X_{10}, P_{6:10})$$

\subsection{Combining Parity and Topology Locality in Wide Stripes}\label{sec:widestripe}

Wide stripes are used to minimize redundancy.
There are $n$ chunks, $k$ data chunks, and $m=n-k$ which is 3 or 4 parity chunks \cite{HCY+21}.
For the ten examples given by Table 1: $1.18 \leq \mbox{redundancy}=n/k  \leq 1.5$.
The notation used in the paper is given in Table \ref{tab:notation}.

\begin{table}[h]
\begin{footnotesize}
\begin{center}
\begin{tabular}{|c|c|}\hline
Notation            & Description                                   \\ \hline \hline
$n$                 &total \# of chunks of a stripe                 \\ \hline
$k$                 &total \# of data stripes of a chunk            \\ \hline
$r$                 &\# of retrieved chunks to repair a lost chunk  \\ \hline
$z$                 &\# of racks to store a stripe                  \\ \hline
$c$                 &\# of chunks of a stripe in a rack             \\ \hline
$f$                 &\# of tolerable node failures of a stripe      \\ \hline
$\gamma$            &maximum allowed redundancy                     \\ \hline
\end{tabular}
\end{center}
\end{footnotesize}
\caption{Notation for combined locality\label{tab:notation}}
\end{table}

Challenges for wide stripes are:
(i) Requiring $k$ chunks to be retrieved is expensive,
(ii) CPU cache may not be able to hold large number of chunks,
(iii) updating $n-k$ check chunks is expensive.

There is a trade-off between redundancy and repair penalty:
(i) Parity locality incurs high redundancy.
(ii) Topology locality incurs high cross-track repair bandwidth.
The Azure LRC (32,20,2) is given in Figure \ref{fig:AzureLRC}.

\begin{figure}
\begin{footnotesize}
\begin{center}
\begin{tabular}{|c|c|c|c|c|c|c|c|c|c|c|}\hline
$D_1$ & $D_3$ &$D_5$ & $D_7$ & $D_9$ & $D_{11}$ & $D_{13}$  &$D_{15}$ &$D_{17}$    &$D_19$   &$Q_1$[1:20]     \\ \hline
$D_2$ & $D_4$ &$D_6$ &  $D_8$  & $D_{10}$  & $D_{12}$ & $D_{14}$  &$D_{16}$ &$D_{18}$  &$D_{20}$ &$Q_2$[1:20] \\ \hline
$P_1$ [1,2]   & $P_2$ [3,4]   & $P_3$[5,6] & $P_4$ [7,8] & $P_5$ [9,10]  & $P_6$[11,12]
& $P_7$ [13, 14] & $P_8$ [15,16] & $P_9$[17-18] & $P_{10}$[19-20] &-                                           \\ \hline
\end{tabular}
\end{center}
\end{footnotesize}
\caption{\label{fig:AzureLRC}Azure parity locality LRC (n,k,r)=(32, 20,2).}
\end{figure}

\begin{table}[h]
\begin{footnotesize}
\begin{center}
\begin{tabular}{|c|c|c|}\hline
-                &$(n,k,r)$                             &$(16,10,5)$     \\ \hline \hline
Azure-LRC        &$f= n-k+ \lceil k/r \rceil + 1$       & $f=5$          \\ \hline
Xorbas           & $f \leq n-k - \lceil k/r\rceil +1$   & $f=4$          \\ \hline
Optimal LRC      & $f \leq n-k - \lceil k/r\rceil +1$   & $f=4$          \\ \hline
Azure LRC+1      & $f \leq n-k - \lceil k/r\rceil$      & $f=4$          \\ \hline
\end{tabular}
\end{center}
\end{footnotesize}
\caption{Number of tolerable node failures $f$ for different LRCs for (n,k,r)=(16.10,5)}
\end{table}

$(n,k,r,z)=(26,20,5,9)$, where $z=9$ is the number of elements in a column is given in Figure \ref{fig:fig2}


\begin{figure}
\begin{footnotesize}
\begin{center}
\begin{tabular}{|c c c|}\hline
$D_1$      &$D_2$       &$D_3$        \\
$D_4$      &$D_5$       &$P_1$[1:5]   \\ \hline
$D_6$      &$D_7$       &$D_8$        \\
$D_{9}$    &$D_{10}$    &$P_2$[6:10]  \\ \hline
$D_{11}$   &$D_{12}$    &$D_{13}$     \\
$D_{14}$   &$D_{15}$    &$P_3$[11,15] \\ \hline
$D_{16}$   &$D_{17}$    &$D_{18}$     \\
$D_{19}$   &$D_{20}$    &$P_4$[16:20] \\ \hline
$Q_1$[1:20] & $Q_2$[1:20] & - \\         \hline
\end{tabular}
\end{center}
\end{footnotesize}
\caption{\label{fig:fig2}Combined locality (25,20,5,9).}
\end{figure}

\subsection{Practical Design Considerations for Wide LRCs}\label{sec:practical}

The following discussion is based on Kadekodi et. al. 2023 \cite{KSCM23}.
With 4 data blocks $(d_1,d_2,d_3,d_4)$, 2 global parities $c_1$ and $c_2$, and three local parities
$\ell_1 = d_1 \oplus d_2$, $\ell_2 = d_3 \oplus d_4$, $\ell_g = c_1 \oplus c_2$.
The generator matrix is:

$$
G_C =
\begin{pmatrix}
1 &0  &0  &0                         \\
0 &1  &0  &0                         \\
0 &0  &1  &1                         \\
0 &0  &0  &1                         \\
c_{1,1} &c_{1,2} & c_{1,3} & c_{1,4} \\
c_{2,1} &c_{2,2} & c_{2,3} & c_{2,4} \\
1 &1 &0 &0                           \\
0 &0 &1 &1                           \\
c_{1,1}+c_{2,1} & c_{1,2}+c_{2,2} & c_{1,3}+c_{2,3} & c_{1,4}+c_{2,4} \\
\end{pmatrix}
\begin{pmatrix}
d_1 \\ d_2 \\ d_3 \\ d_4 \\ \ell_1 \\ \ell_2 \\ \ell_g
\end{pmatrix}
$$

\begin{table}
\begin{footnotesize}
\begin{center}
\begin{tabular}{|c|c|c|c|c|c|}\hline
Scheme             &n      &k     &r    &p     &Rate             \\ \hline \hline
S1: 24-of-28       &28     &24    &2    &2     &0.857            \\
S2:48-of-55        &55     &48    &3    &4     &0.872            \\
S3: 72-of-80       &80     &72    &4    &4     &0.9              \\
S4: 96-of-105      &105    &96    &5    &4     &0.914            \\ \hline
\end{tabular}
\end{center}
\end{footnotesize}
\caption{\label{sec:comp}
Wide LRC schemes used to compare different LRC constructions with low level of redundancy $k/n > 0.85$.
$p$ is the \# of parity groups and local parities and $r$ the \# of global parities.}
\end{table}

Four different 48-of-55 Azure LRC constructions 
with four local parities and 3 global parities are specifiable as follows.
\hspace{-1mm}
$$c_j = \bigoplus_{12 \times (j-1) + 1}^{12 \times j}, 1 \leq j \leq 4,
\hspace{2mm} rs_1, \hspace{2mm} rs_2, \hspace{2mm} rs_3 $$

Azure-LRC+1 forms a local group of the global parities and protects them using a local parity.
\vspace{-1mm}
$$c_j =\bigoplus_{16 \times (j-1) +1}^{12 \times j}, 1 \leq j \leq 3, \hspace{2mm} c_{rs} = \bigoplus_{j=1}^3 rs_j $$

In the case of 48-of-55 optimal Cauchy LRC the four local parities cover, similarly to Azure parities,
but global parities cover parity blocks (as in Figure 4 (c)).

In the case of uniform Cauchy LRC the $p$ local parities 
are uniformly distributed across the $k$ data blocks
and they are all XORed with the global parity checks.
\vspace{-1mm}
$$
c_1 = \bigoplus _{i=1}^{12} d_i \hspace{5mm}
c_2 = \bigoplus _{i=13}^{25} d_i \hspace{5mm}
c_3 = \bigoplus _{i=26}^{38} d_i \hspace{5mm}
c_4 = [\bigoplus _{i=38}^{48} d_i] \oplus [ \bigoplus_{j=1}^3 rs_j ]. 
$$

The average cost of reconstructing any of the data blocks {\it Average Degraded Read Cost - ARC} 
and two other measures are defined  as follows:
\vspace{-1mm}
$$ ADRC = \sum_{i=1}^k \mbox{cost} ({b}_i / k),
\hspace{5mm}
ARC_1 = \sum _{i=1}^n b_i /n
\hspace{5mm}
ARC_2 =\sum_{i=1,j \neq i}^n \mbox{cost} (b_{i,j} / \binom{n}{2}.$$

Average repair or reconstruction cost takes into account local
and global parity blocks in the computation as ARC$_1$ and 
ARC$_2$ is used as the cost of reconstructing 2 blocks.
Best schemes for various metrics are given in Table \ref{sec:best}.


\begin{table}[h]
\begin{footnotesize}
\begin{center}
\begin{tabular}{|c|c|}\hline
Locality ($\ell$)   &Uniform Cauchy LRC S1: 13, S2: 13, S3: 19, S4: 26               \\ \hline
ADRC                &Azure LRC -  S1: 12, S2=12, S3=18, S4=24                        \\ \hline
ARC1                &Azure LRC - S1: 12.85, S2: 12,76, S3:19, S4: 25.22              \\ \hline
ARC2                &Uniform Cauchy - LRC S1: 27.92; S2: 33.85, S3: 49.22, S4 67.69  \\ \hline
Normalized MTTDL    &S1, 1., S2, 1.01*  S3, 1.0  S4: 1.0                             \\ \hline
\end{tabular}
\end{center}
\end{footnotesize}
\caption{\label{sec:best}Best schemes based on analytic metrics used to compare different LRC constructions.
Results are normalized with respect Uniform Cauchy. *Optimal Cauchy LRC.}
\end{table}

The Markov chain to estimate the MTTDL for a 48-of-55 disk array can be specified as follows.
Given that $i$ denotes the number failed disks a failure at ${\cal S}_i$
leads to a transition to ${\cal S}_{i+1}$ for $i < 5$.
Some of the transition for $5 \leq i \leq 7$ lead to failure.
Transition probabilities are based on AFR.
The MTTDL can be determined as the mean number of visits to each state 
and the holding time: $[(55-i)\delta]^-1$ where $\delta$ is the failure rate.

\section{Schemes to Reduce Rebuild Traffic in Distributed RAID}\label{sec:distr}

A survey of network codes for distributed storage is given in Dimakis et al. 2011 \cite{DRWS11}.
As an example consider a (4,2) MDS code is given in Figure~\ref{fig:MDS42},
which can tolerate any two disk failures.

\begin{figure}
\begin{center}
\begin{footnotesize}
\begin{tabular}{|c|c|c|c|}\hline
Node$_1$   &Node$_2$      &Node$_3$        &Node$_4$          \\ \hline \hline
$A_1$      &$B_1$         &$A_1+B_1$       &$A_2+B_1$         \\ \hline
$A_2$      &$B_2$         &$A_2+B_2$       &$A_1+A_2+B_2$     \\ \hline
\end{tabular}
\end{footnotesize}
\end{center}
\caption{\label{fig:MDS42}An MDS (4,2) code which can tolerate two disk failures.}
\end{figure}

Assuming Node$_1$ fails then
$$B_2 \oplus (A_2 \oplus B_2) \rightarrow A_2 \hspace{5mm} 
(A_2  \oplus B_2) \oplus (A_1 \oplus A_2 \oplus B_2) $$

Assuming Node$_4$ fails then at Node$_1$ we compute $(A_2 \oplus B_2)$ at Node$_2$:
\vspace{-1mm}
$$  (A_2 \oplus B_2) \oplus (B_1\oplus B_2) \rightarrow (A_2 \oplus B_1)$$
\vspace{-1mm}
$$ A_1 \oplus (A_2 \oplus B_2) \rightarrow (A_1\oplus A_2 \oplus B_2)$$

To reproduce broken Node$_1$ and Node$_2$ we can proceed as follows:        \newline

\noindent
$(A_2 \oplus B_2) + (A_1 \oplus A_2 \oplus B_2) \rightarrow A_1) \rightarrow A_1$, \newline            
$A_1  \oplus (A_1 \oplus B_1) \rightarrow B_1$                                     \newline
$(A_1 \oplus B_1)  \oplus (A_2 \oplus B_1) \rightarrow (A_1 \oplus A_2)$           \newline
$(A_1 \oplus A_2) \oplus (A_1 \oplus A_2 \oplus B_2) \rightarrow B_2$              \newline
$B_2 \oplus (A_2 \oplus B_2) \rightarrow A_2$

Rashmi et al. 2013 \cite{RSG+13} use Piggybacking to reduce the volume of transferred data. 
Consider four nodes with two bytes of data each as shown in Figure \ref{fig:Rashmi}

\begin{figure}
\begin{footnotesize}
\begin{center}
\begin{tabular}{|c|c|c|}\hline
node 1    &$a_1$        &$b_1$                     \\ \hline
node 2    &$a_2$        &$b_2$                     \\ \hline
node 3    &$a_1+a_2$    &$b_1+b_2$                 \\ \hline
node 4    &$a_1+2a_2$   &$b_1+2b_2+a_1$            \\ \hline
\end{tabular}
\end{center}
\end{footnotesize}
\caption{A toy example of piggybacking illustrated.\label{fig:Rashmi}}
\end{figure}

If Node 1 fails its recovery would ordinarily require accessing two blocks from nodes 2 and 3 each,
but with piggybacking we only need access the second byte from nodes Node 2, 3, and 4.
Several methods to reduce rebuild traffic in a distributed environment are discussed in this section.

Piggybacking as defined by Muntz and Lui 1990 \cite{MuLu90}
writes blocks reconstructed on demand on the spare disk.
Such small writes tend to be inefficient Holland et al. 1994 \cite{HoGS94}. 
Piggypbacking at the level of tracks on which data blocks resided is considered in Fu et al. \cite{FTHN04a}.
There is a significant increase in disk utilization which affects response times
since rebuild read times equalling disk rotation time are large, but rebuild time is reduced.

\subsection{Pyramid Codes}\label{sec:pyramid}

Pyramid codes have a higher overhead than MDS erasures codes,
but provide a much lower recovery cost Huang et al. 2007 \cite{HuCL07}.
Shown in Figure~\ref{fig:pyramid2} are twelve data blocks subdivided 
into two groups of six blocks with local parities and three global parities.
Microsoft's Azure is a {\it Local Reconstruction Code - LRC} which tolerates all three disk failures,
but the repair of as many as four failed disks is possible Huang et al. 2012 \cite{HSX+12}.

\begin{figure}
\begin{center}
\begin{footnotesize}
\begin{tabular}{|c|c|c|c|c|c|c|}      \hline
$d_1$   &$d_2$   &$d_3$  &$d-4$   &$d_5$   &$d_6$   &$c_{1,1}$                \\ \hline
$d_7$   &$d_8$   &$d_9$  &$d_{10}$  &$d_{11}$  &$d_{12}$  &$c_{1,2}$          \\ \hline
$c_2$   &$c_3$   &$c_4$  &     &     &     &                                  \\ \hline
\end{tabular}
\end{footnotesize}
\end{center}
\caption{\label{fig:pyramid2}A pyramid code with 12 data and 5 check disks.
$c_{1,1}$ and $c_{1,2}$ provide parity protection for the first six and second six disks, respectively.}
\end{figure}

A smaller example to shorten the discussion of data recovery is as follows:
\vspace{-1mm}
$$ \left( x_0, x_1, x_2), p_x, (y_0, y_1, y_2), p_y, c_0, c_1 \right) $$
Consider the failure of $x_0$, $x_2$, and $y_1$.
Repair proceeds by repairing $y_1$ using $p_y$ and $x_0$ and $x_2$
can then be repaired using $c_0$ and $c_1$.
Recovery with four disk failures is possible in 86\% of cases, but is quite complex.

LRC can tolerate arbitrary 3 failures by choosing the following $\alpha$'s and $\beta$'s.

\vspace{-3mm}
\begin{eqnarray}\nonumber
q_{x,0} = \alpha_0 x_0 + \alpha_1 x_1 + \alpha_2 x_2,                    
\hspace{3mm}
q_{x,1} = \alpha^2_0 x_0 + \alpha^2_1 x_1 + \alpha^2 2 x_2,        
\hspace{3mm}
q_{x,2} = x_0 + x_1 + x_2
\end{eqnarray}

\vspace{-3mm}
\begin{eqnarray}\nonumber
q_{y,0} = \beta_0 y_0 + \beta_1 y_1 + \beta_2 y_2,               
\hspace{3mm}
q_{y,1} = \beta^2 y_0 + \beta^2_1 y_1 = \beta^2 y^2    
\hspace{3mm}
q_{y,2} = y_0 + y_1 +y_2
\end{eqnarray}

The LRC equations are as follows:
\vspace{-1mm}
$$p_0 = q_{x,0} + q_{y,0}, \hspace{5mm} p_1 = q_{x_1} + q_{y,1},\hspace{5mm} p_x = q_{x,2} p_y = q_{y,2}$$

The values of $\alpha$s and $\beta$s are selected,
so that LRC can decide all information-theoretically decodable four failures.

{\bf 1. None of four parities fails:}
The failures are equally divided between the two groups. We have the following matrix.

$$
G=
\begin{pmatrix}
1 & 1 & 0 & 0 \\
0 & 0 & 1 & 1 \\
\alpha_i & \alpha_j & \beta_s & \beta_t \\
\alpha^2_i & \alpha^2_j & \beta^2_s & \beta^2_t
\end{pmatrix}
$$
\vspace{-1mm}
$$\mbox{det}(G) = (\alpha_j - \alpha_i) (\beta_t - \beta_s) (\alpha_i + \alpha_j - \beta_s - \beta_t).$$

{\bf 2. Only one of $p_x$ and $p_y$ fails:}
Assume $p_y$ fails. The remaining three failures are two in group $x$ and one in group $y$.
We have three equations with coefficients given by:

$$
G' =
\begin{pmatrix}
1 & 1 & 0 \\
\alpha_i & \alpha_j & \beta s \\
\alpha^2 _i & \alpha^2_j & \beta_2^2
\end{pmatrix}
det(G')= \beta_s(\alpha_j-\alpha_i)(\beta_s - \alpha_j - \alpha_i)
$$
{\bf 3.Both $p_x$ and $p_y$ fail}
The remaining two failures are divided between the two groups.

$$
G" =
\begin{pmatrix}
\alpha_i & beta_s \\
\alpha^2_i & \beta^2_s
\end{pmatrix}
\newline
det(G') = \alpha_i \beta_s  (\beta_s - \alpha_i)
$$
To ensure all cases are decodable all three matrices should be non-singular.

The repair savings of an $(n=16,k=12,r=6)$ LRC compared to that of an $(n=16,k=12)$ RS code
in the Azure production cluster are demonstrated in Huang et al. \cite{HSX+12}.
The $(n=18,k=14,r=7)$ LR code in WAS - Windows Azure Storage
has repair degree comparable to that of an $(9,6)$ RS code,
in the Azure production cluster are demonstrated in this work.
This code has reportedly resulted in major savings for Microsoft.                  \newline
\url{https://www.microsoft.com/en-us/research/blog/better-way-store-data/.}        

\subsection{Hadoop Distributed File System - HDFS-Xorbas}\label{sec:Xorbas}

{\it Hadoop Distributed File System - HDFS} Borthakur 2007 \cite{Bort07},
which employs $n \geq 3$ replications for high data reliability is based on the following assumptions:
1. Hardware failures the norm rather than the exception.
2. Batch processing.
3. Large datasets.
4. Files once created are not changed, i.e., {\it Write-Once, Read-Many - WORM}.
5. Moving computation cheaper than moving data.
6. Portability across heterogeneous platforms required.

HDFS Xorbas was implemented at U. Texas at Austin for Facebook.
This is an LRC build on top of an RS code by adding extra local parities Sathiamoorthy et al. 2013 \cite{SAP+13}.
The experimental evaluation of Xorbas was carried out in Amazon EC2 and a Facebook cluster,
in which the repair performance of $(n = 16, k = 10, r = 5)$ LR code was compared against a $[14, 10]$ RS code.
Xorbas reduced disk I/O and repair network traffic compared to RS codes.
In addition to the four RS parity blocks associated with ten file blocks, 
LRC associates local parities as shown below:

\vspace{-2mm}
\begin{align}\nonumber
S_1 &= c_1 X_1 + c_2 X_2 + c_3 X_3 + c_4 X_4 + c_5 X_5  \\
\nonumber
S_2 &= c_6 X_6 + c_7 X_7 + c_8 X_8 + c_9 X_9 + c_{10} X_{10} \\
\nonumber
S_3 &= {c'}_5 S_1 + {c'}_6 S_2 \mbox{ with }{c'}_5 = {c'}_6 =1
\end{align}

With three local parities the storage overhead is 17/10,
but $S_3$ need not be stored by choosing $c_i, 1 \leq i \leq 10$
satisfy the additional alignment equation $S_1  + S_2 + S_3 = 0$.
It is shown that this code has the largest possible distance ($d = 5$)
for this given locality ($r = 5$) and blocklength ($n = 16$).
This is an LRC $(10,6,5)$ code.

The reliability analysis of the system with the birth-death model (Kleinrock 1975 \cite{Klei75})
is given in Ford et al. \cite{For+10}.
Chunks fail independently with rate $\lambda_i$ and recover with rate $\rho_i$.
The failure transition rates are ${\cal S}_i \rightarrow {\cal S}_{i+1}$: $\lambda_i$, $0 \leq i \leq 4$
and repair transition rates $S_{i} \leftarrow S_{i+1}$ $\rho_i$, $0 \leq i \leq 3$.
Starting with state $s$ and given that the system can tolerate $r$ failures
we are interested in the time to failure (the time to reach ${\cal D}_{r-1}$).

\vspace{-2mm}
\begin{eqnarray}
\mbox{MTTDL} =
\frac{1} {\lambda}
\left(
\sum_{k=0}^{s-r} \sum_{i=0}^k
( \frac{\rho^i}{\lambda^i} ) \frac{1}{ (s-k+i)_{i=1}}
\right)
\end{eqnarray}
where $(a)_{(b)} \stackrel{\text{def}}{=}
a(a-1)(a-2) \ldots (a-b+1)$.
Since recovery takes much less time than time 
to failure ($\rho \gg \lambda$) the MTTF can be approximated as:

\vspace{-1mm}
\[
\mbox{MTTDL} \approx
\frac{\rho^{s-r}}{\lambda^{s-r+1}}
\frac{1}{(s)_{(s-r+1)}} + \mbox{\large O}
\left( \frac{ \rho^{s-r-1} }{\lambda^{s-r}} \right)
\]

A comparison of three redundancy methods is given in Table~\ref{tab:comparison20}

\begin{table}[t]
\begin{footnotesize}
\begin{center}
\begin{tabular}{|c|c|c|c|}\hline
                   &Storage    &Repair      &MTTDL    \\ 
Scheme             &overhead   &traffic     &(days)   \\ \hline \hline 
3-replication      &2x         &1x          &2.31E+10 \\ \hline
RS(10,4)           &0.4x       &10x         &3.31E+13 \\ \hline
LRC(10,6,5)        &0.6x       &5x          &1.21E+15 \\ \hline
\end{tabular}
\end{center}
\end{footnotesize}
\caption{\label{tab:comparison20}Comparison of three redundancy schemes.}
\end{table}

\subsection{Hadoop Adaptively-Coded Distributed File System - HACDFS}\label{sec:HACDFS}

HACDFS uses two codes: a fast code with low recovery cost 
and a compact code with low storage overhead Xia et al. \cite{XSBP15}.
It exploits the data access skew observed in Hadoop workloads 
to decide on the initial encoding of data blocks.
HACDFS uses the fast code to encode a small fraction of the frequently accessed data
and provides overall low recovery cost for the system.
The compact code encodes the majority of less frequently accessed data blocks with a low storage overhead.

After an initial encoding, e.g., with the compact code,
HACDFS dynamically adapts to workload changes by using upcoding and downcoding 
to convert data blocks between the fast and compact codes.
Blocks initially encoded with fast code are upcoded into compact code 
enabling HACDFS system to reduce the storage overhead. 
Downcoding data blocks from compact code to fast code representation lowers 
the overall recovery cost of the HACDFS system. 
The upcode and downcode operations are efficient and only update the associated parity blocks.
HACDFS exploits the data access skew observed in Hadoop workloads.    \newline
\begin{scriptsize}
\url{http://blog.cloudera.com/blog/2012/09/what-do-real-life-hadoop-workloads-look-like/}
\end{scriptsize}

\begin{figure}[h]
\begin{footnotesize}
\begin{center}
\begin{tabular}{|c|c|c|c|c|c|}  \hline
D &D &D &D &D &P   \\ \hline
D &D &D &D &D &P   \\ \hline
P &P &P &P &P &P   \\ \hline
\end{tabular}
\hspace{5mm}
\begin{tabular}{|c|c|c|c|c|c|} \hline
D   &D  &D  &D &D &P   \\ \hline
D   &D  &D  &D &D &P   \\ \hline
P   &P  &P  &P &P &P   \\ \hline
\end{tabular}
\hspace{5mm}
\begin{tabular}{|c|c|c|c|c|c|}  \hline
D   &D  &D  &D &D &P   \\ \hline
D   &D  &D  &D &D &P   \\ \hline
P   &P  &P  &P &P &P   \\ \hline
\end{tabular}
\end{center}
\begin{center}
\begin{tabular}{|c|c|c|c|c|c|}    \hline
D   &D  &D  &D &D &P          \\ \hline
D   &D  &D  &D &D &P          \\ \hline
D   &D  &D  &D &D &P          \\ \hline
D   &D  &D  &D &D &P          \\ \hline
D   &D  &D  &D &D &P          \\ \hline
D   &D  &D  &D &D &P          \\ \hline
P   &P  &P  &P &P &P          \\ \hline
\end{tabular}
\end{center}
\end{footnotesize}
\caption{\label{fig:xia15}Parity encodings to attain reliability in Hadoop.
Upcoding (2x5) arrays to 6x5 results in a reduced redundancy from 18/10=1.8 to 42/30=1.4.
Horizontal parity codes are simply copied,
while the vertical parity bits are the XOR of the parities in the high density code.}
\end{figure}

Fewer data accesses are required for rebuild strips with $2 \times 5$ encoding,
i.e., two strips accesses to rebuild a strip.

\section{Copyset Replication for Reduced Data Loss Frequency}\label{sec:cidon} 

{\it Copyset Replication - CR} is a technique 
to significantly reduce the frequency of data loss events with triplicated data.
It was implemented and evaluated on two open source data center storage systems 
by Cidon et al. in \cite{CRS+13}. 
HDFS in Subsection \ref{sec:Xorbas} and RAMCloud in Subsection \ref{sec:RAMCloud}).

Data replicas are ordinarily randomly scattered across several nodes 
for parallel data access and recovery. 
CR presents a near optimal trade-off between the number of nodes 
on which the data is scattered and the probability of data loss.
It is better to lose more data  less frequently than vice-versa as stated 
by a Facebook engineer in the paper.

\begin{quote}
``Even losing a single block of data incurs a high fixed cost, 
due to the overhead of locating and recovering the unavailable data. 
Therefore, given a fixed amount of unavailable data each year,
it is much better to have fewer incidents of data loss with more data each 
than more incidents with less data. 
We would like to optimize for minimizing the probability of incurring any data loss.
in other words fewer events are at a greater loss of data are preferred.'' 
\end{quote}

Replication characteristics of three data centers are given in Table \ref{tab:rep}.

\begin{table}
\begin{footnotesize}
\begin{center}
\begin{tabular}{|c|c|c|c|c|}\hline
System          &Chunks             &Cluster        &Scatter    &Replication        \\
                &per node           &size           &width      &scheme             \\ \hline \hline
Facebook        &10,000             &1000-5000      &10         &Small group         \\ \hline
RAMCloud        &8000               &100-10,0000    &N-1        &Across all nodes    \\ \hline
HDFS            &10,000             &100-10,000     &200        &Large group         \\ \hline
\end{tabular}
\end{center}
\end{footnotesize}
\caption{\label{tab:rep} Three replication schemes. 
In the case of Facebook and HDFS 2nd and 3rd replica is on same rack.}
\end{table}

In {\it Random Replication - RR} assuming that the primary replica is placed at node $i$
the secondary replica is placed at node $i+1 \leq j \leq i+S$
and if $S=N-1$ the secondary replicas are drawn uniformly from all the nodes in the cluster. 
The scatter width ($S$) is the number of nodes that may store copies of a node's data.

For example, for $N=9$ nodes, degree of replication $R=3$, 
$S=4$ if the primary replica is on Node 1 then the secondary replica is on Nodes (2,3,4,5).
$P_{DL}$ is the ratio \#copysets = 54 and the maximum number of sets $\binom{9}{3}$ or 0.64.
A lower scatter width decreases recovery time,
while a high scatter width increases the frequency of data loss.

Correlated failures such as cluster power outages are poorly handled by RR and (0.5\%-1\%) of nodes do not power up.
There is a high probability that all replicas of at least one chunk in the system will not be available.
According to Figure 1 in the paper as the number nodes exceeds 300 nodes 
with degree of replication ($R=3$) the probability of data loss $P_{DL} \approx 1$.

CR - Copyset Replication splits the nodes into groups of $R$ chunks, 
so that the replicas of a single chunk can only be stored in a copyset. 
Data loss occurs when all the nodes of a copyset fail together,
e.g., with $N=9$ nodes and $R=3$ nodes per copyset: $(1,2,3),(4,5,6),(7,8,9)$,
so that data loss occurs if $(1,2,3)$, $(4,5,6)$, or $(7,8,9)$ are lost.
When a CR node fails there are $R-1$ other nodes that hold its data.

The goal of CR is to minimize the probability of data loss, 
given any $S$ by using the smallest number of copysets
When the primary replica is on node $i$ the remaining $R-1$ replicas 
are chosen from $(i+j),j=1, \ldots S$,
With $S=4$ $(1,2,3), (4,5,6), (7,8,9), (1,4,7), (2,5,8), (3,6,9)$.
Node 5's chunks can be replicated as nodes (4,6) and nodes (2,8).
There are \#copysets=6 and if three nodes fail $P_{DL}=0.07$.
Note that the copysets overlap each other in one node and that copysets cover all nodes equally, 

Two permutations of node numbers can be used to define copysets.
The overall scatter width is $S=P(R-1)$.
The scheme creates $P N /R$ or $ S N / (R(R-1))$ copysets. 

Simulation results are used to estimate $P_{DL}$ versus the number of RAMCloud nodes for $3 \leq R \leq 6$.
As shown in Figure 2 $P_{DL} \approx =0$ for $R=6$ beyond a 1000 nodes, 
and increases linearly from 0 to 50\% for 10,000 nodes.  

Figure 3 in the paper plots $P_{DL}$ with $N=5000$, $R=3$, 
with RR and CR versus $1 \leq S \leq 500$ when 1
For RR $P_{DL} \approx 1 $ for $S \geq 50$, 
which is the case for Facebook HDFS and $P_{DL} > 0.0$ for $S=500$. 
 
Consider the case when the system replicates data on the following copysets:
\vspace{-1mm}
$$(1,2,3), (4,5,6),(7,8,9),(1,4,7),(2,5,8),(3,6,9).$$
That is, if the primary replica is placed on node 3, 
the two secondary replicas can only be randomly on nodes 1 and 2 or 6 and 9. 
Note that with this scheme, each node's data will be split uniformly on four other nodes.
With this scheme there are six copysets and $P_{DL}= \#copysets / 84=0.07$.

Power outage in the case of a 5000-node RAMCloud cluster with 
CR reduces the probability of data loss from 99.99\% to 0.15\%. 
For Facebook's HDFS cluster it reduces the probability from 22.8\% to 0.78\%.

\subsection{More Efficient Data Storage: Replication to Erasure Coding}\label{sec:DiskReduce}

DiskReduce is a modification of HDFS enabling asynchronous compression 
of initially triplicated data down to RAID6.   
This increases cluster's storage capacity by a factor of three
and this is delayed long  enough to deliver the performance benefits of multiple data copies

Also considered in this study was mirrored RAID - RAID1/5
which was mirrored RAID5 within and across files.
A plot of space overhead versus RAID group size $2^i, 1\leq i \leq 5$ shows 
a less than 5\% overhead for $i=5$ for RAID6 and across files.

If the  encoding is delayed for $t$ seconds,
we  can  obtain  the  full  performance of  having  three  copies  from  a  block's  creation  
until  it  is $t$ seconds old. 
The penalty is modeled by $r$, which stands for $100(1-r)\%$ degradation.
Different $r$ gives different expected performance bounds, a
The expected performance achieved by delaying encoding can be bounded as:
\vspace{-1mm}
$$\mbox{Penalty}(t)= 1 \times \Phi(t) + r \times (1−\Phi(t)),$$
where $Phi(t)$ is CDF of block access with regard to blockage, 
which is derived from the trace (Figure 4 in paper),
Plotting Penalty$(t)$ versus $t$ it can be seen in Figure 5 in the paper, 
that for $r=0.8, 0.5, 0.333$ there is very little system performance penalty after an hour.

Based on traces collected from Yahoo!, Facebook, and Opencloud cluster the following issues are explored, 

\begin{enumerate}
\item
The capacity effectiveness of simple and not so simple strategies for grouping data blocks into RAID sets; 
\item
Implications of reducing the number of data copies on read performance and how to overcome the degradation; 
\item
Different  heuristics  to  mitigate  SWP.   
\end{enumerate}

The framework has been built and submitted into the Apache Hadoop project.

The {\it Quantcast File System - QFS} \cite{ORR+13} is a plugin compatible alternative 
to HDFS/MapReduce offering efficiency improvements:                               \newline
(1) 50\% disk space savings through erasure coding instead of replication,        \newline
(2) doubling of write throughput,                                                 \newline
(3) a faster name node,                                                           \newline
(4) support for faster sorting and logging through a concurrent append feature,   \newline
(5) a native command line client much faster than Hadoop file system,             \newline
(6) global feedback-directed I/O device management. 

\section{Clustered RAID5}\label{sec:CRAID}

{\it Clustered RAID - CRAID} proposal Muntz and Lui 1990 \cite{MuLu90} 
was focused on the reduction of load increase in degraded mode.
The {\it Parity Group - PG} size: $G$ is set to be smaller 
than the stripe width set equal to the number of disks ($N$).
The read load increase is given by the declustering ratio $\alpha=(G-1)/(N-1) < 1$,
since each access to the failed disk results in accesses to surviving disks in the PG.
CRAID will result in a reduction of $R_n^{F/J}(\rho)$ due to a reduced $\rho$ 
in degraded mode and that that a $ n= G < N$-way F/J requests are required.

The load decrease in RAID5 is tantamount to reduced rebuild time according to Eq. (\ref{eq:beta}),
which is due to reduced load and the fact that rebuild reading is parallelized.
With dedicated sparing that disk may become a bottleneck even if there is no read redirection,
but curtailing redirection by monitoring the load of the spare disk can be used to reduce rebuild time.
Distributed sparing is a possible solution.

The load in clustered RAID5 and RAID6 is obtained using decision trees in Thomasian 2005b \cite{Thom05b}.
Let $\bar{T}_{\cdot}$ denotes read, write, and {\it Read-Modify-Write - RMW} mean disk service times.
RMW is a read followed by a write after a disk rotation and 
can be substituted by independent single reads and writes.
With arrival rate $\Lambda$ to a RAID5 with $N$ disks 
disk utilizations factors due to read and write requests with one failed disk are:

\vspace{-5mm}
\begin{align}\label{eq:increase}
\rho^{RAID5/F1}_{read} &=
\frac{ \Lambda f_{read} }{N-1}
\left[ \frac{N+G-2}{N} \bar{T}_{read} \right]                         \\
\nonumber
\rho^{RAID5/F1}_{write} &=
\frac{ \Lambda f_{write}} {N-1}
\left[ \frac{G-2}{N} \bar{T}_{read} +\frac{2}{N} \bar{T}_{write} + \frac{2(N-2) }{N} \bar{T}_{RMW} \right]
\end{align}

Six properties for ideal layouts for CRAID  are presented in Holland et al. 1994 \cite{HoGS94},
but not all six properties are attainable as shown in the study:                                     \newline
(i) Single failure correcting, the strips in the same stripe are mapped to different disks.          \newline
(ii) Balanced load due to parity, all disks have the same number of parity strips.                   \newline
(iii) Balanced load in failed mode, 
the reconstruction workload should be balanced across all disks. \newline
(iv) Large write optimization, 
each stripe should contain $N-1$ contiguous strips, where $N$ is the parity group size.              \newline
(v) Maximal read parallelism is attained,
i.e., the reading of $n \leq N $ disk blocks entails in accessing $n$ disks.                         \newline
(vi) Efficient mapping, the function that maps physical to logical addresses is easily computable.

According to Alvarez et al. 1998 \cite{ABSC98} for declustering with $n \geq 2$ disk drives,
$b$ parity groups, $k \leq n$ strips per group, and $r$ strips per disk the equality $b k = n r$ holds.
According to Theorem 5 in the paper there is a placement-ideal layout with parameters $n$, $k$, $r$,
{\it if and only if - iff} $k|r$ and $(n-1)|k(r-1)$ and one of the following holds:
(1) $k=n$, (2) $k=n-1$, (3) $k=2$, (4) $k=3$ and $n=5$ (5) $k=4$ and $n=7$.

The {\it Permutation Development Data Layout - PDDL} 
is a mapping function described in Schwarz et al. 1999 \cite{ScSB99}.
It has excellent properties and good performance 
both in light and heavy loads like the PRIME Alvarez et al. 1998 \cite{ABSC98}
and the DATUM data layout Alvarez et al. 1997 \cite{AlBC97}, respectively.

In what follow we discuss four other CRAID organizations.

\subsection{Balanced Incomplete Block Design - BIBD}\label{sec:BIBD}

The {\it Balanced Incomplete Block Design - BIBD}
incurs a small cost for mapping PGs with balanced parities Holland et al. 1994 \cite{HoGS94}.
Four parameters for a declustering layout are given as follows:
$n \geq 2$ is the number of disks, $k \leq n$ is the  PG size $G$,
$b$ is the number of PGs,
the number of stripes (rows) per disk is $r$.
An additional parameter $L$ which is the number of PGs
common to any pair of disks is specified in Ng and Mattson 1994 \cite{NgMa94}, 
Because of the following two equalities, only three out of five variables are free,
\vspace{-1mm}
$$ b k = n r \mbox{  and  } r (k-1) = L ( n-1 ).$$
For $N=10$, $k=G=4$, and $L=2$, $r= 2 \times 9 /3=6$, and $b= N \times r / G= 10 \times 6 / 4 = 15$.
Figure \ref{fig:BIBD} is a BIBD layout given in Ng and Mattson \cite{NgMa94} extracted from Hall 1986 \cite{Hall86}.

\begin{figure}[b]
\begin{footnotesize}
\begin{center}
\begin{tabular}{|c|*{10}{c}|} \hline
Disk \#     &  1  &   2 &   3 &   4 &  5 &  6 &  7 &  8 &  9 &10  \tabularnewline \hline\hline
            &  1  &   1 &   2 &   3 &  1 &  2 &  3 &  1 &  2 & 3  \tabularnewline
            &  2  &   4 &   4 &   4 &  5 &  6 &  7 &  5 &  6 & 7  \tabularnewline
Parity      &  3  &   6 &   5 &   5 &  8 &  8 &  8 &  9 & 10 &11  \tabularnewline
Groups      &  4  &   7 &   7 &   6 & 10 &  9 &  9 & 12 & 12 &12  \tabularnewline
            &  8  &   9 &  10 &  11 & 11 & 11 & 10 & 14 & 13 &13  \tabularnewline
            & 12  &  13 &  14 &  15 & 13 & 14 & 15 & 15 & 15 &14  \tabularnewline \hline
\end{tabular}
\end{center}
\end{footnotesize}
\caption{\label{fig:BIBD}BIBD data layout of a segment with $N=10$ disks and PG size $G=k=4$.}
\end{figure}

Each set of six consecutive stripes constitute a segment.
The $r=6$ strips in each column of the segment can be rebuild
by accessing three strips per PG for $G=k=4$ or $r(k-1)=18$ strips.

The following disk accesses are required to reconstruct the six strips
on failed Disk\#3 specified as strip\#(disk\#1.disk\#2,disk\#3):            
2(1,6,9), 4(1,2,4), 5(4,5,8), 7(2,7,10), 10(5,7,9), 14(6,8,10),            \newline
Two strips are read from each one of surviving disks denoted by D\#i.      \newline
D\#1(2,4), D\#(4,7), D\#4(4,5), D\#5(5,10), D\#6(2,14), D\#7(7,10), D\#8(5,14), D\#9(2,10), D\#10(7,14).
The on-demand reconstruction load is balanced and there is a 3-fold increase in rebuild bandwidth.

One of the strips, say the first (or the last), in all  PGs, may be set consistently to be the parity.
Given $b=15$ PGs and $N=10$ disks there should be 1.5 parities per disk on the average.
Inspection shows that five disks have one parity and another five two parities per segment.
Ten rotations of segments balance parity loads with 15 parities
in each column or an average of 1.5 parities per segment.

An example for BIBD parity placement for $N=5$ and $G=4$ is given in Holland et al. 1994 \cite{HoGS94}.
BIBD designs are not available for all $N$ and $G$ values,
e.g., layouts for $N=33$ with $G=12$ do not exist,
but designs with $G=11$ and $G=13$ can be used instead.

\subsection{CRAID Implementation Based on Thorp Shuffle}\label{sec:Thorp}

The Thorp shuffle to implement CRAID is proposed in Merchant and Yu \cite{MeYu96}.
It is stated that unlike BIBD it can generate layouts for arbitrary $N$ and $G$,
but provided examples are restricted to $N/G=2^i, 0 \leq i \leq 3$ with $G=8$.

The Thorp shuffle cuts a deck of cards into two equal half decks A and B and
then the following is repeated until no cards are left:
``Pick deck A or B randomly and drop its bottom card and
then drop the bottom card from the other half deck atop it" Thorp 1973 \cite{Thor73}.

A shuffle-exchange network where the propagation through the network
is determined by random variables was used for this purpose.
With $N$ disks and $B$ strips per disk the overall time complexity 
of computing block addresses is O($\mbox{log}(B) + \mbox{log}(N)$).

The Thorp shuffle method has the disadvantage of incurring high computational cost 
per block access Holland et al. 1994 \cite{HoGS94}, 14 $\mu$s (microseconds) on an SGI computer,  \newline
\begin{scriptsize}
\url{https://en.wikipedia.org/wiki/Silicon_Graphics}                                              \newline
\end{scriptsize}
but disk capacities which affect $B$ have increased by orders of magnitude in the last quarter century,
making the Thorp shuffle more costly to implement,
In addition it seems sufficient load balancing can be attained by the following simpler methods.

\subsection{Nearly Random Permutation - NRP}\label{sec:NRP}

The {\it Nearly Random Permutation - NRP} in Fu et al. 2004 \cite{FTHN04}
incurs less mapping cost by permuting individual stripes (rows).
This is different from the scheme in Section \ref{sec:Thorp} given this name. 
To build the mapping consecutive PGs of size $G < N$
are placed consecutively in the matrix with $N$ columns in as many rows as disk capacity allows,
so that PG$_i$ occupies strips $iG:iG+(G-1)$.
In the case of RAID5 (resp. RAID6) the $P$ (resp. $P$ and $Q$) parities
are consistently placed as the first or last two strips in each PG.

According to this placement P parities appear on half of the disks 
as shown by ``Initial allocation'' in Figure \ref{fig:NRP}.
This imbalance is alleviated by randomizing strip placement,
so that approximately the same number of parity blocks are allocated per disk.
Load balance can be attained by rotating columns in segments of $K$ rows defined below
and placing parity blocks on odd numbered disks and use this layouts alternatively.

Algorithm 235 in Durstenfeld 1964 \cite{Durs64} for randomly permuting:

\begin{quote}
Consider an array $A$ with $N$ elements. Set $n=N$. \newline
L: Pick a random number $k$ between $1$ and $n$.    \newline
If $k \neq n$ then $A_n \leftrightarrow A_k$.       \newline
Set $n=n-1$ and go back to L if $n <2 $.
\end{quote}

If $N\;mod(G)=0$ then the random permutation is applied once
and otherwise it is repeated at $K = \mbox{GCD}(N,G)$ rows.
A permutation obtained using an easy to compute polynomial function 
$f(I)$ with $I = j K, j \geq 0 $ is used as a seed.
The same permutation is applied to stripes $j K, j K + 1, \ldots, jK +(K-1)$.

A pseudo-{\it Random Number Generator - RNG}
yields a random permutation of $\{ 0,1,\dots,N-1 \}$ as:
$\underline{P_I} = \{P_0, P_1, \ldots, P_{N-1} \}$.\newline
For example, consider an example $N=10$ and $G=4$ so that $K=\mbox{GCD}(10,4)=2$. The permutation
\vspace{-1mm}
$${\bf P}= \{ 0, 9, 7, 6, 2, 1, 5, 3, 4, 8 \}$$
should be applied to two rows at a time 
in Figure~\ref{fig:NRP} as show in Figure~\ref{fig:NRP2}.
This assures that a PG straddling two rows is mapped onto different disks,
since initially all strips were on different disks
(this is the case for $D_6$, $D_7$, $D_8$, $P_{6:8}$).
Since data strips in a PG are assigned to different disks they can be accessed in parallel.

\begin{figure}
\begin{footnotesize}
\begin{center}
\begin{tabular}{|c|*{10}{c}|} \hline
Disk \#      &  1  &    2   &   3   &   4   &  5   &   6   &   7   &    8    & 9 &10
\tabularnewline \hline\hline
Initial     &$D_1$   &$D_2$  &$D_3$ &$P_{1:3}$ &$D_4$  &$D_5$   &$D_6$  &$P_{4:6}$   &$D_7$ &$D_8$
\tabularnewline \hline \hline
allocation &$D_9$  &$P_{7:9}$  &$D_{10}$   &$D_{11}$  &$D_{12}$ &$P_{10:12}$ &$D_{13}$ &$D_{14}$  &$D_{15}$ &$P_{13:15}$
\tabularnewline \hline \hline
\end{tabular}
\end{center}
\end{footnotesize}
\caption{\label{fig:NRP}The two rows show the preliminary allocation of strips
before they are permuted using nearly random permutation method ($N=10, G=4, K=2$).}
\begin{footnotesize}
\begin{center}
\begin{tabular}{|c|c|c|c|c|c|c|c|c|c|c|} \hline
Disk\#       &  1 &  2   &   3   &   4   &  5   &   6   &   7   &    8    & 9 & 10
\tabularnewline \hline\hline
Final        &  $D_1$  & $D_5$  &  $D_4$  &  $P_{4:6}$ &  $D_7$  & $D_6$  & $P_{1:3}$ &  $D_3$ &  $D_8$   &  $D_2$
\tabularnewline
allocation   &  $D_9$  & $P_{10:12}$ &  $D_{12}$ & $D_{14}$ &   $D_{15}$ & $D_{13}$ & $D_{11}$  & $D_{10}$ & $P_{13:15}$ & $P_{7:9}$
\tabularnewline \hline
\end{tabular}
\end{center}
\end{footnotesize}
\caption{\label{fig:NRP2}Permuted data blocks with nearly random permutation method (N=10, G=4).
Only first two rows are shown.}
\end{figure}

\subsection{Shifted Parity Group Placement}\label{sec:shifted}

When $N$ is divisible by $G$ or $N\mbox{mod}G=0$ 
the following approach can be used to balance parity loads.
For $N=12$, $G=4$, $L=\mbox{LCM}(N,G)=12$, $p=L/G=3$ PGS in each row.
PGs in one row need be shifted $G-1=3$ times in following rows.

If $G$ is coprime with $N$, 
i.e., if $m=\mbox{GCD}(N,G)=1$ then $i G \mbox{mod} N, i \geq 1$ generates 
all values $0:N-1$ in a periodic manner.
For $N=10$, $G=3$, $L=30$, $K=3$, $p=10$, $m=\mbox{GCD}(10,3)=1$, with 10 PGs in $K=3$ rows.
When parities are the first element of a PG they will appear in columns: 
\vspace{-1mm}
$$(1,4,7,10,3,6,9,2,5,8).$$
For $N=11$, $G=4$, $L=44$, $K=4$, $m=1$, $p=L/G=11$ PGs in four rows. Shifting not required.



For $N=15$, $G=6$, $L=30$, $K=2$ $r=N\mbox{mod}G=3$, $r-1=2$ shifts to two rows.
Figure \ref{tab:table6} $N=10$, $G=4$, $L=20$, $K=2$, $r=N\mbox{mod}G=2$,
hence $r-1=1$ shift applied to the second set of two rows.

\vspace{-2mm}
\begin{figure}[b]
\begin{footnotesize}
\begin{center}
\begin{tabular}{|c|c|c|c|c|c|c|c|c|c|c|} \hline
Disk \#     &1  &  2  &  3  &  4  &  5 &  6   &   7   &  8   &  9  & 10
\tabularnewline \hline\hline
 Initial    & $D_1$  &  $D_2$  &  $D_3$  &  $P_{1:3}$ &  $D_4$  &  $D_5$  &  $D_6$  &  $P_{4:6}$  &  $D_7$ &  $D_8$
\tabularnewline
 allocation & $D_9$  & $P_{7:9}$  & $D_{10}$   & $D_{11}$ & $D_{12}$ & $P_{10:12}$ & $D_{13}$ & $D_{14}$  & $D_{15}$ & $P_{13:15}$
\tabularnewline \hline \hline
 Shifted     & $D_{17}$  & $D_{18}$  & $P_{17:18}$ & $D_{19}$ & $D_{20}$  & $D_{21}$ & $P_{19:21}$  &$D_{22}$  &$D_{23}$ &$D_{24}$
\tabularnewline
allocation) & $P_{22:24}$ &$D_{25}$ & $D_{26}$ & $D_{27}$  &$P_{25:27}$ & $D_{28}$ &$D_{29}$ &$D_{30}$ &$P_{28:30}$ &$D_{16}$
\tabularnewline \hline
\end{tabular}
\end{center}
\end{footnotesize}
\vspace{-1mm}
\caption{\label{tab:table6}Shifted allocation of PGs with $N=10$, $G=4$.}
\end{figure}

\section{Miscellaneous Topics  Related to RAID5}\label{sec:misc}

The following topics are covered in this section.
Distributed sparing in RAID5 in Subsection \ref{sec:distrsparing}.                       
Permanent Customer Model - PCM versus VSM in Subsection \ref{sec:PCM}.        
Rebuild processing in Heterogeneous Disk Arrays - HDAs in Subsection \ref{sec:prioritize}. 
Optimal single disk rebuild in EVENODD, RDP, X-code in Subsection \ref{sec:singlediskrebuild}. 

\subsection{Distributed Sparing}\label{sec:distrsparing}

Distributed sparing in RAID5 allocates sufficient spare areas 
on surviving disks to accommodate a single disk failure, 
i.e., to hold the data blocks of a failed disk as shown in Figure 1 in \cite{ThMe97}.
Empty strips are placed diagonally in parallel with parity strips at each disk. 
The advantage of distributed sparing with respect to dedicated sparing 
is all disks contribute to the processing of the workload in normal mode.

Rebuild reading time is slowed down because surviving disks 
in addition to being read are written reconstructed RUs. 
An iterative solution is used in \cite{ThMe97} 
to equalize the rate at which RUs are being read and written
to prevent the overflow of DAC' buffer capacity.
Once a spare disk becomes available the contents of spare areas are copied onto it in copyback mode 
This mode is also required with restriping discussed 
in Section \ref{sec:HRAID} to return the system to its original state.  

\subsection{Rebuild Processing Using the Permanent Customer Model}\label{sec:PCM}

The {\it Permanent Customer Model - PCM} is proposed and analyzed in Boxma and Cohen 1991 \cite{BoCo91}
postulates a circulating permanent customer in an M/G/1 queue, 
which also serves external customers arriving according to a Poisson process.
The permanent customer once completed rejoins a FCFS queue from which all customers are served. 
PCM is adopted in Merchant and Yu 1996 \cite{MeYu96} to model rebuild processing. 
FCFS scheduling is not compatible with modern disk scheduling policies such as SATF.

VSM for rebuild processing outperforms PCM by attaining 
a lower response times for external disk requests, 
since it processes rebuild read requests at a lower priority than external requests.

In PCM rebuild reads spend more time in the FCFS queue 
and this reduces the probability of consecutive rebuild reads with respect to VSM.
The probability that rebuild requests are intercepted
by user requests arriving with rate $\lambda$ with VSM and PCM rebuild are given in \cite{FTHN04a}:

$$
P_{VSM} = 1 - exp( - \lambda \overline{x}_{RU} ), \hspace{2mm}
P_{PCM} = 1 - exp( - \lambda ( \overline{x}_{RU} + W_{RU} )),
$$
$x_{RU}$ is the time to read an RU and $W_{RU}$ is the mean waiting time in the queue for rebuild reads in PCM.
Since $\overline{x}_{RU} + W_{RU}  > \overline{x}_{RU}$ implies $P_{VSM} < P_{PCM}$.

Both effects are verified via simulation in Fu et al. 2004a \cite{FTHN04a},
where it is shown that VSM outperforms PCM on both measures.

Given that rebuild time with VSM takes less time than rebuild time with PCM
fewer external requests are processed while rebuild processing is in progress.

\subsection{Rebuild in Heterogeneous Disk Arrays}\label{sec:prioritize}

{\it Heterogeneous Disk Array - HDA} postulates a DAC, 
which can control multiple RAID levels, 
e.g., RAID1 and RAID5 sharing space in a disk array Thomasian and Xu 2011 \cite{ThXu11b}. 
The choice of RAID1 and RAID5 is based on their suitability,
e.g., RAID1 for OLTP workloads and RAID5 for parallel access. 
Data allocation in early version of HDA used a two level allocation 
into areas dedicated to RAID1 and RAID5  which were also allocated dynamically
as such storage was exhausted Thomasian et al. \cite{ThBH05}. 
Another aspect of HDA is that VAs with higher dependability levels 
are allocated as RAID$(4+k),k > 1$ or with 3-way replication.

The study is concerned with alternative data allocation method of 
{\it Virtual Disks - VDs} constituting {\it Virtual Arrays - Array - VAs}.
Two dimensional disk bandwidth and capacity per VD are considered in this study,
but the latter is of less relevance for terabyte disks.

The VA width $W$ is selected to limit the physical disk load per VD
below a certain percentage of its bandwidth expressed in IOPS.
Allocations are carried out in degraded mode, 
as if one of VDs of a VA has failed to avoid overload if a disk fails.


Given that disk layout is known there is no need to read unallocated disk areas.
It is also possible to prioritize the rebuild processing of VAs used by more critical applications, 
e.g., OLTP for e-commerce generates more revenue than data mining. 

URSA minor cluster based storage in Abd-el-Malek et at. 2005 \cite{Abd+05}  is described as follows:
``No single encoding scheme or fault model is optimal for all data. 
A versatile storage system allows them to be matched to access patterns, 
reliability requirements, and cost goals on a per-data item basis. 
Ursa Minor is a cluster-based storage system that allows data-specific selection of, 
and on-line changes to, encoding schemes and fault models. 
Thus, different data types can share a scalable storage infrastructure and 
still enjoy specialized choices, rather than suffering from "one size fits all."
 Experiments with Ursa Minor show performance benefits of 2–3x 
when using specialized choices as opposed to a single, more general, configuration. 
Experiments also show that a single cluster supporting multiple workloads simultaneously is 
much more efficient when the choices are specialized for each distribution rather than forced 
to use a "one size fits all" configuration. When using the specialized distributions, 
aggregate cluster throughput nearly doubled.'' 

The IBM {\it RAID Engine and Optimizer - REO} project described 
in Kenchamana-Hosekote et al. 2007 \cite{KeHH07} has similarities to HDA
in that it can accommodate different RAID levels on shared disks. 
REO works for any XOR-based erasure code
and can systematically deduce a least cost reconstruction strategy for a read to lost pages 
Two plausible measures for an I/O plan are:
(1) The number of distinct disk read and write commands needed to execute an I/O plan. 
(2) The number of cache pages input to and output from XOR operations in executing an I/O plan.
Minimizing XOR leads to lower memory bandwidth usage. 

\subsection{Optimal Single Disk Rebuild in RAID6 Arrays}\label{sec:singlediskrebuild}

An optimal data recovery method was proposed in the context of RDP in Xiang et al. 2010 \cite{XXLC10}.
The method is optimal in the number of accessed pages and balancing disk loads,
but erroneously assumes that that symbols are as large as strips.
The method was extended to the EVENODD layout in Xiang et al. 2011 \cite{XXL+11},
whose implementations uses small symbol sizes like RDP as discussed in Section \ref{sec:RDP}.
 
The method is applicable to X-code described in Subsection \ref{sec:xcode}
since it applies coding at strip level size Xiang et al. 2014 \cite{XLL+14}.
A similar method to minimize page accesses was informally proposed in Thomasian and Xu 2011 \cite{ThXu11}.

The method should be evaluated in reducing computational cost for reconstruction
because blocks would be reused in caches. 
More efficient recovery methods should be similarly applicable to Grid files in Section \ref{sec:gridfiles}.

\section{Evaluation of RAID5 Performance}\label{sec:RAIDperfeval} 

With the advent of faster CPUs, the performance of applications 
with stringent response time requirements, such as OLTP, 
rely heavily on storage performance, which is especially a problem for disks. 
Potential storage accesses are obviated by caching, 
e.g., higher levels of B+ tree indices are held in main memory.
Storage performance can be improved by utilizing larger caches 
and improved cache management policies. 
Further improvement is possible adopting improved data organizations on disk 
and disk scheduling as discussed in Subsection \ref{sec:sched}
and more detail in Chapter 9 in Thomasian 2021 \cite{Thom21}. 

OLTP workloads generate accesses to small randomly placed blocks,
which are costly since they incur high positioning time,
as in the case of the TPC-C order-entry benchmark.  \newline
\begin{scriptsize}
\url{https://www.tpc.org/tpcc/}
\end{scriptsize}

Disks have an onboard cache for fetching sectors preceding/following 
requested sectors on a track Jacob et al. 2008 \cite{JaNW08}.
The DAC provides a much larger cache for data from the RAID disks.
Part of the DAC cache is NVRAM and duplexed NVRAM is as reliable as disk storage 
for logging according to Menon and Cortney 1993 \cite{MeCo93},
Logging for database recovery is discussed in Chapter 18 
on Crash Recovery in Ramakrishnan and Gehrke \cite{RaGe02}.
This allows a fast-write capability and deferred destaging of dirty blocks.
Dirty blocks can be overwritten several times before they are destaged to disk, thus reducing the disk load.
Batch destaging allows disk scheduling to minimize disk utilization.

Results obtained by investigating I/O traces by Treiber and Menon 1995 \cite{TrMe95} 
were incorporated into the analysis in Thomasian and Menon 1997 \cite{ThMe97},
e.g., occurrences of two blocks a short distance apart on a track, treated as one disk access.    
Note similarity to proximal I/O to increase disk bandwidth in Schindler et al. 2011 \cite{ScSS11}.
Large database or filesystem buffers are provided 
in the main memory of computers running such applications.
Disk cache miss ratio analysis and design considerations are discussed in Smith 1985 \cite{Smit85}.

Self-similarity was first observed in nature by B. Mandelbrot.   \newline
\begin{scriptsize}
\url{https://en.wikipedia.org/wiki/Self-similarity}               \newline
\end{scriptsize}           
It is discussed in detail by Peitgen et al. 2004 \cite{PeJS04}.
The fractal structure of cache references model by a statistically self-similar underlying process, 
which is transient in nature is discussed in McNutt 2002 \cite{McNu02}.

IBM z16 mainframe supports up to 40 terabytes (TB=$10^{12}$ bytes) main memory. \newline
\begin{scriptsize}
\url{https://www.ibm.com/common/ssi/ShowDoc.wss?docURL=/common/ssi/rep_sm/1/897/ENUS3931-_h01/index.html} \newline
\end{scriptsize}
The very large capacity of caches requires very long traces 
making trace-driven simulation difficult if not impossible. 
Online measurement of page miss ratios versus memory size 
using hardware and a software is explored in Zhou et al. 2004 \cite{ZPS+04}.

Two examples of main memory databases are IMS FastPath and 
Oracle's TimesTen Lahiri et al. 2013 \cite{LaNF13}.
The sizes of largest databases at this time are as follows.   \newline 
\begin{scriptsize} 
\url{https://www.comparebusinessproducts.com/fyi/10-largest-databases-in-the-world}
\end{scriptsize}

\section{Analysis of RAID5 Performance$^*$}\label{sec:RAID5perfanal} 

An M/G/1 queueing model is adopted in this section with Poisson arrivals 
with exponential interarrival times with mean $\bar{t}=1/\lambda$ ({\bf M}) standing for Markovian) 
and generally distributed disk service times ({\bf G}) with their $i^{th}$ moment denoted by $\overline{x^i}$. 

\subsection{Components of Disk Service Time w/o and with ZBR$^*$}\label{sec:diskservice}

The CPU checks if a referenced block by a program 
is cached in the database or file buffer in main memory.
If not it issues an I/O request and the DAC checks if it caches the block.
If not the requests is past on to the disk where it is sought on the onboard disk cache 
before accessing the disk taking into account faulty reallocated sectors Ma et al. 2015 \cite{Ma++15}.
Disk are discussed in Part III of Jacob et al. 2008 \cite{JaNW08}.

Disk service time is the sum of seek (s), latency ($\ell$), and transfer time (t),
Given the mean and variance of service time components 
we have the following mean and variance for disk service time:

\vspace{-2mm}
\begin{eqnarray}\label{eq:diskservice}
\bar{x}_{disk} = \bar{x}_s + \bar{x}_\ell + \bar{x}_t, \hspace{5mm}
\sigma^2_{disk} = \sigma^2_s + \sigma^2_\ell + \sigma^2_t.
\end{eqnarray}

The latter equality requires the independence of random variables,
especially latency and transfer time which is true for accesses to small blocks.
The second moment of disk service time required in the analysis is given as follows: 
\vspace{-1mm}
$$\overline{x^2}_{disk} = \sigma^2_{disk} + (\bar{x}_{disk})^2.$$ 

Seagate Cheetah 15k.5 model ST3146855FC parameters extracted in 2007 
at CMU's {\it Parallel Data Laboratory - PDL} are given as follows: \newline
\begin{scriptsize}                                              
\url{https://www.pdl.cmu.edu/DiskSim/diskspecs.shtml}               \newline
\end{scriptsize}
146 GB, maximum logical block number b=286749487 (number of 512 B sectors), 
c=72,170 cylinders, number of surfaces 4, 15015 RPM, $h=4$ heads,
The disk rotation time for 7200 RPM disks is $T_{rot} \approx 60,000/RPM=8.33$ ms.
The mean number of sectors per track is: $b/(c \times h) \approx 993$,
so that the transfer time for 4 KB blocks with 8 sectors is negligible. 

\begin{framed}
\subsection{Seek Distance Distribution without and with ZBR$^*$}\label{sec:seekdistance}

Most analytic studies make the false assumption that disk are fully loaded, which is not true in practice.
Before striping in RAID was introduced it was observed in a 1970s study that disk arms rarely move,
which might have been the case for reading successive blocks of a file in batch processing. 
The probability of not moving the disk arm is to set tp $0 \leq p \leq 1 $
and with probability $1-p$ the remaining disk cylinders are accessed uniformly.
With these two assumptions the seek distance is given in Lavenberg 1983 \cite{Lave83} by Eq. \ref{eq:dist} 

\begin{eqnarray}\label{eq:dist}
P_D [d] = 
\begin{cases}
p, \mbox{  for  }d=0, \\ 
(1-p)\frac{2(C-d)}{C(C-1)}, \mbox{  for  } 1 \leq d \leq C-1
\end{cases}
\end{eqnarray}

There are $2(C-d)$ ways to move $1 \leq d \leq C-1$ cylinders, 
which are normalized  by $\sum_{d=1}^{C-1} 2 (C-d) = C(C-1)$.
For uniform distribution across all cylinders:
\vspace{-2mm}
\begin{eqnarray}\label{eq:dist2}
p=P_D[0]=\frac{1}{C} \mbox{  and  }P_D[d]=\frac{2(C-d)}{C^2}, 1 \leq d \leq C-1.
\end{eqnarray} 

The seek time characteristic $t_{seek}(d)$ 
is the seek time versus seek distance in cylinders or tracks ($d$). 
Curve-fitting to experimental results yields equations 
of the following forms Hennessey and Patterson 2019 \cite{HePa19}: 
\vspace{-1mm}
$$t_{seek} (d) = a + b \sqrt{d-1}, \hspace{5mm} 1 \leq d \leq C-1$$  
\vspace{-1mm}
$${t}_{seek} (d) = a + b \sqrt{d-1} + c (d-1) \hspace{5mm} 1 \leq d \leq C-1$$  

The average seek time can be obtained by plugging in the average seek distance $C/3$ for uniform accesses, 
but it is argued in Thomasian and Liu 2005 \cite{ThLi05})
that a more accurate method to obtain seek time:
\vspace{-1mm}
$$\overline{x^i_s} = \sum_{d=1}^{C-1} P_D (d) t^i_{seek} (d), 
\hspace{5mm}\sigma^2_s = \overline{x^2_s} - (\bar{x_s})^2$$

The analysis in Merchant and Yu 1996 \cite{MeYu96} postulates a seek-time formula of the form:
$t_{seek} (f) = a + b \sqrt{f}$, where $f$ is the fraction of traversed cylinders.
It is implicitly assumed that the seek distance is uniform $0 \leq f \leq 1$, hence $\bar{f}=1/2$. 
This is not true according to Eq. (\ref{eq:dist2}) when disk accesses are uniform over all disk cylinders,
since $P_D[0]=1/C$ and drops from $P_D[1]= \approx 2 / C$ to $P_D [d=C-1] = 2 / C^2$
and the mean seek distance is then $\bar{d} \approx C/3$.
This issue is discussed in Thomasian and Li 2005 \cite{ThLi05}, 
in pointing out the same error in Varki et al. 2004 \cite{VMXQ04}.

The analysis of disk seek times for disks with ZBR should 
take into account the variability of track capacities as in Thomasian et al. 2007 \cite{ThFH07}.
The number of sectors on cylinder $c$ is $s_c, 1 \leq c \leq C$. 
With the assumption that disk is fully loaded and its sectors are uniformly accessed 
the probability of accessing cylinder $c$ is proportional to the number of sectors it holds:
\vspace{-1mm}
$$P_c= s_c / C_d \mbox{  where the disk capacity is } C_d = \sum_{c=1}^C s_c.$$
Starting with the probability of seek distance $d$ starting at cylinder $c$,
requires $C>d$ and $c+d <C$. This requires unconditioning on $c$:
\vspace{-1mm}
$$P_D (d|c) = P_c (c+d) + P_c (c-d), \hspace{5mm}   P_D(d)= \sum_{c=1}^{C} P_D(d|c) P_c $$

\end{framed}

Rebuild analysis requires the {\it Laplace-Stieltjes Transform - LST} of seek time and transfer time.
Provided that the {\it Rebuild Unit - RU} is a track after a seek is completed
there is no latency due to {\it Zero Latency Access - ZLA},
which allows track reading to start at the first encountered sector.
Given that transfer time equals disk rotation time, the LST is: ${\cal L}^*_t (s) = T_R/s$.

Latencies to access small blocks are uniformly distributed over tracks.
The LST and $i^{th}$ moment of latency which is uniformly distributed over $(0,T_R )$ are given as: 
\vspace{-1mm}
$${\cal L}^*_{\ell} (s) = )(1- e^{- sT_R ) / (s T_R}), \hspace{5mm} \overline{x^i_\ell} \approx T^i_R /(i+1).$$  

Given that the transfer time of $j$ sectors from cylinder $c$ is $x_t (j) = j T_{rot} / s_c$, 
it follows that the mean over all cylinders is:
\vspace{-1mm}
$$\overline{x_t^i} (j) = \sum_{c=1}^C x_t^i (j) P_c , \hspace{5mm} \sigma^2_t = \overline{x^2_t} - (\bar{x_t})^2 .$$
$j=8$  sectors is assumed in the discussion.

\begin{framed}
\subsection*{Shingled Magnetic Recording - SMR}

SMR is a technique for writing data to HDDs whereby the data tracks partially overlap 
to increase the areal density radially and hence increase disk capacity.
While ZBR increases recording capacity linearly, while SMR increases this density radially.         \newline
\begin{scriptsize}
\url{https://en.wikipedia.org/wiki/Shingled_magnetic_recording}  \newline
\end{scriptsize}
The emergence of SMR has been attributed to slowdown in disk capacity 
after 2010 after fast growth in the period 2000-2010.
An influential paper on disk developments in earlier disk storage technology is Gray and Shenoy 2000 \cite{GrSh00}.

Design issues in SMR are presented in Amer et al. 2010 \cite{ALM+10}.
Theoretical justifications for SMR are given in Sanchez 2007 \cite{Sanc07},
which in Section 1.2 introduces the trilemma: Thermal stability, 
Writeability, and Media {\it Signal to Noise Ratio - SNR}.
A technical description of SMR drives is given in Feldman and Gibson 2013 \cite{FeGi13}.

The performance of SMR drive is less predictable in that clean drives can be written quickly, 
but if the drive has too many writes queued, 
or has insufficient idle time to reorganize or discard overwritten data, 
then write speeds can be significantly lower than the usual disk bandwidth, e.g., 50-150 MB/s.
Because of their access time variability SMR disks should be excluded from RAID5 disk arrays. \newline
\begin{scriptsize}o
\url{https://en.wikipedia.org/wiki/Hard_disk_drive_performance_characteristics}
\end{scriptsize}

\end{framed}

\section{RAID5 Performance in Normal Mode$^*$}\label{sec:normal}

Let $f_r$ and $f_w=1-f_r$ denote the fraction of logical read and write requests, respectively.
Each read request results in a {\it Single Read - SR} access if the data is not cached 
and each logical write may require two SR and two {\it Single Write - SW} accesses:
\vspace{-1mm}
$$f_{SR} = (f_r+ 2 f_w)/(3 f_r+ 2 f_w),  \hspace{5mm}f_{SW}=1-f_{SR}.$$
The moments of disk service time are given as follow:

\vspace{-2mm}
\begin{eqnarray}
\overline{x^i}_{disk} = f_{SR} \overline{x^i_{SR}} + f_{SW} \overline{x^i_{SW}}.
\end{eqnarray}

Given that the arrival rate to the RAID5 disk array is $\Lambda$            .
then RAID striping results in uniform accesses to $N+1$ disks. 
Due to striping the arrival rate to the $N+1$ disks are balanced: $\lambda = \Lambda / (N+1)$,
so that the disk utilization factor is: 
\vspace{-1mm}
$$\rho=\lambda \bar{x}_{SR} + 2 f_w (\bar{x}_{SR} + \bar{x}_{SW}).$$

Given the mean and variance of disk service time with FCFS scheduling
the  coefficient of variation of disk service time is: $c^2_X=\sigma^2_{disk}/ (\bar{x}_{disk})^2.$
The mean and variance of waiting time in M/G/1 queues 
is given by the Pollaczek-Khinchine formula in Kleinrock 1975 \cite{Klei75}:

\vspace{-2mm}
\begin{eqnarray}\label{eq:MG1}
W = 
\frac{ \lambda \overline{x^2}_{disk} }{2(1-\rho)} = 
\frac{\rho \bar{x}(1+c^2_X)}{2(1-\rho)},\hspace{5mm}
\overline{W^2} = 2 W^2 + \frac{ \lambda \overline{x^3}_{disk} } {3(1-\rho)}.
\end{eqnarray}

The prioritized processing of SR requests on behalf of logical reads 
is only affected by disk utilization due to other logical read requests.
The fraction of such requests is  ${f'}_{SR} = f_r / (f_r + 4 f_w)$
and moments of $x_{disk}$ should be recomputed with ${f'}_{DR}$. 
The waiting of SR requests with nonpreemtive priorities is given in Kleinrock 1976 \cite{Klei76}. 

\vspace{-2mm}
\begin{eqnarray}\label{eq:priority}
W_{pr} = \frac{\lambda \overline{x^2}_{disk} }{2(1-\rho_{SR})}
\mbox{  with  }\rho_{SR} = {f'}_r \lambda \bar{x}_{SR}.
\end{eqnarray}

The read response without and with priorities is
$R_{SR} = \bar{x}_{SR} + W$ and $R_{SR} = \bar{x}_{SR} + W_{pr}$.

A write is considered completed when both the data and parity blocks are written to disk,
which is an instance of F/J processing discussed in Section \ref{sec:degraded} 
given by Eq. (\ref{eq:asymmetric}) in Section \ref{sec:degraded}.
Due to the fast write capability write response times are of little interest.

Since for the FCFS policy the waiting time is independent from service time then: 
$E[\tilde{W}_{SR} \tilde{x}_{SR}] =  \bar{W}_{SR} \times \bar{x}_{SR}$ and the second moment of SR requests is:

\vspace{-2mm}
\begin{eqnarray}
\overline{R^2}_{SR} = \overline{W^2}_{SR}  +\overline{x^2}_{SR} +2 W \bar{x}_{SR}.
\end{eqnarray}

$\sigma^2_{SR} = \overline{R^2}_{SR} - (R_{SR})^2$ is used in the Fork/Join analysis in Section \ref{sec:degraded}. 

\section{Methods to Obtain Response Times for Fork/Join Requests$^*$}\label{sec:degraded}

In the case of a RAID5 with $N+1$ disks an $N$-way (resp. $N-1$-way) F/J request 
is required to process read (resp. write) requests involving failed disks in degraded mode of operation. 
The expected mean of $n$-way F/J requests 
with Poisson arrivals and general service times is not known, 
but can be approximated with the maximum of $n$-way requests as follows:
$R_n^{F/J} (\rho) \approx R_n^{max} (\rho)$ as discussed below.

Before rebuild processing starts F/J requests on behalf of read requests 
constitute one-half of read requests processed by surviving disks, 
but this fraction gets smaller as rebuild progresses due to ``read redirection'' Muntz and Lui 1990 \cite{MuLu90}.
Read redirection simply reads reconstructed data from the spare disk,
but the term is a misnomer since redirection is also used 
for write requests to a failed disk when $d_{old}$ is already reconstructed on a spare to compute the parity block.
Otherwise an F/J request to surviving disks in the parity group.
When $p_{old}$ is not reconstructed 
yet there is no need update it by computing and writing $p_{new}$,
since this is done more efficiently later as part of rebuild.


Given the response time distribution in M/M/1 queues $R(\rho)$, 
which was given by Eq. (\ref{eq:Rdistr}) the only exact mean response time 
known is for 2-way F/J requests is given by Flatto and Hahn 1984 \cite{FlHa84}.

\vspace{-1mm}
\begin{eqnarray}\label{eq:FlHa84}
R_2^{F/J} (\rho) = \left[ H_2 - \frac{\rho}{8} \right] R(\rho)  = \frac{12-\rho}{8} R(\rho),
\mbox{   where the Harmonic sum is: } H_K \stackrel{\text{def}}{=}  \sum_{k=1}^K \frac{1}{k}.
\end{eqnarray}

Note that $R_2^{FJ} (0) = R_2^{max} (0) = \frac{H_2}{\mu}$ 
The following approximate equation for $2 \leq n \leq 32$-way M/M/1 F/J queueing systems 
is derived in Nelson and Tantawi 1988 \cite{NeTa88}.

\vspace{-2mm}
\begin{eqnarray}\label{eq:NeTa88}
R_n^{F/J} (\rho) =  \left[ \frac{H_n}{H_2} + (1- \frac{H_n}{H_2} ) \alpha (\rho) \right] R_2^{F/J} (\rho),
\end{eqnarray}
Curve-fitting to simulation results yielded $\alpha (\rho ) \approx (4/11) \rho $.

This equation was used in Menon 1994 \cite{Meno94} for estimating F/J response times in degraded mode, 
which is not the case in this study because of interfering requests.

\begin{figure}[t]
\begin{center}
\includegraphics[scale=0.650,angle=00]{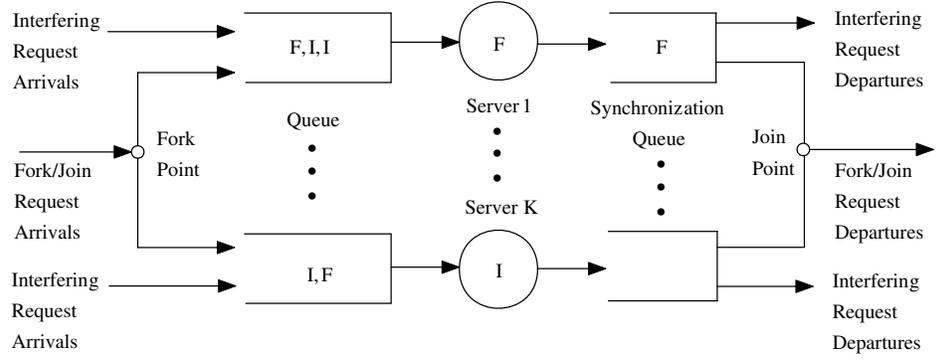}
\caption{Fork/join processing with interfering requests processing \label{fig:FJ}}
\end{center}
\end{figure}

The processing of F/J requests with interfering requests is shown in Figure \ref{fig:FJ}.
Simulation results for the mean F/J response time versus server utilization 
by varying the fraction of F/J requests,
while the total is kept fixed for different service time distributions 
is given in Thomasian and Tantawi 1994 \cite{ThTa94}.
The graphs show that as the fraction of F/J requests decreases with respect to interfering requests,
their mean response times increases and tends to the expected value of the maximum.
\vspace{-1mm}
$$R_{n}^{F/J} \approx R_{n}^{max}.$$

An approximation for the expected value of the maximum of $n$ i.i.d. random variables 
is given as a function of the mean $\mu_x$ and standard deviation $\sigma_X$ in order statistics, 
see e.g., David and Nagaraja 2003 \cite{DaNa03}:

\vspace{-2mm}
\begin{eqnarray}\label{eq:Xmax}
\overline{X}_n^{max} \approx \mu_X + \sigma_X G(n). 
\end{eqnarray}
$G(n) = H_n - 1$ and $\sigma_X = \mu_X $ for the exponential distribution: 
$\overline{X}_n^{max}= \mu_X H_n$, which is also given by Eq. (\ref{eq:FJ}). 

Eq. (\ref{eq:ThTa94}) in Thomasian and Tantawi 1994 \cite{ThTa94} was motivated by Eq. (\ref{eq:Xmax}):

\vspace{-2mm}
\begin{eqnarray}\label{eq:ThTa94}
R_n^{F/J} (\rho) = R (\rho) + F_n \sigma_R (\rho) \alpha_n (\rho),
\end{eqnarray}
where $R(\rho)$ and $\sigma_{R} (\rho) $ are the mean 
and standard deviation of response time at utilization factor $\rho$.
In a limited study surface-fitting to simulation results was used 
to obtain $\alpha_n (\rho)$ for a few distributions.

The {\it Probability Distribution Function - PDF} of the maximum of $n$ random variables 
with PDF $F_Y(y)$ is also the $n^{th}$ order statistic Trivedi 2002 \cite{Triv02}: 
\vspace{-1mm}
$$F_{Y_n^{max}} (y)  = [F_Y (y)]^n.$$ 

Given that response times in an M/M/1 queueing system are exponentially distributed 
applying the formula for the maximum of $n$ exponentials in Trivedi \cite{Triv02}: 

\vspace{-2mm}
\begin{eqnarray}\label{eq:FJ}
R_n^{max} (\rho) = H_n R (\rho).
\end{eqnarray}

It is shown in Nelson and Tantawi 1988 \cite{NeTa88} 
that $R_n^{max} (rho)$ is an upper bound to $R_n^{F/J}(\rho)$ for the exponential distribution.
This effect is observed via simulation in Thomasian and Tantawi 1994 \cite{ThTa94} for other distributions
and that $R_n^{max}(\rho)$ is  a good approximation to $R_n^{F/J}(\rho)$ 
when F/J requests constitute a small fraction of processed requests.
This is the case for RAID5 in degraded mode as rebuild progresses 
and most requests to the failed disk are satisfied by redirection.

A simple method to obtain the maximum of $n$ i.i.d. random variables 
is to approximate it with the two-parameter {\it Extreme Value Distribution - EVD} 
Johnson et al. 1995 \cite{JoKB95}, Kotz and Nadarajah 2000 \cite{KoNa00}. 
\vspace{-1mm}
$$ F_Y (y) =  P(Y < y) = exp (- e^{- \frac{y-a}{b} } ), \hspace{3mm}
\overline{Y} = a + \gamma b, \hspace{3mm} \sigma^2_Y = \frac{\pi^2 b^2}{6}, $$
where $\gamma = 0.577721 $ is the Euler constant.
Matching the first two moments: $\bar{Y} =R(\rho)$ and $\sigma_Y=\sigma_R$:
we obtain $a$ and $b$ given the first two moments of $Y$: 
\vspace{-1mm}
$$b=  \sqrt{\frac{6}{\pi}} \sigma_Y \mbox{   and   } a=\bar{Y} - \gamma b.$$  
The maximum of $n$ random variables with EVD has a simple form:

\vspace{-2mm}
\begin{eqnarray}\label{eq:meanEVD}
\overline{Y^n}_{max} = ( a + \gamma b) + b \mbox{ln} (n)
= \overline{Y} +  \frac{ \sqrt{6} } { \pi}  \sigma_Y.
\end{eqnarray}


It follows from simulation results in Thomasian et al. 2007 \cite{ThFH07}  
that a better approximation to $Y_n^{max}$ is obtained if the second summation term is divided by 1.27.

The coefficient of variation of response times of M/G/1 queues for accesses to small randomly placed disk blocks 
for given disk characteristics with $c^2_R < 1$ observed in Thomasian and Menon 1997 \cite{ThMe97}.

\vspace{-2mm}
\begin{eqnarray}
c^2_R (\rho) = \frac{\sigma^2_R (\rho)}{R^2 (\rho)}=
\frac{c^2_X + \frac{\rho s_X }{3(1-\rho)} +\frac{\rho(1+c2_X)}{ 4(1-\rho)^2} }
{1 +\frac{\rho (1+c^2_X }{1-\rho}+\frac{\rho(1+c2_X)}{ 4(1-\rho)^2} },
\end{eqnarray}
where $s_X= \overline{x^3}_{SR}/(\bar{X}_{SR})^3$. 
Where $c_R = c_X$ for $\rho=0$ and that as $\rho \rightarrow 1$ $c_R \rightarrow 1$, 
i.e., an exponential distribution, hence $c_R < 1$ for $C_X <1$.

Approximating the response time distribution with with a distribution, 
which is an exponential in form has the advantage that its maximum can be computed easily by integration.
The number of stages in the Erlang distribution is: $k=\lceil 1/c^2 \rceil$ 
and the mean delay per stage is $1/\mu = R(\rho) / k$ 
$R_n^{max} (\rho) $ for an $n$-way  $k$-stage Erlang distribution is then:

\vspace{-2mm}
\begin{eqnarray}\label{eq:balanced}
R^{max}_n (\rho) = 
\int_0^\infty \left[ 1 -  \prod_{i=1}^n 
\left( 1 - e^{ - \mu_i t }
\sum_{j=0}^{k_i -1}
\frac{ (\mu_i t )^j }{ j!}
\right)
\right]  dt  .
\end{eqnarray}

Provided disks have an XOR capability updating parities results 
in a 2-way F/J request with unequal service times:
$d_{new}$ and the block address is sent to the disk
and after $d_{old}$ is read $d_{diff}$ is computed and $d_new$ 
overwrites $d_{old}$ after a disk rotation.
$d_{diff}$ and the address of $P_{old}$ are then sent via the DAC to the parity disk 
where it is XORed with $p_{old}$ to compute $P_{new}$,
which overwrites $p_{old}$ after one disk rotation.
The mean write response ties is the maximum of data and parity writes.

\vspace{-2mm}
\begin{eqnarray}\label{eq:asymmetric}
R^{max}_{2} (\rho) = \sum_{i=1}^2 R_i (\rho)  -
\sum_{m=0}^{k_1 - 1}
\sum_{n=0}^{k_2 - 1}
\binom{m+n}{m}
\frac{\mu_1^m \mu_2^n }
{(\mu_1 + \mu_2 )^{m+n+1} } ,
\end{eqnarray}
where $\mu_1= k_1/ R_1 (\rho) $ and $\mu_2 = k_2 / R_2 (\rho)$.

The more flexible Coxian distribution used by Chen and Towsley 1993 \cite{ChTo93}.
can be applied to any response time distribution regardless of the value of $c_R$. 

\section{Rebuild Processing in RAID5}\label{sec:RAID5rebuild}

The discussion in this section is based on HDDs where rebuild time is a problem, 
but while it is less of a problem in Flash SSDs as discussed in Section \ref{sec:flash} 
RAID availability is critical for some applications such as OLTP and e-commerce.
RAID with one disk failure is vulnerable to data loss 
if a second  disk fails or an unreadable disk sector is encountered.
The possibility of a second disk failure is high because of high rebuild time.
RAID repair requires rebuilding the contents of a failed disk 
and this should be carried out concurrently with 
the processing of external requests due to the high cost of downtime. \newline
\begin{scriptsize}
\url{https://queue-it.com/blog/cost-of-downtime/}                    \newline
\end{scriptsize}
Rebuild time is lengthened due to the interference of external disk requests
and conversely response times are increased by rebuild reads.

Reconstructed data is written onto a spare disk in dedicated sparing 
or spare areas on $N$ surviving disks in distributed sparing 
Thomasian and Menon 1994/1997 \cite{ThMe94,ThMe97}.
Restriping which is overwriting check strips was proposed in Rao et al. 2011 \cite{RaHG11}.
In parity sparing two RAID5 arrays are merged and 
parity blocks on one RAID5 are used as spare areas Reddy et al. \cite{ReCB93}.

RAID5 rebuild time in is investigated via simulation in Holland et al. 1994 \cite{HoGS94},
where it is concluded that {\it Disk-Oriented Rebuild - DOR} 
is superior to {\it Stripe-Oriented Rebuild - SOR},
which proceeds by reading disk strips one stripe at a time.
SOR introduces unnecessary synchronization delays at the cost of reducing buffer requirements with respect to DOR,
where {\it Rebuild Units - RUs} are read from disks independently when the disk is idle.

Before the introduction of ZBR the RU size was set to be fixed size tracks. 
Fixed RU sizes can also be adopted with ZBRs.
RUs are read into the DAC cache where corresponding RUs are XORed 
to produce successive RUs on the failed disk.
The analysis is complicated by the fact disks continue processing external disk requests,
which interfere with rebuild reads.
Reading of successive RUs on surviving disks starts when the disk is idle.

With ZBR rather than {\it Constant Angular Velocity - CAV} disks 
a higher data transfer rate is achieved at outer disk tracks.
With ZBR the RAID stripe unit (strip) size can be varied across tracks
and then RU sizes could be set to be a fraction or multiple of strip sizes.

Rebuild time decreases with increased RU size due 
to the fact that fewer seeks are incurred in rebuild processing,
but large RU sizes will result in an unacceptable increase in disk response times.
Seek times an can be reduced by utilizing several anchor points 
to reduce seek distances from the point 
where the last external requests was served Holland et al. 1994 \cite{HoGS94}. 
Seek affinity can be exploited in copying the contents of a failing disk 
or the surviving mirrored disk where the other disk has failed.

While RAID5 during rebuild is vulnerable to data loss due to additional disk failures,
most rebuild failures are due to {\it Latent Sector Errors -  LSEs} Dholakia et al. 2008 \cite{DEH+08}.
A queueing model to estimate rebuild time in RAID5 with HDDs is discussed in the next section.

\section{Vacationing Server Model for Analyzing Rebuild Processing$^*$}\label{sec:RAID5rebuildanalysis}

The {\it Vacationing Server Model - VSM} with multiple vacations with two types 
based on the analysis in Doshi 1985 \cite{Dosh85}, Takagi 1991 \cite{Taka91} 
was adopted in Thomasian and Menon 1994, 1997 \cite{ThMe94,ThMe97} to analyze rebuild processing.
The analysis yields the increased response time of external disk requests due to rebuild processing
and the effect of external disk requests on expanding rebuild processing time.
The accuracy of the analyses was verified using 
a random-number driven simulations in aforementioned studies. 

VSM starts reading successive {\it Rebuild Units - RUs} after the processing of external disk requests is completed, 
or a busy period ends using queueing theory terminology. 
Rebuild reads are resumed at the point where the last read was halted as there was an external disk requests.

\begin{figure}[h]
\begin{center}
\includegraphics[scale=0.55,angle=00]{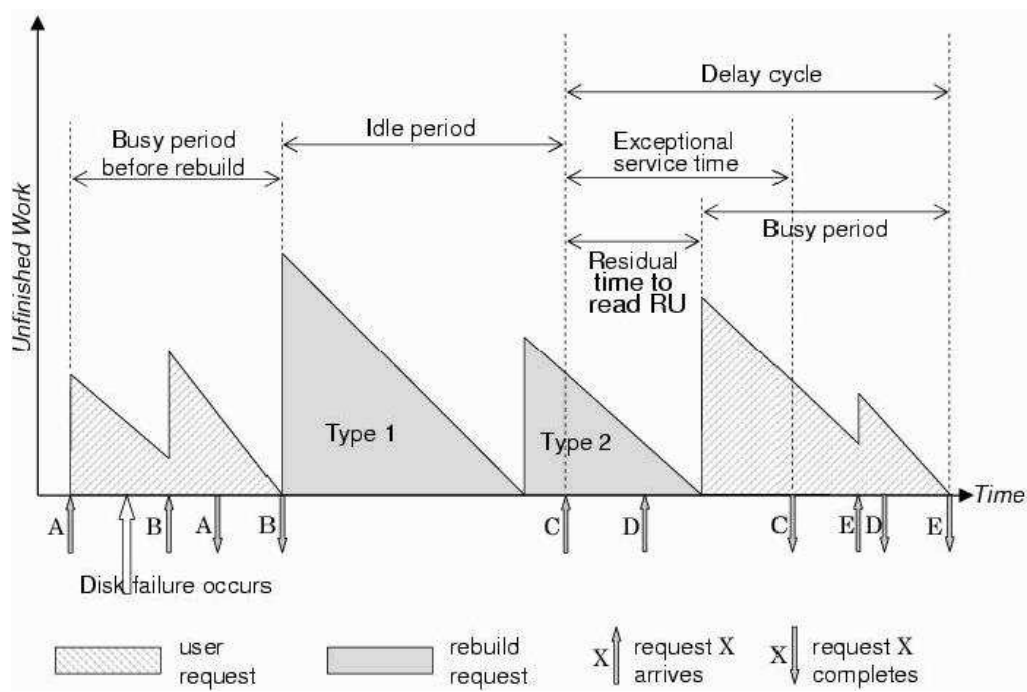}
\caption{Key VSM parameters associated with rebuild processing.}
\label{fig:rebuild}
\end{center}
\end{figure}

The initial RU read request takes longer, 
since it involves a seek and latency (estimated in Section \ref{sec:alternative})
as shown in Figure \ref{fig:rebuild}).
The reading of succeeding RUs takes less time, since a seek is not required,
e.g., a disk rotation if the RU size is a track.

The delay encountered by an external request requests with VSM follows from the fact that 
{\it Poisson Arrivals See Time Averages - PASTA} Wolff 1982 \cite{Wolf82}:
The mean waiting time for external disk requests is the sum of three delays:              \newline
(1) the mean waiting time in the queue,
which is a product of mean queue size ($\bar{N}_q$) and mean service time ($\bar{x}$).    \newline
(2) if the server is busy with probability $\rho$
in which case there is an additional delay given by the mean residual service time
($\bar{x}_r=\bar{x^2}/(2 \bar{x})$ Kleinrock 1975 \cite{Klei75}).                         \newline
(3) the system is idle with probability $1-\rho$,
in which case the delay is the mean residual vacation (RU reading) time ($\bar{v}_r=\overline{v^2}^2 / (2 \bar{v})$,
where the $\overline{v^i}$ is the $i^{th}$ moment of vacation time. 

\vspace{-1mm}
$$W = \bar{N}_q \bar{x} + \rho \bar{x}_r + (1-\rho) \bar{v}_r \mbox{ due to Little's result:}
\bar{N}_q = \lambda W,  \hspace{2mm}\rho = \lambda \bar{x}$$

Simplifying the above equation using the two substitutions yields:
\vspace{-2mm}
\begin{eqnarray}\label{eq:VSM}
W_{VSM} = W_{M/G/1} + \bar{v}_r = \frac{ \lambda \bar{x^2} } { 2 (1- \rho) } + \bar{v}_r.
\end{eqnarray}

An analysis of VSM with multiple vacations of two types is repeated in Section \ref{sec:VSM}.

There is little variation in the completion time of disk rebuild reading times 
(see e.g., Figure 4 in Thomasian and Menon 1997 \cite{ThMe97}).
The spare disk is not a bottleneck unless the system is clustered RAID, 
which is discussed in Section \ref{sec:CRAID}, 
so that rebuild time can be approximated with rebuild read time. 

Assuming that the RU is a track, the number of disk tracks: $N_{track}$
and the mean number of tracks read per cycle: $\bar{n}_{track}$,  
where the cycle time ($T_{cycle}$) is defined and below by Eq. (\ref{eq:cycletime}).

\vspace{-2mm}
\begin{eqnarray}\label{eq:rebuildtime}
T_{rebuild} (\rho) = \frac{N_{track}}{\bar{n}_{track}} \times T_{cycletime}.
\end{eqnarray}


The M/G/1 queue alternates between busy periods with mean $\bar{g}$ 
during which external disk requests are served 
and idle periods given by mean interarrival time ($1/\lambda$) during which rebuild reads are processed,
although the the processing of 
the first external request is delayed until a rebuild request is completed, 
unless preemption is allowed as discussed in Thomasian 2005b \cite{Thom05b}.
The utilization factor $\rho$ is the fraction of time the server is busy, which can be expressed as follows:
\vspace{-2mm}
\begin{eqnarray}\label{eq:meanbusy}
\rho = \frac{\overline{g}}{\overline{g}+1/\lambda} \Longrightarrow \overline{g} = \frac{\bar{x}_{disk}}{1-\rho}.
\end{eqnarray}

The delay cycle is different from a busy period in that it starts with a special request
whose mean is the sum of mean residual rebuild time ($\bar{v}_r$) and $\bar{x}_{disk}$ 
as shown in Fig. \ref{fig:rebuild}.
The mean duration of the delay cycle using the intuitive argument 
expressed by Eq.~(\ref{eq:meanbusy}) is then Thomasian 2018 \cite{Thom18}:

\begin{eqnarray}\label{eq:dc}
\rho =
\frac{
\bar{T}_{dc} (\rho) - \bar{v}_r}
{ 
\bar{T}_{dc} (\rho) + 1/\lambda
}
\hspace{2mm}
\Longrightarrow
\hspace{2mm}
\bar{T}_{dc} (\rho) = \frac{\bar{x}_{disk} + \bar{v}_r }{ 1- \rho}.
\end{eqnarray}

The cycle time is defined as the time between 
the start of successive busy periods and is given by Eq. (\ref{eq:cycletime}): 

\vspace{-2mm}
\begin{eqnarray}\label{eq:cycletime}
\bar{T}_{ct} (\rho) = \bar{T}_{dc} (\rho) + \frac{1}{\lambda}.
\end{eqnarray}

Given their LSTs of busy period and delay cycle Kleinrock 1975 and 1976 \cite{Klei75,Klei76} leads to their moments. 
The variances obtained in this manner can be used to obtain 
to obtain a better estimate of rebuild time as given by Eq. (\ref{eq:Xmax})

\section{M/G/1 VSM Analysis of Multiple Vacations with Two Types$^*$}\label{sec:VSM}

The analysis in this section is based on Doshi 1985 \cite{Dosh85}, 
which is summarized in Takagi 1991 \cite{Taka91} and 
was adopted in Thomasian 1994/1997 \cite{ThMe94,ThMe97}.
The analysis assumes that the RU size is a track.

The first RU/track read at the start of an idle period requires a seek followed by the reading of a track,
while successive rebuild reads just take a disk rotation, as shown in Figure \ref{fig:rebuild}.
Due to ZLA reading of a track starts with the first encountered sector, 
i.e., the rotational latency is negligible. 
Track and cylinders skew in reading successive tracks is negligible Jacob et al. \cite{JaNW08}.

\begin{framed}
\subsection{The Laplace Stieltjes Transform for Seek Time$^*$}\label{sec:LST}

LSTs are discussed in Appendix I in Kleinrock 1975 \cite{Klei75} 
and Appendix D in Trivedi 2002 \cite{Triv02}. 
To obtain the LST of seek time required for type 1 vacations
we approximate the distribution in Eq. (\ref{eq:dist}) by a continuous function,
whose {\it probability density function - pdf} is: 

\vspace{-2mm}
\begin{eqnarray}\label{eq:4p}
f_X (x) = p \delta (x) + (1-p) \frac {2 (C-x) } { C(C-2)},
\end{eqnarray}
where $\delta (x)$ is the unit impulse at $x=0$.
$(C-2)$ is used instead of $(C-1)$ in the denominator, 
so that $\int_{x=1}^{C-1} \frac{2(C-x)}{(C(C-2)} dx=1$. 

Motivated by Abate et al. 1968 \cite{AbDW68} to derive the LST of seek time 
by first using a piecewise linear approximation to the seek time characteristic:
seek time $T_k$ versus seek distance $C_k$, where $C_1=1$ and $C_{K+1}=C-1$. 
\vspace{-1mm}
$$t(x)=\alpha_k x - \beta_k  \hspace{5mm} C_k \leq x \leq C_{k+1}.  $$
\vspace{-1mm}
$$ \alpha_k = \frac{C_{k+1} - C_k} {T_{k+1}- T_k} \mbox{   and  }\beta_k= \alpha_k T_k - C_k.$$
Given that $x$ denotes the number of moved tracks: 

\vspace{-2mm}
\begin{eqnarray}\label{eq:5}
x= \sum_{k=1}^K (\alpha_k t- \beta_k ) [ u(t - C_k)- u(t- C_{k+1} ],
\end{eqnarray}
where $u(x)$ is the unit step function Kleinrock 1975 \cite{Klei75}.
\vspace{-1mm}
$$ 
u(x)= 
\begin{cases}
1 \mbox{ for }x \geq 0 \\
0\mbox{  for  } x < 0. 
\end{cases}
$$

The LST of seek time using Eq. (\ref{eq:4p}) and Eq. (\ref{eq:5}) is, but $p=1/C$ was used in the study: 
\begin{eqnarray}
{\cal L}^*_s (s) = \frac{p}{s} + \frac{2(1-p)}{C(C-2)}
\sum_{k=1}^K  \alpha_k (C+\beta_k ) \frac{ e^{-s T_k} - e^{-sT_{k+1}} }{s} \\
\nonumber
-\alpha_k^2  \left( \frac {  T_k e^{-s T_k} - T_{k+1} e^{-s T_{k+1}} } {s}  
-     \frac{ e^{-s T_k} -         e^{-s T_{k+1}} } {s^2}  
\right)  
\end{eqnarray}

The LST of disk service time according to Eq. (ref{eq:diskservice}) 
is the product of the LSTs of it components:
\vspace{-1mm}
$${\cal L}^*_{disk} (s) =  {\cal L}^*_s (s) \times {\cal L}^*_{\ell} (s) \times {\cal L}_t (s) .$$  

The moments of disk service time can be obtained by taking the derivatives of ${\cal L}^*_{disk}$, 
such as Eq. (\ref{eq:vacmoments}).


\end{framed}

Reading of RUs is modeled as two vacation types: 
Type 1 vacations start once the disk is idle and 
require a seek to access the next RU to be read.
Type 2 vacations are reads of consecutive RUs until an external request arrives.

The PDFs of two vacation types are denoted by $V_i (t),i=1,2$
In the case of Poisson arrivals the probability of no arrival (resp. an arrival) is
$e^{-\lambda t}$ (resp. $1-e^{-\lambda t}$) in time interval $(0,t)$.
Let $p_i$ denote the probability that a request arrives during the $i^{th}$ vacation.
Unconditioning on $V_i (t)$

\vspace{-2mm}
\begin{eqnarray}
p_i= \left[ 1- \int_0^\infty e^{-\lambda t}dV_i (t) \right] 
\prod_{j=1}^{i-1} \int_0^\infty e^{-\lambda T} dV_j(t)
=[1-  V_i^* (t)] \prod_{j=1}^{i-1} V_j^* (\lambda).
\end{eqnarray}
where $V^*_j(\cdot)$ is the LST of the $j^{th}$ vacation time. 
The mean number of tracks read per idle period is:

\vspace{-2mm}
\begin{eqnarray}\label{eq:ntrack} 
\bar{n}_{track} = \bar{J} = \sum_{j=1}^\infty j p_j = 
1 + \frac { V^*_1 \lambda) }{1 - V^*_2 ( \lambda) }
\end{eqnarray}

The split seek-option skips track reading following a seek
if there is an arrival during the rebuild seek Thomasian and Menon 1994 \cite{ThMe94}.
Preemption during the latency and even track transfer phases 
are studied via simulation in Thomasian 1995 \cite{Thom95}.
While preemption results is an improvement in disk response times, 
rebuild time is elongated since more requests are processed 
and this results in an increased number of requests during rebuild processing (with increased response times).
The cumulative sum of response times is an appropriate metric in comparing rebuild options. 

The probability of an arrival during the first and $i^{th}$ vacation is:
\vspace{-1mm}
\begin{eqnarray}
p_1= 1 - V^*_1 (\lambda) \hspace{5mm}
p_i = [1 - V^*_2 (\lambda) ] V_1^* (\lambda) ] V^*_1 (\lambda) [V^*_2 (\lambda)]^{i-2}
\end{eqnarray}

The probability that the $i^{th}$ vacation occurs is $q_i = p_i / \bar{J}$,
where $\bar{J}$ was given by Eq. (\ref{eq:ntrack}).
The probability $q_i, i=1,2$ that an external arrival occurs during a type one or two vacation is:
\vspace{-1mm}
$$q_1 = \frac{1}{ 1 + V^*_1 (\lambda) / (1-V^*_2\lambda) }, \hspace{5mm}q_2= 1 - q_1.$$
It follows that overall LST for vacations is:

\vspace{-2mm}
\begin{eqnarray}\label{eq:LTvac}
V^* (\lambda) = \frac{ (1- V^*_2 (\lambda) ) V_1^*(s) + V^*_1 (\lambda) V_2^* (s) }
{ 1- V_2^* (\lambda) + V_1^* (\lambda) }.
\end{eqnarray}   

Derivatives of the LST Eq. (\ref{eq:LTvac} set to zero yield the $j^{th}$ moment of vacation/rebuild time:

\vspace{-2mm}
\begin{eqnarray}\label{eq:vacmoments}
\overline{v^j} = 
(-1)^j \frac{dV^j(\lambda)}{d\lambda} |_{\lambda=0} = 
\sum_{i=1}^\infty q_j \overline{v_i^j}  =
\frac{ (1 - V^*_2 (\lambda) ) \overline{v}_{1^j} + V^*_1 (\lambda) \overline{v}_{2^j} }
{ 1 - V_2^* (\lambda) + V_1^* (\lambda) }.
\end{eqnarray}

The mean residual vacation/rebuild time is the second moment divided by twice the mean, 
as noted earlier: $\overline{v_r} =  \overline{v^2} / ( 2 \overline{v} )$.

The variation in disk utilization due to redirection in computing $T_{rebuild} (\rho_i)$
is taken into account by computing rebuild processing time over $k$ intervals.
A read only workload to simplify the discussion results in the doubling of disk loads,
but otherwise the load decrease as rebuild progresses are given by Eq. (\ref{eq:degraded}).
The rate of rebuild processing accelerates as disk utilizations drop due to redirection.

\vspace{-2mm}
\begin{eqnarray}
T_{rebuild} (\rho) = \sum_{i=1}^k T_{rebuild} (\rho_i), \hspace{2mm}
\rho_i = [ 2 - \frac{i}{k}] \rho.
\end{eqnarray}

RAID5 rebuild time at an initial disk utilization $\rho$ can be approximated by Eq. (ref{eq:beta}), 
where $\rho$ is the disk utilization when rebuild starts with all read requests,
which results in highest increased load in degraded mode.

\vspace{-2mm}
\begin{eqnarray}\label{eq:beta}
T_{rebuild} (\rho) = \frac{T_{rebuild} (0) }{1- \beta \rho},\mbox{   where  }\beta \approx 1.75.  
\end{eqnarray}

\section{An Alternate Method to Estimate Rebuild Time$^*$}\label{sec:alternative}

Memory bandwidth does not constitute a bottleneck in determining disk rebuild time,
as shown in Section 5.2 of Dholakia et al. 2008 \cite{DEH+08}, 
so we only consider disk bandwidth. 
Rebuild time is estimates  as the ratio of utilized disk capacity ($U$)
and mean rebuild bandwidth: $b_d$ is presented in this section.

With varying track circumference the number of sectors do not vary track to track.
Zoning subdivides disk tracks into $I$ bands with the same number of sectors,
There are $n_i$ tracks with $c_i$ sectors per track in the $i^{th}$ band.

The sector size in FBA disks used to be 512 bytes, 
but there was a transition to 4096 byte sectors circa 2010.  
It follows that the mean disk transfer rate is:                        
\vspace{-1mm}
$$t_d = \sum_{i=1}^I  n_i  \times c_i \times s  / T_{rot}.$$              
 
Given that the RU size is $s_{RU}$ the transfer time per RU is $s_{RU}/t_d$,
but the first RU according to VSM requires a seek ($\bar{x}_s$)
plus rotational latency estimated as follows.
Let $f$ denote the size of RU as a fraction of the tracks size,
then if the R/W head lands inside the RU then we need a full disk rotation 
to transfer an RU ($f \times T_{rot}$) and otherwise the cost is $(1-f)^2 T_{rot}/2$, hence:
\vspace{-1mm}
$$\bar{x}_{\ell} = (1+f^2)T_{rot}/2$$
The following RU accesses incur a track or cylinder skew Jacob et al. 2008 \cite{JaNW08}.

The number of bytes transferred per cycle is the product 
of the number of RUs transferred ($\bar{n}_{RU}$) and RU size $(s_{RU})$.
The cycle time is the sum of interarrival time ($1/\lambda$),
the delay cycle for processing external disk requests given by Eq. (\ref{eq:dc}),
\vspace{-1mm}
$$\bar{T}_{dc} = \frac{ \bar{x}_{disk}+\bar{v}_r}{1-\rho}$$ 
Given $\bar{x}_{RU} = s_{RU} / t_d$ the mean transfer rate for RUs is:

\vspace{-2mm}
\begin{eqnarray}\label{eq:bd}
b_d = \frac { \bar{n}_{RU} \times s_{RU} }
{ \bar{T}_{dc} + \bar{x}_{seek} + \bar{x}_{\ell} + \bar{n}_{RU} \times \bar{x}_{RU} }
\end{eqnarray}

An upper bound to rebuild time can be obtained by setting $\bar{n}_{RU}=1$,
which is also the assumption made in Dholakia et al. 2008 \cite{DEH+08}.
In fact the number of consecutively transferred RUs with no intervening seeks 
can be derived as follows. 
\vspace{-1mm}
$$p_n = (1- e^{-\lambda x_{RU}}) (e^{-\lambda x_{RU}})^{n-1} , n \geq 1
\mbox{  hence  }\bar{n}_{RU} = (1 - e^{-\lambda s_{RU}})^{-1}$$

In the case of ZBR outer disk tracks have have higher capacities
and a small fraction of disk tracks are utilized for critical data, 
since this allows shorter seeks or short strocking.
With a fraction $U$ of disk capacity: $C_d$ utilized rebuild time is:

\vspace{-3mm}
\begin{eqnarray}
T_{rebuild} = U C_d / b_d.
\end{eqnarray} 

Dholakia et al. 2008 \cite{DEH+08} estimates $b_d$ in Eq. (13) as follows:  
\vspace{-1mm}
$$b_d = s_{req} / ( 1/r_{io} +s_{req} / t_d) ,$$
where $r_{io}$ is the request rate and $s_{req}$ is the request size. 

A major shortcoming of most studies is the assumption that disks are fully occupied   
and disk requests are uniformly distributed over all disk cylinders.
The analysis of seek distances in Thomasian et al. 2007 \cite{ThFH07} 
and outlined in Section \ref{sec:diskservice} can be modified to take 
into account only accesses to outer disks.
A mapping of files onto disk cylinders and their access rates can be used 
to obtain more accurate estimates of seek times given the seek time characteristic,
but such studies are precluded by striping which scatters data. 

\section{RAID Reliability Analysis}\label{sec:RAIDrel}

RAID5 disk arrays with $N+1$ disks continue their operation without data loss 
but in degraded mode with a single disk failure.
Denoting disk reliability versus time ($t$) as $r(t)$ (with $r(0)=1$)
the reliability of RAID5 at time $t$ with no repair:
\vspace{-1mm}
$$R_{RAID5} (t) = r^{N+1} (t)   + (N+1) (1-r (t) ) r^N (t) = (N+1) r^{N-1} (t) - N r^N (t)$$
Generally the reliability for RAID(4+k) with $N+k$ disks with no repair is:

\vspace{-2mm}
\begin{eqnarray}\label{eq:RAID(4+k)}
R_{RAID(4+k)} = \sum_{i=0}^k  \binom{N+k}{i} r^{N+k-i}(t)  (1-r(t))^i.
\end{eqnarray}

Due to its mathematical tractability the time to disk failure in Gibson 1992 \cite{Gibs92}
is approximated by the negative exponential distribution: $r(t)=e^{-\delta t}$, 
where $\delta$ is the disk failure rate.
\footnote{We are using $\delta$ here since we are reserving $\lambda$ as the request rate in queueing systems.}
The Weiball distribution, see e.g., Trivedi 2002 \cite{Triv02}), 
is a better fit and was utilized in Elerath and Schindler 2014 \cite{ElSc14}.

RAID5 reliability analysis is given in Section 5.4.1 in \cite{Gibs92}.
\footnote{Out of print, but also not available at \url{https://www2.eecs.berkeley.edu/Pubs/Dissertations/}}            
A similar analysis in the context of mirrored disks is given by Example 8.34 in \cite{Triv02}, 
where a surviving disk is being copied into a spare.

In addition to the $N+1$ disks in RAID5 a spare disk is available 
so that rebuild processing can be started right away after a disk failure. 
The disk {\it Mean Time to Failure - MTTF}$=1/\delta$ 
and the RAID5 {\it Mean Time to Repair - MTTR}$=1/\mu$.
There will be data loss if a second disk fails before rebuild is completed.
A much more common cause of rebuild processing 
is the stalling of rebuild processing due to LSEs as discussed in Section \ref{sec:DEH}.

The evolution of the system is modeled by a {\it Continuous Time Markov Chain - CTMC}, 
see e.g., Kleinrock 1975 \cite{Klei75} and Trivedi 2002 \cite{Triv02}.
The CTMC has three states ${\cal S}_i, i=0,2$, 
where $i$ is the number of disks failed disks.
We have the following transitions:

\vspace{-1mm}
$$
{\cal S}_0 \xrightarrow{(N+1) \delta} {\cal S}_1 , \hspace{5mm} 
{\cal S}_0 \xleftarrow{\mu} {\cal S}_1 , \hspace{5mm}
{\cal S}_1 \xrightarrow{N \delta} {\cal S}_2
$$

The system evolution is described by the following set of linear 
{\it Ordinary Differential Equations - ODEs}, 

\vspace{-1mm}
$$\frac{dP_0(t)}{dt} = - (N+1) \delta P_0(t) + \mu P_1(t), \hspace{2mm}
\frac{dP_1(t)}{dt} = - (\delta + \mu ) P_1(t) (N+1) \delta P_0(t), \hspace{2mm}
\frac{dP_2(t)}{dt} = N \delta P_1 (t). $$ 


The {\it Laplace Stieltjes Transforms - LSTs} 
$L^* (s) = \int_0^\infty P(t) e^{-st} dt$ for the DOEs are given by Eq. (\ref{eq:LT})  

\vspace{-2mm}
\begin{align}\label{eq:LT}
s {\cal L}^*_0 (s) - P_0(0) &= -(N+1) \delta {\cal L}^*_0 (s) + \mu {\cal L}^*_1 (s),\hspace{5mm}P_0(0)=1. \\
\nonumber
s {\cal L}^*_1 (s) - P_1)0) &= -(\delta + \mu ) {\cal L}^*_1 (s) + (N+1) {\cal L}^*_0 (s)\hspace{5mm}P_1(0)=0. \\
\nonumber
s {\cal L}^*_2 (s) - P_2(0) &= N \delta {\cal L}**_1 {s}\hspace{5mm}P_2(0)=0. 
\end{align}


Solving for ${\cal L}^*_2(s)$ leads to  

\vspace{-2mm}
\begin{eqnarray}\nonumber
{\cal L}^*_Y (s) = s {\cal L}^*_2 (s) = \frac{2 \delta ^2} {s^2 + ((2N+1)\delta + \mu)s +2N \delta^2}  
= \frac{2 \delta^2} {\zeta-\eta} \left( \frac{1}{s+\zeta} - \frac{1}{s+\eta} \right)         
\end{eqnarray}

The reliability is the sum of probabilities that the system is functional. 

\vspace{-2mm}
$$ R (t) = P_0 (t) + P_1 (t) = 1- P_2 (t)\mbox{ so that } f_Y(t) = - \frac{dR(t)}{dt} = \frac{dP_2(t)}{dt}. $$

\vspace{-3mm}
\begin{eqnarray}\label{eq:RAID5t}
R_(t) = \frac{\zeta e^{\eta t} - \eta e^{\zeta t}}{\zeta - \eta}
\mbox{ where: }
\zeta,\eta =  \frac{1}{2}[-(2N+1)\delta+\mu \pm \sqrt{\delta^2+ \mu^2+2(2N+1)\delta \mu}]. 
\end{eqnarray}


Using RAID5 reliability $R(t)$ given by Eq. (\ref{eq:RAID5t}) 
and using the equation to obtain the expected value we have:

\vspace{-2mm}
\begin{eqnarray}
\mbox{MTTDL}_{RAID5} = \int_{t=0}^\infty R(t) dt =
\frac{1}{\zeta - \eta} [- \frac{\zeta}{\eta} +\frac{\eta}{\zeta}] = 
- \frac{\zeta+\eta}{\zeta \eta} = - \frac{1}{\zeta} - \frac{1}{\eta}
\end{eqnarray}

Substitutions leads to the following equation:

\vspace{-2mm}
\begin{eqnarray}\label{eq:R5MTTDL}
\mbox{MTTDL} = \frac{ (2N+1) \delta + \mu }{ N(N+1) \delta^2 }
\approx \frac{\mu}{N(N+1)\delta^2} = \frac{MTTF^2}{N(N+1)\mbox{MTTR}}.
\end{eqnarray}

A simple method to derive Eq. (\ref{eq:R5MTTDL}) is to multiply 
the number of times the normal state: ${\cal S}_0$ and the degraded state: ${\cal S}_1$ are visited, 
before there is a transition to the failed state: ${\cal S}_2$.
Given that the probability of an unsuccessful rebuild is $p_f = N \delta / (N \delta + \mu)$,
then the probability of $k$ successful rebuilds is: 
\vspace{-1mm}
$$p_k = (1-p_f)^{k-1} p_f, k \geq 1 \mbox{  with  mean  }\bar{k}= 1/p_f=  1 +\mu/(N \delta) 
\approx \frac{MTTDL}{N \times MTTR}.$$

Chen et al. 1994 \cite{Che+94} generalized the formula to RAID$(4+k),k=1,2$ 
with $N$ disks in total and parity group size $G$

\vspace{-2mm}
\begin{eqnarray} 
\mbox{MTTDL}_{\mbox{RAID5}} \approx \frac{\mbox{MTTF}^2}{N(G-1)\mbox{MTTR}}, \hspace{5mm}
\mbox{MTTDL}_{\mbox{RAID6}} \approx \frac {\mbox{MTTF}^3 }{N(G-1)(G-2)\mbox{MTTR}^2}.
\end{eqnarray}

Generalizing the formula and setting $N=G$ (one parity group) we have:

\vspace{-2mm}
\begin{eqnarray}\label{eq:MTTDL}
\mbox{MTTDL}_{\mbox{RAID}(4+k)} \approx
\frac{\mbox{MTTF}^{k+1} (k-1)! }{ n! \mbox{MTTR}^k }
\end{eqnarray}

Chen's formula assumes that the repair rate remains fixed regardless of the number of failed disks, 
so that one disk is repaired at a time.
Given $N$ disks and with $k$ failed disks we have the transitions.
\vspace{-1mm}
$${\cal S}_{N-i} \rightarrow {\cal S}_{N-(i+1)}: \hspace{2mm} (N-i)\delta \hspace{5mm}
{\cal S}_{N-(i+1)} \leftarrow {\cal S}_{N-i}: \hspace{2mm} \mu $$

With two failed disks in RAID6 corresponding strips in both disks are reconstructed together. 
A repair rate proportional to the number of disks is postulated in Angus 1988. 
and furthermore the number of failed disks is allowed to exceed $n-k$ in Angus 1988 \cite{Angu88}.
\vspace{-1mm}
$${\cal S}_{N-i} \rightarrow {\cal S}_{N-(i+1)}: \hspace{2mm} (N-i)\delta \hspace{5mm}
{\cal S}_{N-(i+1)} \leftarrow {\cal S}_{N-i}: \hspace{2mm} i \mu $$

Using the previous notation the Angus MTTDL formula, which was shown to be 
more accurate against simulation results is as follows:

\vspace{-2mm}
\begin{eqnarray}
MTTDL_{Angus} = \frac{ \mbox{MTTF}^{n-k+1} }{k \binom{n}{k} \mbox{MTTR}^{n-k} }
\times \sum_{i=0}^{n-k} \binom{n}{i} ( \frac{MTTR}{MTTF} )^i \approx 
\frac{ \mbox{MTTF}} { k \binom{n}{k} } 
( \frac {\mbox{MTTF} }{ \mbox{MTTR}} )^{n-k}.
\end{eqnarray}

It is shown in Table \ref{tab:Resch} via simulation in Resch and Volvivski 2013 \cite{ReVo13} 
that Chen's formula overestimates MTTDL by a fact $\approx(n-k)!$,
but this is attributable to modeling assumptions. 
Setting MTTR=1 smaller values of MTTF are used for simulation efficiency.

\begin{table}
\centering
\begin{footnotesize}
\begin{tabular}{|c|c|c|c|c|c|c|c|}\hline
$n$    &$k$    &\mbox{MTTF}    &Simul     &Chen     & S/C   & Angus     & S/C   \\ \hline \hline
10     &10     &2000           &1.988E29  &2.000E2  & 0.94  & 1.988E2   &0.944  \\ \hline
10     &9      &2000           &4.488E4   &4.444E4  & 1.010  & 4.467E4  & 1.005 \\ \hline
10     &8      &1500           &9.446E6   &4.688E6  & 2.105  &9.438E6   &1.001  \\ \hline
10     &7      &500            &7.786E7   &1.240E7  & 6.278  &7.591E7   &1.0126 \\ \hline    
10     &6      &200            &6.407E7   &2.511E7  & 25.513 &6.441E7   0.9996  \\ \hline  
\end{tabular}
\end{footnotesize}
\caption{\label{tab:Resch}Validation of Chen and Angus MTTDL formulas versus simulation.}
\end{table}

Both CTMC models differ from the more accurate CTMC model for RAID6 where both RAID6 are rebuilt concurrently,
i.e., the same transition rate applies for transitions ${\cal S}_i \rightarrow {\cal S}_0$, for $i=1,2$.   



A RAID5 system is considered operational with no or one disk failures, 
but the two states differ substantially, e.g., the maximum read bandwidth is halved when a disk fails.
Performability also defined a computation before failure 
is a better of reliability, see Trivedi 2002 \cite{Triv02}.
Performability of four RAID1 variants discussed in Section \ref{sec:mirhyb}
are derived in Thomasian and Xu 2008 \cite{ThXu08}.


In reliability modeling RAID5 can be substituted by a single component with the following reliability, 
where MTTDL is given by Eq. (\ref{eq:R5MTTDL}).

\begin{eqnarray}\label{eq:appr}
R_{\mbox{RAID5}}^{appr} = e^{-t/\mbox{MTTDL}}
\end{eqnarray}

RAID5 reliability using Eq. (\ref{eq:R5MTTDL} and approximate MTTDL by Eq. (ref{eq:appr} 
is compared in Figures 5.4a and 5.4b in Gibson 2002 \cite{Gibs92}.

The failure of a RAID5 with $N+1$ disks due to a second disk failure is:
\vspace{-1mm}
$$\delta_2 = \delta^2 \binom{N+1}{2} c \mbox{MTTDL}_{RAID_5}.$$
where $c$ is the correlation factor and $c>1$ if disk failures are correlated. 


\subsection{RAID5/6 Reliability Analysis With Unrecoverable Errors}\label{sec:7.2}

Based on the discussion inn Dholakia et al. 2008 \cite{DEH+08}.
we extend the previous discussion to include 
the effect of unrecoverable sector failures in addition to disk failures on MTTDL.

We consider a CTMC with two states representing the number of failed disks
and DF and UF states representing a data loss due to a disk failure (DF)
and an unrecoverable sector failure (UF).

$$
{\cal S}_0 \rightarrow {\cal S}_1 :    \hspace{2mm} N \delta  \hspace{3mm}
{\cal S}_1 \rightarrow {\cal S}_{UF} : \hspace{2mm} \mu_1 P_{uf}^{(1)}  \hspace{3mm}
{\cal S}_1 \rightarrow {\cal S}_{DF} : \hspace{2mm} (N-1) \delta   \hspace{3mm}
{\cal S}_1 \leftarrow {\cal S}_0 : \hspace{2mm} \mu_1 (1- P_{uf}^{(1)} 
$$

$P_{uf}^{k},k=1,2$ are given by Eq. (\ref{eq:puf}) for RAI5 and RAID6 arrays, respectively.
The state transition matrix for RAID5 is given as follows.

$$
{\bf Q}= 
\begin{pmatrix}
- N \delta            &N \delta             &0               &0                 \\
\mu (1 - P_{uf}^{(1)} &-\mu -(N-1) \delta   &(N-1)\delta     &\mu P_{uf}^{(1)}  \\
0                     &0                    &0               &0                 \\
0                     &0                    &0               &0                 \\
\end{pmatrix}
$$

Since there are no transitions from the DF and UF states, 
the submatrix $Q_T$ representing states 0 and 1 is: 
\vspace{-1mm}
$$
\begin{pmatrix}
- N ]\delta &N \delta  &0         &0 \\
\mu (1 - P_{uf}^{(1)}  &-\mu -(N-1) \delta \\
\end{pmatrix}
$$

The vector of the average time spent in the transient states 
before the Markov chain enters either one of the absorbing states DF and UF, 
is obtained as follows Trivedi 2002 \cite{Triv02}]
\vspace{-1mm}
$$\tau {\bf Q}_T = {\bf P}_T(0)\mbox{  where }\tau-[\tau_0, \tau_1]\mbox{ and }{\bf P}_T (0) = [1,0.0]$$ 

Solving the above equations leads to
\vspace{-1mm}
$$\tau_0 = 
\frac{ (N-1) \delta + \mu}
{ N(N-1) \delta +\mu P_{uf}^{(1)} } \hspace{5mm} 
\tau_1 = \frac{1}{(N-1) \delta + \mu P_{uf}^{(1)}}
$$

The MTTDL is given by

\vspace{-2mm}
\begin{eqnarray}\label{eq:MTTDLRAID5}
\mbox{MTTDL}_{RAID5} = \tau_0 + \tau_1 = 
\frac{ (2N-1)\delta + \mu}
{N \delta [(N-1)\delta +\mu P_{uf}^{(1)} ] }.
\end{eqnarray}

Numerical results have shown that that RAID5 plus IDR attains the same MTTDL as RAID6.
While IDR incurs longer transfers, this overhead is negligible compared
to the extra disk access required by RAID6 with respect to RAID5.
There is also the cost of an extra disk.

In the case of RAID6 the probability $P_{recf}$ that a given segment of the failed disk 
cannot be reconstructed is upper-bounded by the probability that two 
or more of the corresponding segments residing in the remaining disks are in error. 
As segments residing in different disks are independent, the upper bound to $P_{recf}$ is given by
\vspace{-1mm}
$$P^{UB}_{recf} = \sum_{j=2}^{N-1} \binom{N-1}{j}P_{seg}^j (1- P_{seg}^{N-1-j} \approx \binom{N-1}{2} P^2_{seg}.$$

Given $N_d$ segments per disk drive their reconstruction 
is independent of the reconstruction of the other disk segments.
The probability that an unrecoverable failure occurs, 
because the rebuild of the failed disk cannot be completed is given by:

\vspace{-1mm}
$$P_{uf}^r = 1 - (1- p^{UB}_{recf} ) ^n_d$$

It is assumed that rebuild time in the degraded and critical mode 
are exponentially distributed with rate $\mu_1$ and $\mu_2$.
Let ${\cal S}_i, i=0,1,2$ denote states the operating states of RAID6 with $i$ failed disks,
${cal S}_{DF}$ and ${\cal S}_{UF}$ failed states due to disk failure and unreadable  sectors.
We have the following transitions due to failures and repairs
\vspace{-1mm}
$$
{\cal S}_0 \rightarrow {\cal S}_1 :    \hspace{2mm} N \delta  \hspace{3mm}
{\cal S}_1 \rightarrow {\cal S}_2 :    \hspace{2mm} (N-1) \delta   \hspace{3mm}
{\cal S}_1 \rightarrow {\cal S}_{UF} : \hspace{2mm} \mu_1 p_{uf}^{(2)}  \hspace{3mm}
{\cal S}_2 \rightarrow {\cal S}_{DF} : \hspace{2mm} (N-2) \delta   \hspace{3mm}
{\cal S}_2 \rightarrow {\cal S}_{UF} : \hspace{2mm} \mu_2 P_{uf}^{(2)} 
$$
\vspace{-1mm}
$$
{\cal S}_1 \leftarrow {\cal S}_0 : \hspace{2mm} \mu_1 (1- P_{uf}^r ) \hspace{3mm}
{\cal S}_2 \leftarrow {\cal S}_0 : \hspace{2mm} \mu_2 (1- P_{uf}^{(2)} ).
$$

Similarly to RAID5 we have the following submatrix ${\bf Q}_T$ for states (0,1,2): 
\vspace{-1mm}
$$\tau {\bf Q}_T = {\bf P}_T(0)\mbox{  with  }{\bf P}_T (0) = [1,0.0]$$ 
\vspace{-1mm}
$$
\begin{pmatrix}
-N \delta                & N \delta     & 0                      \\
\mu_1 (1-P_{uf}^{(r)}    &-(N-1)\delta  &- \mu (N-1)\delta       \\ 
\mu_2(1- P_{uf}^{(2)}    &0             &-(N-2)\delta -\mu_2     \\  
\end{pmatrix}
$$

$$
\tau_0 = \frac{ [(N-1) \delta +\mu_1] [(N-1) \delta +\mu_2]}{N \delta V}, \hspace{2mm}
\tau_1 \frac{(N-2)\delta +\mu_2 }{V} \hspace{2mm} \tau_2 - \frac{N-1)\delta }{V}
$$

\vspace{-1mm}
$$V = [(N-1)\delta+ +\mu P_{uf}^{(r)}]
[(N-2)\delta + \mu_2 P_{uf}^{(r)}]  + \mu_1 \mu_2 P_{uf}^{(r)} (1- P_{uf}^{(2)} ) $$

\vspace{-1mm}
$$MTTDL_{RAID6} = \tau_0 + \tau_1+ \tau_2$$

Given $P_{uf}^{r)}=P_{uf}^{(1)}=0$ then $V=[(N-1)(N-2)\delta^2$
It is to be noted that both failed disks are rebuild concurrently,
but this is not the case with the Chen and Angus models.

\section{Disk Scrubbing and IntraDisk Redundancy to Deal with Latent Sector Errors}\label{sec:DEH}

The main cause of rebuild failures are {\it Latent Sector Errors - LSEs} rather than whole disk failures.
{\it IntraDisk redundancy - IDR} in Dholakia et al. 2008 \cite{DEH+08} and 
disk scrubbing are two less costly methods to deal with LSEs. 

The probability of uncorrectable disk failures ($P_{uf}$) due segment errors is used in \cite{DEH+08} 
to extend RAID5 reliability analysis in Section \ref{sec:RAIDrel} to take into account LSEs.

IDR divides strips or SUs into segments of length $\ell = n+m$ sectors.
There are $n$ data and $m$ checks sectors per segment.
There are $S$ bits per sector and the probability that a bit is in error is $P_{bit}$.
Given that the probability of an uncorrectable sector error is $P_s$ 
the probability that a segment is in error when no coding is applied ($m=0$) is $P_{seg}$ 
\vspace{-1mm}
$$P_S = 1 - (1-P_{bit})^S, \hspace{5mm} P_{seg}^{none} = 1 - (1 - P_S )^\ell .$$

A single parity sector can be used to correct a single sector in error, 
since the sector in error is identified with a {\it Cyclic Redundancy Check - CRC},    \newline
\begin{scriptsize}
\url{https://en.wikipedia.org/wiki/Cyclic_redundancy_check} \newline
\end{scriptsize}
so that the sectors in error can be identified.

Disk circa 2010 adopted 4096 B  rather than 512 B sectors,
but the section size $S=512$ in the 2008 study.
The {\it Error Correcting Code - ECC}                      \newline
\begin{scriptsize}
\url{https://en.wikipedia.org/wiki/Error_correction_code}  \newline
\end{scriptsize}
in the latter case is 100 rather 50 bytes
with an efficiency of $4096/(4196)=97.3\%$ versus $512/562= 88.7\%$.
More details about 4 KB sectors known as Advanced Format are given below.     \newline
\begin{scriptsize}
\url{https://en.wikipedia.org/wiki/Advanced_Format}
\end{scriptsize}

There are $n_d = C_d / (\ell S) $ segments on a disk with capacity $C_d$.
The probability of an uncorrectable failure for a RAID$(4+k)$ with $k$ failed disks,
so that it in critical mode:

\vspace{-3mm}
\begin{eqnarray}\label{eq:puf}
P_{uf}^{(k)} = 1 - ( 1- P_{seg} ) ^ { \frac{ (N-k) C_d } {\ell S} }
\end{eqnarray}
Note that the exponent denotes the number of disk segments in critical rebuild mode.

Sector errors can be classified as independent and correlated.
In the case of independent errors there is a probability $P_s$ 
that a sector has an {\it Unrecoverable Failure - UF}.
Correlated errors have an average burst length $\bar{B}$ and 
there are $\bar{I}$ successive error-free sectors on the average.  
Both are modeled with a geometric distribution:                                          
\vspace{-1mm}
$$a_j = P[I=j] =\alpha (1-\alpha)\alpha^{j-1},\mbox{   hence  }\bar{I}=1/\alpha$$          
We also have $P[B=j]=b_j$ and $\bar{B}= \sum_{j \geq 1} j b_j$.
Utilized in further discussions is the following parameter.
\vspace{-1mm}
$$G_n \overset{\Delta}{=} P[\mbox{burst\_length }\geq n]= \sum_{j \geq n} b_j$$ 

The independent model is a special case of correlated model where 
\vspace{-1mm}
$$b_j=(1-P_s)P_s^{j-1}\mbox{ and }\bar{B}=1/(1-P_s)$$

The probability $P_s$ that an arbitrary sector has an unrecoverable error is: 
\vspace{-1mm}
$$ P_S = \frac{ \bar{B} }{\bar{B}+\bar{I}}$$
It follows 
\vspace{-1mm}
$$\alpha = \frac{P_s}{\bar{B}(1-P_s)} \approx \frac{P_s}{\bar{B}} 
\mbox{  hence }P_s \leq \frac{\bar{B}}{\bar{B}+1}$$ 

For independent errors we have:
\vspace{-3mm}
\begin{eqnarray}
P{seg}^{none} = 1- (1- P_s)^\ell \approx \ell P_s, \hspace{5mm}
P_{seg}^{none} \approx (1+\frac{\ell-1}{\bar{B}})P_s
\end{eqnarray}
for bursty errors
\vspace{-3mm}
\begin{eqnarray}
P_{seg}^{none} = 1 - (1-P_s)(1- \frac{1}{\bar{B}} ) \approx (1+\frac{\ell-1}{\bar{B}})P_s
\end{eqnarray}
The probability that a segment is in error can be specified with the summation below,
but only one term need to be considered for small $P_s$ and the correlated model 
\vspace{-1mm}
$$P_{seg} = \sum_{i \geq 1}c_i P_s^i  \hspace{5mm}P_{seg} = c_1 P_s + O(P_s^2)$$

\begin{framed}
\subsection{Schemes to Implement IntraDisk Redundancy - IDR}\label{sec:IDRschemes}

Three IDR methods are proposed and evaluated in \cite{DEH+08}.

\subsubsection*{Reed-Solomon Check}\label{sec:RS}

{\it RS - Reed-Solomon} codes are discussed in coding theory textbooks in Section \ref{sec:intro}.
An error occurs when over $m$ bits among $\ell=n+m$ bits are in error.

\vspace{-2mm}
\begin{eqnarray}
P_{seg}^{RS} = \sum_{j=m+1}^ell P_s^j (1-P_s)^{\ell-j} \approx \binom{\ell}{m+1} P_s^{m+1}.
\end{eqnarray}

In the case of a correlated model for small values of $P_s$.

\vspace{-2mm}
\begin{eqnarray}
P_{seg}^{RS} \approx \left[ 1 + \frac{1}{\bar{B}} 
\left( (\ell - m - 1)G_{m+1} - \sum_{j=1}^m G_j \right) \right] P_s.
\end{eqnarray}

This is the best we can do since for a code with $m$ parity symbols 
for a codeword of $n$ symbols any $m$ erasures can be corrected,
but RS codes are computationally expensive.

\vspace{-2mm}
\begin{eqnarray}
P_{seg}^{RS} = 
\sum_{j=m+1}^\ell \binom{\ell}{j} P_s^j  (1- P_s)^{\ell-j} \approx \binom{\ell}{m+1} P_s^{m+1}.
\end{eqnarray}
In the case of correlated model for small values of $P_s$ results in Appendix B are used to show:

\vspace{-2mm}
\begin{eqnarray}
P_{seg}^{RS} \approx
\left[ 1 + \frac{ (\ell -m -1)G_{m+1} - \sum_{j=1}^m G_j }{\bar{B}} \right] P_s. 
\mbox{where }\bar{B}=\sum_{j=1}^\infty G_j, \hspace{2mm} b_j = G_j - G_{j-1}. 
\end{eqnarray}

\subsubsection*{Single Parity Check - SPC}

The simplest coding scheme is one in which a single parity sector is computed by
using the XOR operation on $\ell - 1$ data sectors to form a segment with $\ell$ sectors. 
The independent model yields:

\vspace{-2mm}
\begin{eqnarray}
P_{seg}^{SPC} = \sum_{j=2}^{\ell} \binom{\ell}{j} P_s^j (1-P_s)^{\ell - j} \approx
\frac{\ell(\ell-1)}{2} P_s^2 
\end{eqnarray}

In the case of correlated model for small values of $P_s$

\vspace{-2mm}
\begin{eqnarray}
P_{seg}^{SPC} \approx
\left[ 1 + \frac{1}{\bar{B}} \left(  (\ell -2)G_2 - 1 \right)\right] P_s. 
\end{eqnarray}

\subsubsection*{Interleaved Parity Check - IPC} 

In this scheme $\ell-m$ contiguous sectors are placed in a rectangle with $m$ columns. 
The XOR of the sectors in a column is the parity sector which is placed in an additional row.
This scheme with $\ell/m$ sectors per interleave can correct a single error per interleave,
but a segment is in error if there is at least one interleave with two uncorrectable segment errors.

An IPC scheme with $m$ ($m \leq \ell/2$) interleaves per segment, 
i.e. $\ell/m$ sectors per interleave, has the capability of correcting a single error per interleave. 
Consequently, a segment is in error if there is at least one interleave 
in which there are at least two unrecoverable sector errors. 
Note that this scheme can correct a single burst of $m$ consecutive errors occurring in a segment. 
However, unlike the RS scheme, it in general does not have 
the capability of correcting any $m$ sector errors in a segment.
The probability of an interleave error is then:
\vspace{-1mm}
$$P_{interleave\_error} = 
\sum_{j=2}^{\ell/m} \binom{\ell/m}{j} P_s^j (1-P_s)^{\ell/m-j}
\approx \frac{\ell(\ell-m)}{2m^2} P_s^2$$

\vspace{-3mm}
\begin{eqnarray}
P_{seg}^{IPC} = 1  - (1-P_{interleave\_error})^m \approx \frac{\ell (\ell- m)}{2m} P_s^2
\end{eqnarray}

In the case of correlated model:

\vspace{-3mm}
\begin{eqnarray}
P_{seg}^{IPC} \approx \left[ 1 + \frac{1}{\bar{B}} 
\left( (\ell-m-1) G_{m+1} - \sum_{j=1}^m G_j \right) \right]  P_s
\end{eqnarray}

It follows $P_{seg}^{RS} \approx P_{seg}^{IPC}$,
but of course IPC  is preferable since it is easier to implement.

\end{framed}

\section{Effect of IDR on Disk Performance}\label{sec:IDRperf}

The main focus in \cite{DEH+08} is to study the effect of IDR on RAID5 reliability, 
but performance is also investigated using simulation and 
{\it IO Equivalents - IOEs} introduced in Hafner et al. 2004 \cite{HDKR04}.

\vspace{-2mm}
\begin{eqnarray}\label{eq:IOE}
IOE(k) = 1 + 
\frac{\mbox{time to transfer 4 KB}}{\mbox{average time per 4 KB}}= 
1 + \frac{\mbox{4 KB/1024}}
{\mbox{Avg. media transfer rate in MB/sec}} \times \mbox{Avg IO/sec.}
\end{eqnarray}

\begin{table}[h]
\caption{Results for IOE.\label{tab:IOE}}
\begin{footnotesize}
\begin{center}
\begin{tabular}{|c|c|c|c|} \hline 
                          &10K PM               &15K RPM             & Units    \\ \hline \hline
IOs/sec.                  &250                  &333                 &IO/sec.   \\ \hline
Avg. media transfer rate  &53.25                &62                  & MB/sec.  \\ \hline
IOE for k KB              &$1+k/55$             &$1+k/48$            &IOE       \\ \hline
\end{tabular}
\end{center}
\end{footnotesize}
\end{table}

The approximation $IOE(k)= 1+ k/50$ is used in the study.

When a single-sector {\bf A} is written using the IPC scheme, for example,
it is written by a single I/O request also containing 
the corresponding intra-disk parity sector {\bf PA}.
The average length of an I/O request is minimized 
when the intra-disk parity sectors are placed in the middle of the segment.
The analysis in Section 8.2 of the paper determines 
the increase in IOE as 8.5\% for small and 3.6\% for long writes

The following parameters were used by the simulator: 
$1/\delta$=500K hours, disk capacity $S=300$ GB, $P_{bit}=10^{-14}$,
number of disks $N=8/16$ for RAID5/6, rebuild time $1/\mu=17.8$ hr, sector size $S=512$ bytes,
$\ell = 128$ sectors, $m=8$ interleaves.


Both random number and trace-driven simulations results are reported  in this study.
FCFS scheduling is used in the study, but other policies such as SSTF, Look, C-Look are tried 
and shown to have a small effect on the relative performance due to IDR.  

\subsection*{Numerical Results}

SATA drives (see Section 20.3) in Jacobs et all. 2008 \cite{JaNW08})
with disk capacity $C_d = 300$ GB, $P_{bit}=10^{-14}$.
so that 512 bytes sectors and $P_s = 4096 \times P_{bit} = 4.096 \times 10^{-11}$.
The segment length was set to $\ell=128$ sectors with $m=8$ interleaving
burst length distribution with at most 16 sector and $\bar{B}= 1.0291$ and \newline
\vspace{-1mm}
\begin{footnotesize}
$$\underline{b}= [0.9812, 0.016, 0.0013(2), 0.0003(2), 
0.0002 0.0001(2), 0, 0.0001, 0, 0.0001(2), 0, 0.0001(2)]$$
\end{footnotesize}

\begin{table}
\centering
\begin{footnotesize}
\parbox{0.45\linewidth}{
\begin{tabular}{|c|c|c|} \hline
Errors     & Independent    & Correlated                              \\ \hline
None       & $5.2 \times 10^{-9}$  & $5   \times 10^{-9}$             \\ \hline
RS         & $6.2 \times 10^{-81}$ & $2.5 \times 10^{-12}$            \\ \hline
SPC        & $1.3 \times 10^{-17}$ & $8.5 \times 10^{-14}$            \\ \hline
IPC        & $1.6 \times 10^{-18}$ & $9.5 \times 2.5 \times 10^{-12}$ \\ \hline
\end{tabular} 
\caption{\label{tab:[pseg}Approximate $P_{seg}$ for two error models}
}
\hspace{2.5mm}
\parbox{0.45\linewidth}{
\begin{tabular}{|c|c|c|} \hline
Scheme &Independent & Correlated \\ \hline
None &   $1.5  \times 10*{-1}$  &  $1.5 \times 10^{-1}$ \\ \hline
RS   &   $2    \times 10^{-73}$ &  $7.9  \times 10^{-5}$ \\ \hline
SPC  &   $4.3  \times 10^{-10}$ &  $3.1 \times 10^{-3}$  \\ \hline
IPC  &   $6.1  \times 10^{-11}$ &  $7.95 \times 10^{-5}$ \\ \hline
\end{tabular}  
\caption{\label{tab:Puf}Approximate $P_{uf}$ for RAID5 with $N=8$.}
}
\end{footnotesize}
\end{table}

\subsection{Comparison of Disk Scrubbing and IDR with IPC}\label{sec:comparison}

The conclusions of quite complex analysis of disk scrubbing in Iliadis et al. \cite{IHHE11} 
are summarized in this section and the analysis is used in comparing it with the IDR scheme.

Closed-form expressions were derived for operation with random, 
uniformly distributed I/O requests, 
for the probability of encountering unrecoverable sector errors, 
the percentage of unrecoverable sector errors that disk scrubbing detects and corrects, 
the throughput and the minimum scrubbing period achievable. 

The probability of an error due to a write is $P_w$
and writes constitute a fraction $r_w$ of disk accesses with rate $h$.
The probability of error in reading a sector without scrubbing is $P_e = r_w P_w$.
The probability of sector failure for deterministic scrubbing 
and random (exponentially distributed) scrubbing period are:
\vspace{-1mm}
$$P_s^{(det)} = [1 - (1- e^{- h T_s})/(h T_S)] P_e \mbox{ and }P_s^{exp)} =  [ h T_S / (1+ h T_S)] P_e,$$
where $T_S$ denotes the mean scrub period.
Given that $h T_s \ll 1 $ we have the following approximations for deterministic and exponential scrubbing:
\vspace{-1mm}
$$ P_s^{(det)}  \approx P_e h T_s \mbox { and } P_s^{(exp)} \approx \frac{1}{2} P_s P_e h T_s $$
Deterministic scrubbing is preferable, since random scrubbing has double the value for $P_s$.



For heavy-write workloads and for typical scrubbing frequencies and loads, 
the reliability improvement due to disk scrubbing does not reach 
the level achieved by the IPC-based IDR scheme, which is insensitive to the load. 
Note that IPC-based IDR for SATA drives achieves the same reliability 
as that of a system with no sector errors, with a small (about 6\%) increase in capacity. 

The penalty of disk scrubbing on I/O performance can be significant, 
whereas that of the IPC-based IDR scheme is minimal,
but of course is some disk space overhead.

Given IOE(k) in Eq. (\ref{eq:IOE}) the maximum rate of read/write operations to the disk is:

\begin{eqnarray}\label{eq:sigma}
\hat{\sigma}_{max}  = [IOE(K) t_{seek} ] ^{-1}
\end{eqnarray}

Given a small write scenario with $k$ sectors per chunk, $S_D$ sectors per disk, 
with $G_S$ sectors scrubbed a time, and $p=1$ for RAID5 and $p=2$ for RAID6 

\vspace{-2mm}
\begin{eqnarray}
\sigma \leq \sigma_{max} (k, T_S) = 
\frac { \hat{\sigma}_{max} (k,T_S ) }
{ 1+ (1+2p) r_w ] }
\end{eqnarray}

The scrubbing $T_S$ period should not be smaller than a critical value $T^*_S$ given below:

\vspace{-2mm}
\begin{eqnarray}
T_S \geq T_S^* =  \frac{ S_D IOE(G_S) }
{ [ 1 + (1+2p) r_w) ] G_S  IOE(k) [\sigma_{max} (k) - \sigma ]}
\end{eqnarray}

In Figure 6 in \cite{IHHE11} it is shown that in the area of practical interest, 
the reliability offered by scrubbing still inferior to that achieved 
by the intradisk redundancy scheme.

Iliadis et al. \cite{IHHE11} argument that that IDR is preferable to disk scrubbing
contradicts that of Schroeder et al. 2010 \cite{ScDG10}. 

\section{Combining Scrubbing with SMART and Machine Learning}\label{sec:combined}


SMART has been applied to predict HDD failures as discussed in 
Hughes et al. 2002 \cite{HMKE02}, Eckart et al. 2008 \cite{ECHS08}.
Once a disk is determined to be failing its contents are copied onto a spare 
with possibly occasional assist to restore unreadable segments using the RAID5 paradigm.  \newline
\begin{scriptsize}
\url{https://en.wikipedia.org/wiki/Self-Monitoring,_Analysis_and_Reporting_Technology}
\end{scriptsize}

In monitoring disks with SMART there are 30 Attributes corresponding 
to different measures of performance and reliability Allen 2004 \cite{Alle04}.  
Attributes have a one-byte normalized value ranging from 1 to 253, 
e.g., SMART 187 is the number of read errors that could not be recovered using hardware ECC. 
If one or more of the Attribute values less than or equal to its corresponding threshold, 
then either the disk is expected to fail in less than 24 hours 
or it has exceeded its design or usage lifetime.         \newline
\begin{scriptsize}
\url{https://www.backblaze.com/blog/hard-drive-smart-stats/}
\end{scriptsize}
Western Digital {\it Network Attached Storage  - NAS} HDDs are automatically given a warning label 
in Synology's {\it DiskStation Manager -DSM after they were powered on for three years.
This is also done Westen Digital device Analytics - WDDA} and Seagate's IronWolf is a smilar offering. \newline
\begin{scriptsize}
\url{https://en.wikipedia.org/wiki/Network-attached_storage}   \newline
\url{https://arstechnica.com/gadgets/2023/06/clearly-predatory-western-digital-sparks-panic-anger-for-age-shaming-hdds/}
\end{scriptsize}

Nimble storage which is part of HPE has advertised predictive storage, 
which makes 86\% of problems disappear.
A 6-nines (99.9999\%) availability is guaranteed over the installed base
and 54\% of problems resolved are determined to be outside storage.     \newline
\begin{scriptsize}
\url{https://m.softchoice.com/web/newsite/documents/partners/hpe/Nimble-Six-Nine-White-Paper.pdf} \newline
\end{scriptsize}
InfoSight developed by Nimble applies data science to identify, predict, 
and prevent problems across infrastructure layers.                       \newline
\begin{scriptsize}
\url{https://en.wikipedia.org/wiki/Data_science}
\end{scriptsize}


{\it Machine Learning - ML} methods are used by Murray et al. 2005  \cite{MuHK05}
to predicting HDD failures by monitoring internally the attributes of individual drives.  
An algorithm based on the multiple-instance learning framework and 
the naive Bayesian classifier (mi-NB) are specifically designed for the low false-alarm case.  
Other methods compared are {\it Support Vector Machines - SVMs}, 
unsupervised clustering,  and non-parametric statistical tests (rank-sum and reverse arrangements).   \newline
\begin{scriptsize}                                          
\url{https://en.wikipedia.org/wiki/Support_vector_machine}                                            \newline
\url{https://www.section.io/engineering-education/clustering-in-unsupervised-ml/}                     \newline
\url{https://www.ncbi.nlm.nih.gov/pmc/articles/PMC4754273/}                                           \newline
\end{scriptsize}
The failure-prediction performance of the SVM, rank-sum and mi-NB algorithm is considerably better 
than the threshold method currently implemented in drives, while maintaining low false alarm rates.   

Goldszmidt 2012 \cite{Gold12} describes,  characterizes,  and evaluates 
D-FAD - Disk Failure Advance Detector of {\it Soon-To-Fail - STF} disks.   
The input to D-FAD is a single signal from every disk, 
a time series containing in each sample the {\it Average Maximum Latency - AML}  
and the output is an alarm according to a combination of statistical models. 
The parameters in these models are automatically trained from a population 
of healthy and failed disks ML techniques.
Results from an {\it Hidden Markov Model - HMM}
and a threshold (peak counter) were fused using logistic regression to sound alarms. \newline
\begin{scriptsize}
\url{https://en.wikipedia.org/wiki/Hidden_Markov_model}                              \newline
\url{https://en.wikipedia.org/wiki/Logistic_regression}                              \newline
\end{scriptsize}
When applied to 1190 production disks D-FAD predicted 15 out of 17 failed disks 
(88.2\% detection rate), with 30 false alarms (2.56\% false positive rate). 

Amvrosiadis et al. 2012 \cite{AmOS12} state that the goal of a scrubber 
is to minimize the time between the occurrence of an LSE and its detection/correction 
or {\it Mean Latent Error Time - MLET}, 
since during this time the system is exposed to the risk of data loss. 
In addition to reducing the MLET, a scrubber must ensure 
to not significantly affect the performance of foreground workloads.

Staggered scrubbing provides a lower MLET exploiting 
the fact that LSEs occur temporal and spatial bursts.
Rather than sequentially reading the disk from beginning to end
staggered scrubbing quickly probes different regions of the drive hoping that 
if a region has a burst of errors the scrubber will detect it quickly and then immediately scrub the entire region. 

Staggered scrubbing potentially reduces scrub throughput and increase the impact on foreground workloads. 
At the time the articles was written there was no experimental evaluation that quantifies this overhead, 
and staggered scrubbing is currently not used in practice.
The second design question is deciding when to issue scrub requests. 
The scrubbers employed in commercial storage systems today 
simply issue requests at a predefined rate, e.g. every $r$ msec. 
Larger request sizes lead to more efficient use of the disk, 
but also have the potential of bigger impact on foreground traffic, 
as foreground requests that arrive while a scrub request is in progress get delayed



RAIDShield developed at EMC characterizes, monitors, 
and proactively protects against disk failures eliminating 88\% of triple disk failures in RAID6,
which constituted 80\% of all RAID failures Ma et al. 2015 \cite{Ma++15}.
A method using the joint failure probability quantifies and predicts vulnerability to failure.
Simulation shows that most vulnerable RAID6 systems can improve the coverage to 98\% of triple errors.

The use of ML to make storage systems more reliable by detecting sector errors in HDDs and SSDs 
is explored by Mahdisoltani et al. 2017 \cite{MaSS17} 
Exploration of a widerange of ML techniques shows that sector errors 
can be predicted ahead of time with high accuracy. 
Possible use cases for improving storage system reliability through 
the use of sector error predictors is provided
The mean time to detecting errors can be greatly reduced by adapting 
the speed of a scrubber based on error predictions.

A scrub unleveling technique that enforces a lower rate scrubbing 
to healthy disks and a higher rate scrubbing to disks subject to LSEs 
is proposed by Jiang et al. 2019 \cite{JiHZ19}.
A voting-based method is introduced to ensure prediction accuracy. 
Experimental results on a real-world field dataset have demonstrated 
that our proposed approach can achieve lower scrubbing cost together 
with higher data reliability than traditional fixed-rate scrubbing methods. 
Compared with the state-of-the-art, our method can achieve 
the same level of Mean-Time-To-Detection - MTTD with almost 32\% less scrubbing.

The first experimental comparison of sequential versus staggered scrubbing in production systems 
determined that staggered scrubbing implemented with the right parameters 
can achieve the same (or better) scrub throughput as a sequential scrubber, 
without additional penalty to foreground applications. 
A detailed statistical analysis of I/O traces is used to define policies for deciding when to issue scrub requests, 
while keeping foreground request slowdown at a user-defined threshold. 
The simplest approach, based on idle waiting and 
using a fixed scrub request size outperforms more complex statistical approaches.


Scrubbing schemes to deal with LSEs have several limitations according to Zhang et al. 2020 \cite{ZWW+20}.
(1) schemes use ML to predict LSEs, but only a single snapshot of training data for prediction; 
ignoring sequential dependencies between different statuses of a HDD over time. 
(2) accelerating scrubbing at a fixed rate based on the results of a binary classification model 
results in unnecessary increases in scrubbing cost; 
(3) naively accelerating scrubbing over full disks neglects partial high-risk areas; 
(4) they do not employ strategies to scrub these high-risk areas in advance based on I/O accesses patterns, 
in order to further increase the efficiency of scrubbing.

{\it Tier-Scrubbing - TS} scheme combines a {\it Long Short-Term Memory  - LSTM} based 
{\it Adaptive Scrubbing Rate Controller - ASRC}, 
a module focusing on sector error locality to locate high-risk areas in a disk, 
and a piggyback scrubbing strategy to improve the reliability of a storage system. 
Realistic datasets and workloads evaluated from two data centers demonstrate 
that TS can simultaneously decreases {\it Mean-Time-To-Detection - MTTD} by about 80\% 
and scrubbing cost by 20\%, compared to a state-of-the-art scrubbing scheme.



Disk fault detection models based on SMART data with ML algorithms 
require a large amount of disk data to train the models, 
but the small amount of training data in traditional ML algorithms 
greatly increases the risks of overfitting or weak generalization, 
which weaken the performance of the model and seriously affect the reliability of the storage systems.

A novel {\it Small-Sample Disk Fault Detection - SSDFD} optimizing method, 
with synthetic data using {\it Generative Adversarial Networks - GANs} \cite{Goo+14}. 
is proposed in Wang 2022 \cite{Wang22} is 
The proposed approach utilizes GAN to generate failed disk data conforming to the failed disk data distribution 
and expands the dataset with the generated data, then the classifiers are trained. 
Disk SMART attributes vary with the usage and are time-related; 
LSTM is adopted since it is good at learning the characteristics of time series data 
as the GAN generator to fit the distribution of SMART data, 
and use the multi-layer neural network as the discriminator to train the GAN-model 
to generate realistic failed disk data. With sufficient generated failed disk data samples, 
ML algorithms can detect disk faults more precisely than before with the small original failed disk data samples.

\section{Undetected Disk Errors and Silent Data Corruption - SDC}\label{sec:UDE}

Bairavasundaram et al. 2007 \cite{BGPS07} is a major study revealing interesting aspects 
of {\it Undetected Disk Errors - UDEs} or LSEs.
The following observations were made:
(1) a high degree of temporal locality between successive LSE occurrences;
(2) the vast majority of disks developed relatively few errors during first three years,
but these few errors can cause significant data loss if not detected proactively bu disk scrubbing. 

{\it Silent Data Corruption - SDC} manifests itself as UDEs
The causes of UDEs and their effects on data integrity 
is discussed by Hafner et al. 2008 \cite{HDBR08}.
Techniques to address the problem at various software layers in the I/O stack
and solutions that can be integrated into the RAID subsystem are discussed at:.     \newline
\url{https://en.wikipedia.org/wiki/Data_corruption}.

UDE types are {\it Undetectable Write/Read Errors - UWEs/UREs}.
UWEs are in the form of dropped or phantom writes and 
off-track or misdirected writes resulting in stale data.
UREs are in the form of {\it Error Correcting Code - ECC} miscorrects
or off-track reads, which result in reading of stale or corrupted data.

Presented in this work are technologies for addressing data corruption problems caused by UDEs,
as well as other causes of data corruption.
Technologies are divided into classes depending on the software layer and the type of problem.

\begin{description}

\item[Checking at middleware or application software layers:]
Some file systems checksum 4-8 KB data chunks and store them separately.
When the data chunk is read the checksum is recomputed and compared.
A mismatch indicates corruption of either the data or the checksum,
but the former is more likely because of its larger size.

This method was utilized by ZFS Karlsson 2006 \cite{Karl06}
and the {\it IRON - Internal RObustNess} file system Prabhakaran et al. 2005 \cite{Pra+05},
The IRON file system prototype which incurs minimal time and space overheads 
via in-disk checksumming, replication, and parity is shown to greatly enhance file system robustness.

\item[Detection of Errors by the Storage System:]
Errors introduced along the datapath between the storage system and the application may not be detected.
This class has several subclasses with respect to different powers of detection and correction 
with associated performance impacts and performance/cost tradeoffs.

\end{description}

By itself data and parity scrubbing cannot detect UDEs.
A parity scrub provides an additional step beyond a data scrub.

\begin{description}

\item[Data scrubbing.]
A data scrub in addition to reading all of the data blocks recomputes the parity
and compares this computed parity with the stored parity on disk.
A miscompare implies a data corruption error.
Data scrubs and parity scrubs do not always provide for recovery,
and a comparison does not always provide a guarantee that no data corruption has occurred.

\item[Metadata options.]
When metadata is colocated with data and metadata contains checksums,
a read of the data and metadata can detect data corruption errors,
especially those introduced by firmware, software, or the memory bus.
When the metadata contains the data address, a read can detect off-track writes 
but only at the offset target location, where data was incorrectly overwritten.
These methods are usually combined with a data scrub 
to enhance its effectiveness as discussed in Sundaram 2006 \cite{Sund06}. \newline
\begin{scriptsize}
\url{https://atg.netapp.com/wp-content/uploads/2018/08/TECH_ONTAP-Private_Lives_of_Disk_Drives.pdf}
\end{scriptsize}

\item[SCSI Write Verify Disk Command.]
This command rereads data after it is written to ensure that it was written correctly.
This works for dropped write errors but cannot necessarily detect off-track writes,
because the same possibly incorrect track is accessed.
This is an obvious performance impact, since a full disk rotation is required.
This method has a simple recovery algorithm which rewrites the data to the same or a different location on disk.

\item
[Verification handled by fresh data in write cache]
When the data needs to be evicted from the cache,
the disk is first read and compared with the cache copy
and the eviction occurs only if the data compare is correct.
This has the same simple recovery algorithm as the write-verify approach.

\item[\bf Idle Read After Write - IRAW]
proposed by Riedel and Riska 2008 \cite{RiRi03}.
is an improvement over {\it Read After Write - RAW} to reduce this performance penalty 
The idea is to retain the written content in the disk cache
and verify it once the disk drive becomes idle.
Trace-driven evaluation of IRAW shows its feasibility and that disk idleness
can be utilized for WRITE verification with minimal effect on user performance.

\item[\bf Metadata as checksum or version number]
The second copy of the metadata is stored in memory for fast comparison and access,
but must be flushed to disk to maintain detection capabilities after system crashes.

\item[\bf Metadata stored as RAID]
It then has a locality of reference 
that can be used to mitigate some of the disk I/O overheads of separate locations.
The advantages of this approach are lower I/O and bandwidth penalties for most operations;
the disadvantage is potential loss of detection under certain failure scenarios.

\end{description}


SDCs may go undetected until a system or application malfunction occurs.
A major problem with SDC is that data errors propagate during data rebuild.
Developed in Li and Shu 2010 \cite{LiSh10} 
a reconstruction method in the presence of silent data corruption.
This method outperforms others when periodic validation is carried out.

\section{Mirrored and Hybrid Disk Arrays Description and Reliability Comparison}\label{sec:mirhyb}

Mirrored disks are classified as RAID1 in Patterson et al. \cite{PaGK88}
Chen et al. 1994 \cite{Che+94}.  
{\it Basic Mirroring - BM} is an early form of mirroring,
which was utilized in Tandem (now part of HP) NonStop SQL system as duplexed disk Tandem 1987 \cite{TaDG87}
and by EMC's (now Dell/EMC) Symmetrix disk array with capability to emulate IBM drives. \newline
\begin{scriptsize}
\url{https://en.wikipedia.org/wiki/EMC_Symmetrix}    \newline
\end{scriptsize}  

Four mirrored disk organizations including BM are described and 
their reliability and performance analyzed in Thomasian and Blaum 2006, 
and Thomasian and Xu 2008 \cite{ThBl06,ThXu08}.
Hybrid disk arrays which store XORs of multiple disks instead 
of just mirroring are described and evaluated in Thomasian and Tang 2012 \cite{ThTa12}

Most performance studies of RAID1 are carried out in the context of BM organization. 
Mirroring provides the opportunity to reduce disk access time for read requests
by accessing the disk one of $k$ disks in $k$-way replication,
which provides the lower seek distance and hence time
assuming that the position of disk arms is known to OS.
Derived in Gray and Bitton 1988 \cite{BiGr88} is the expected seek distance 
with $k$-way replication for reads and writes, 
which are the minimum and maximum of $k$ uniformly distributed seek distances.

\begin{eqnarray}\label{eq:Ik}
E[\mbox{min of k-way seeks}] \approx \frac{C}{2k+1}, \hspace{5mm} 
E[\mbox{max of k-way seeks}] \approx C(1 - I_k),\mbox{  where:  }I_k= 
\frac{2k}{2k+1} \frac{2k-2}{2k-1} \ldots \frac{2}{3}. 
\end{eqnarray}

Loads across disk pairs can be balanced via striping 
at the higher level in the hierarchical RAID0/1,
where RAID0 is at the higher level and RAID1 at the lower level.
The load due to read requests to disk pairs is balanced via uniform routing 
(with equal probabilities) or round-robin routing.
In what follows the latter is shown to exhibit shorter waiting times 
under simplifying queueing assumptions.

With Poisson arrivals of disk requests the latter improves response time,
since the arrival process to each disk is the sum of tow exponential interarrivals or Erlang-2,
which has a {\it Coefficient of Variation - CV}$=1/2<1$ for Poisson arrivals Kleinrock 1975 \cite{Klei75}.
The mean waiting time on a GI/M/1 queue with Erlang-2 arrivals and exponential service time with mean $\bar{x}=1/\mu$ is:
\vspace{-1mm} 
$$W_{GI/M/1} = \frac{\sigma\bar{x}}{1- \sigma} \mbox{ with } \sigma= {\cal A}^* (\mu - \mu \sigma),\mbox{ where }
{\cal A}^* (s) = [\frac{2 \lambda } {s+2 \lambda } ]^2 $$  
\vspace{-1mm}
$$\sigma^2 - (1+4 \rho)\sigma + 4 \rho^2 =0, \hspace{3mm} \sigma = \frac{1}{2}(1+4 \rho - \sqrt{1+8 \rho} )$$

A smaller CV does not insure a smaller mean waiting time $W_{GI/M/1}$ 
which can be shown by a counterexample given in Thomasian 2014b \cite{Thom14b}.

RAID1 is especially suited in doubling disk bandwidth for random disk accesses in OLTP,
while RAID5 incurs SWP. 
RAID1 seems to be less efficient than RAID5 in writing large blocks,
which can be carried out as full stripe writes,
but this can be done in parallel on striped RAID 0/1 arrays as well.

When one of two of disk pairs in BM fails the load of the surviving disk is doubled.
This may lead to overload leading to high queueing delays and this led to novel RAID1 configurations 
starting with {\it Interleaved Declustering - ID} 
in the case of Teradata {\it Data Base Computer - DBC}/1012. \newline
\begin{scriptsize}
\url{https://en.wikipedia.org/wiki/DBC_1012}                               \newline
\end{scriptsize}
Thomasian and Blaum 2006 and Thomasian and Xu 2008 \cite{ThBl06,ThXu08}:

\begin{description}

\item[Basic Mirroring - BM:]
Also known as duplexing this is the simplest form of replication, 
where the same data appears at both disks.  
RAID0/1 is a striped array with $M=N/2$ pairs of mirrored disks.
When a disk fails BM has the worst performance in terms of unbalanced disk loads,
i.e., the read load of the surviving disk is doubled.
Up to $M$ disk failures can be tolerated as long as they are not pairs.
According to Table \ref{tab:MTTDL} BM has the highest reliability among 
the four RAID1 organizations considered here.
This is because the probability that the second disk failures 
that may lead to data loss by being a pair of the first failed disk is lowest.

\item[Interleaved Declustering - ID:] 
Teradata DBC/1012 database machine partitioned its $N$ disks into $c$ clusters
with $n=N/c$ disks per cluster as shown in Figure \ref{fig:ID}.
Each disk has a primary and a secondary area. 
The data in primary area is partitioned and placed onto $n-1$ secondary areas in the same cluster. 
A failed disk results in $n/(n-1)$ load increase at the $n-1$ surviving disks in a cluster. 
Two disk failure in a cluster will lead to data loss.
The setting $c=N/2$ or $n=2$ is tantamount to BM.

\item[Group Rotate Declustering - GRD:]
There are two mirrored RAID0 arrays with $M=N/2$ disks or RAID1/0,
but strips on the right side are rotated with respect 
to the strips on the left side as shown in Figure \ref{fig:GRD} Chen and Towsley 1996 \cite{ChTo96}.
When a disk on the left side fails its load is evenly distributed over the disks on the right side.
Two disk failures on one side will lead to data loss.

\item[Chained Declustering:]
The primary data on the $i^{th}$ disk is replicated 
on the secondary area of the ${i+1}^{st}\mbox{mod}(N)$ disk as shown in Figure \ref{fig:CD}
Hsiao and DeWitt 1990 \cite{HsDe90}.
Similarly to ID primary and secondary data can be placed in outer or inner cylinders,
or upper and lower disk tracks when there are an even number of tracks per cylinder.
Two successive disk failures leads to data loss,
but the reliability is lower than BM since there are twice as many data loss opportunities as in BM:
when following the $i^{th}$ disk failure the $i \pm 1 \mbox{mod}(N)$ disk fails.

\end{description}

\begin{figure}
\centering
\begin{footnotesize}
 \begin{tabular}{|c c c c||c c c c|}
 \hline
 \multicolumn{4}{|c||}{Primary Disks} &  \multicolumn{4}{c|}{Secondary Disks}\\  \hline \hline
 $D_1$ & $D_2$ & $D_3$ & $D_4$ & $D_5$ & $D_6$ & $D_7$ & $D_8$ \\ \hline
 $A$ & $B$ & $C$ & $D$ & $A'$ & $B'$ & $C'$ & $D'$ \\  \hline
 $E$ & $F$ & $G$ & $H$ & $E'$ & $F'$ & $G'$ & $H'$ \\  \hline
 $I$ & $J$ & $K$ & $L$ & $I'$ & $J'$ & $K'$ & $L'$ \\  \hline
 $M$ & $N$ & $O$ & $P$ & $M'$ & $N'$ & $O'$ & $P'$ \\ \hline
 \end{tabular}
\end{footnotesize}
\caption{\label{fig:BM}RAID0/1 with $N=8$ disks.}
\end{figure}

\hspace{1mm}

\begin{figure}
\centering
\begin{footnotesize}
\begin{tabular}{|c c c c||c c c c|}  \hline
\multicolumn{4}{|c||}{Primary Disks} &  \multicolumn{4}{c|}{Secondary Disks}\\  \hline \hline
 $D_1$ & $D_2$ & $D_3$ & $D_4$ & $D_5$ & $D_6$ & $D_7$ & $D_8$ \\ \hline
 $A$ & $B$ & $C$ & $D$ &   $A'$ & $B'$ & $C'$ & $D'$ \\  \hline
 $E$ & $F$ & $G$ & $H$ &   $H'$ & $E'$ & $F'$ & $G'$ \\  \hline
 $I$ & $J$ & $K$ & $L$ &   $K'$ & $L'$ & $I'$ & $J'$ \\  \hline
  $M$ & $N$ & $O$ & $P$ &  $N'$ & $O'$ & $P'$ & $M'$  \\ \hline
 \end{tabular}
\end{footnotesize}
\caption{\label{fig:GRD}Group Rotate Declustering with $N=8$ disks.}
\end{figure}

\hspace{1mm}

\begin{figure}
\centering
\begin{footnotesize}
 \begin{tabular}{|c c c c||c c c c|} \hline
 \multicolumn{4}{|c||}{Cluster 1} &  \multicolumn{4}{c|}{Cluster 2}\\  \hline
 $D_1$ & $D_2$ & $D_3$ & $D_4$ & $D_5$ & $D_6$ & $D_7$ & $D_8$ \\ \hline \hline \hline
 $~A~$ & $~B~~$ & $~C~~$ & $~D~~$ & $~E~~$ & $~F~~$ & $~G~~$ & $~H~~$ \\  \hline
 $b_3$ & $a_1$ & $a_2$ & $a_3$ & $f_3$ & $e_1$ & $e_2$ & $e_3$ \\  \hline
 $c_2$ & $c_3$ & $b_1$ & $b_2$ & $g_2$ & $g_3$ & $f_1$ & $f_2$ \\  \hline
 $d_1$ & $d_2$ & $d_3$ & $c_1$ & $h_1$ & $h_2$ & $h_3$ & $g_1$ \\ \hline
 \end{tabular}
\end{footnotesize}
\caption{\label{fig:ID} Interleaved Declustering with $N=8$ disks, $c=2$ clusters, and $n=4$ disks per cluster.
Capital letters denote primary data and small letters subsets of secondary data.}
\end{figure}

\hspace{1mm}

\begin{figure}[h]
\centering
\begin{footnotesize}
 \begin{tabular}{|c c c c c c c c|}  \hline
 $D_1$ & $D_2$ & $D_3$ & $D_4$ & $D_5$ & $D_6$ & $D_7$ & $D_8$ \\ \hline \hline
 $\frac{1}{2}A$ & $\frac{1}{2}B$ & $\frac{1}{2}C$ & $\frac{1}{2}D$ & $\frac{1}{2}E$ & $\frac{1}{2}F$ & $\frac{1}{2}G$ & $\frac{1}{2}H$ \\  \hline
 $\frac{1}{2}h$ & $\frac{1}{2}a$ & $\frac{1}{2}b$ & $\frac{1}{2}c$ & $\frac{1}{2}d$ & $\frac{1}{2}e$ & $\frac{1}{2}f$ & $\frac{1}{2}g$ \\ \hline
 \end{tabular}
\end{footnotesize}
\caption{\label{fig:CD} Chained declustering with $N=8$ disks.
Primary (resp. secondary) blocks are in capital (resp. small) letters.
The read load is evenly distributed among the primary and secondary copies.}
\end{figure}

The maximum number of disk failures that can be tolerated without data loss 
for mirroring with $N=2M$ disks is $I=M$, but $I=c$ for the ID organization.
Setting $r=r(t)$ to simplify notation RAID1 reliability expressions can be expressed as follows:

\vspace{-2mm}
\begin{eqnarray}\label{eq:rel}
R_{RAID} (N) = \sum_{i=0}^I A(N,i) r^{N-i} (1-r)^i .
\end{eqnarray}
$A(N,0)= 1$ by definition and $A(N,i)=0$ for $i > M$.

In the case of BM up to $M$ disk failures can be tolerated,
as long as one disk in each pair survives:

\vspace{-2mm}
\begin{eqnarray}\label{eq:BM}
A (N,i) = \binom{M}{i} 2^i , \hspace{2mm} 0 \leq i \leq M.
\end{eqnarray}

In the case of ID with $c$ clusters and $n=N/c$ disks per cluster,
we can have only one disk failure per cluster.

\vspace{-2mm}
\begin{eqnarray}\label{eq:ID}
A (N,i) = \binom{c}{i} n^i , \hspace{2mm} 0 \leq i \leq c.
\end{eqnarray}



In the case of GRD up to $M$ disks can fail as long as they are all on one side.

\vspace{-2mm}
\begin{eqnarray}\label{eq:GRD}
A (N,i) = 2 \binom{M}{i} , \hspace{2mm} 0 \leq i \leq M.
\end{eqnarray}

The expression for $A(N,i)$ for CD is derived in 
Thomasian and Blaum 2006 \cite{ThBl06}:

\vspace{-2mm}
\begin{eqnarray}\label{eq:CD}
A (N,i) = \binom{N-i-1}{i-1} + \binom{N-i}{i}, \hspace{2mm} 1 \leq i \leq M.
\end{eqnarray}



Hybrid disk arrays combine mirroring with parity coding.
LSI RAID has disks holding the XOR of neighboring disks Wilner 2001 \cite{Wiln01}.
\vspace{-1mm}
$$D_A, (D_A \oplus D_B), D_B, (D_B \oplus D_C), D_C, (D_C \oplus D_D), D_D, (D_D \oplus D_A)$$ 

Three disk failures can be tolerated as long as the middle disk is not a data disk.

SSPIRAL proposed in Amer et al. 2008 \cite{ASPL08} is a generalization 
to three disks participating in parities as follows:

{\bf $$D_A, D_B, D_C, D_D, 
(D_A \oplus D_B \oplus D_C), 
(D_B \oplus D_C \oplus D_D), 
(D_C \oplus  D_D \oplus D_A), 
(D_D \oplus D_A \oplus D_B)$$ }

Up to four disk failures can be tolerated as long as it is not a data disk 
and the parities it participates in.


A minor correction to enumerations used 
in SSPIRAL paper to obtain reliability expressions is made in Thomasian and Tang 2012 \cite{ThTa12}.

Mirrored pairs in Tandem computers came with two processors cross-connected to both disks,
instead of dedicated processor/disk pairs.
Guven that $R_c$ and $R_d$ are the CPU and disk reliability, 
the former configuration has a higher reliability.
\vspace{-1mm}
$$R_{cross-connected} = 1 - (1- R_c R_d)^2 \approx 2 R_c R_d - 4 R_c^2 R_d^2,
\hspace{5mm} R_{dedicated} = [1- (1-R_c)^2][1-(1-R_d)^2] \approx 4 R_c R_d - 2 R_c R_d^2 - 2 R_d R^2_cd$$

\begin{framed}
\subsection{Shortcut Method to Compare the Reliability of Mirrored  Disks and RAID(4+k)}\label{sec:shortcut}

We provide a simple method to compare the reliability of various RAID configurations
rather than plotting disk reliability $R(t)$ versus $t$. 
As in Thomasian 2006 \cite{Thom06} this is accomplished 
by expressing disk reliability as $r = 1 − \varepsilon$,
e.g., for MTTF=$10^6$ hours or 114 years, 
a good approximation for relatively short time $(t_s)$ 
\vspace{-1mm}
$$R(t_s) = e^{-\delta t_s} \approx 1 - \delta t_s \mbox{ after three years }R(3)=1 -3/114=0.975, \mbox{ hence :} \varepsilon=0.025.$$

RAID with $n$-way replication fails with $n$ disk failures and 
this is indicated by the power $n$ in the reliability expression (we use $\bar{r}=1-r$.
In other words the number of failures tolerated is $n-1$ 
\vspace{-1mm}
$$ R_{n-way} = 1  - (1-r)^n = 1 - \bar{r}^n = 1- \varepsilon^n $$

The expressions for the RAID1 configurations are as follows:

\vspace{-2mm}
\begin{eqnarray}\label{eq:BM2}
R_{BM} \approx r^N + N r^{N-1} \bar{r} + \binom{N}{2} r^{N-2} (\bar{r})^2 \approx 1 - 0.5 N \varepsilon^2.
\end{eqnarray}

For GRD the reliability expression can be written directly,
since it is as if we have two logical disks each comprising $M=N/2$ disks.

\vspace{-2mm}
\begin{eqnarray}\label{eq:GRD2}
R_{GRD} = 2 r^{N/2} - r^N  \approx 1 -  0.25 N^2 \varepsilon^2.
\end{eqnarray}

\vspace{-2mm}
\begin{eqnarray}\label{eq:ID2}
R_{ID} \approx
r^N +  c (\frac{N}{c}) r^{N-1} \bar{r} + \binom{c}{2} (\frac{N}{c})^2 r^{N-2} (\bar{r})^2
= 1 - 0.5 N ( \frac{N}{c} -1 ) \varepsilon^2.
\end{eqnarray}

We only need the first three terms in the reliability expression to obtain $R_{CD}$
With two disk failures there are N configurations leading to data loss:
\vspace{-1mm}
$$A(N,2) \approx \binom{N}{2} - N = 0.5N (N-3)$$

\vspace{-2mm}
\begin{eqnarray}\label{eq:CD2}
R_{CD} \approx r^N + N r^{N-1} \bar{r} + 0.5 N(N-3) r^{N-2} (\bar{r})^2 = 1- N \varepsilon^2 .
\end{eqnarray}

Reliability of RAID5 and  $RAiD(4+k)$ is given as $R_k$.
In the case of RAID5 we have:

\vspace{-1mm}
$$R_{RAID5} = r^N + N \bar{r} r^{N - 1}  = (1-\varepsilon)^N + N \varepsilon (1-\varepsilon)^{N-1} \approx 1- 0.5N(N-1) \varepsilon^2$$

\vspace{-2mm}
\begin{eqnarray}
R_{RAID(4+k)} \approx 1 - \binom{N}{k+1} \varepsilon^{k+1}  +  \binom{N}{k+2} \varepsilon^{k+2} - \ldots
\end{eqnarray}

\end{framed}

A summary of results of shortcut reliability analysis results in \cite{ThTa12} are given in Table \ref{tab:MTTDL}.

\begin{table}[h]
\centering
 \begin{tabular}{|c|c|c|c|c|c|c|c|c|}\hline
 RAID5 & BM & CD & GRD & ID & RAID6 & LSI & RAID7 & SSP\\ \hline
& $\frac{163}{280 \delta}$                       
& $\frac{379}{840 \delta}$                       
& $\frac{3}{8 \delta }$                          
& $\frac{61}{168 \delta}$                        
& $\frac{ 73}{ 168 \delta}$                      
&  $\frac{82}{105 \delta}$                       
& $\frac{533}{840 \delta}$                       
& $\frac{701}{840 \delta}$ \\ \hline              
  \scriptsize {$0.268 \delta^{-1}$}              
& \scriptsize {$0.582 \delta^{-1}$}              
& \scriptsize {$0.451 \delta^{-1}$}              
&  \scriptsize {$0.375 \delta^{-1}$}             
& \scriptsize {$0.363 \delta^{-1}$}              
& \scriptsize {$0.435 \delta^{-1}$}              
& \scriptsize {$0.781 \delta^{-1}$}              
& \scriptsize {$0.635 \delta^{-1}$}              
& \scriptsize {$0.8345 \delta^{-1}$} \\ \hline   
$\binom{N}{2} \varepsilon^2$                     
& $\frac{N \varepsilon^2}{2}$                    
& \scriptsize{$N\varepsilon^2$}                  
&  $\frac{N(N-1) \varepsilon^2}{4}$              
& $\frac{N(N-c) \varepsilon^2}{2c}$              
& $\binom{N}{3} \varepsilon^3$ 
& $( \binom{N}{3} - \frac{N}{2})$\scriptsize{$\varepsilon^3$}     
& $\binom{N}{4} \varepsilon^4$                  
& $ \frac{1}{5}  \binom{N}{4} \varepsilon^4$  \\ \hline 
\end{tabular}
\caption{\label{tab:MTTDL}MTTDLs as a ratio and a fraction of the MTTF ($\delta^{-1}$)
and the first term in asymptotic reliability expression with $\varepsilon$
denoting the unreliability of a single disk and 
the power is the minimum number of disk failures leading to data loss.}
\end{table}

RAID6 results apply to EVENODD, X-code, and RDP, all three of which are 2DFT. 

Hybrid disk arrays which combine replication and parity coding provide 
a higher reliability at the same redundancy level.
RAID1 provides 2-way, LSI RAID 3-way, and SSPIRAL 4-way replication.
The update penalty is proportional to the number of ways.

\section{Reliability Analysis of Multilevel RAID Arrays}\label{sec:multilevel}

Two multilevel RAID disk arrays are considered in this section:
mirrored RAID5 (RAID1/5) and RAID5 consisting of mirrored disks (RAID5/1).
RAID1/0 and RAID0/1 was discussed in the previous section.
Note that the number of disks is the same in both cases,
but one configuration provides higher reliability than the other
and this is verified with the shortcut reliability analysis method in Thomasian 2006 \cite{Thom06}.

\subsection{Reliability of Mirrored RAID5 - RAID1/5 Reliability}\label{sec:RAID15}

Consider mirroring of RAID5 arrays and no repair via mirroring, from one side to the other.
If more than one disk on each side fails, 
that side is considered failed although the remaining disks are intact. 
\vspace{-1mm}
$$R_{RAID1/5}= 2 R_5(t) - R_5^2 (t)$$

Substituting $R_{RAID5}$ given by Eq. (ref{eq:RAID5t}):

\vspace{-2mm}
\begin{eqnarray}
R_{RAID1/5} = 
2 [ \frac{\zeta e^{\eta t} - \eta e^{\zeta t}}{\zeta - \eta} ]   
- [\frac{\zeta e^{\eta t} - \eta e^{\zeta t}}{\zeta - \eta}]^2 =  \\
\nonumber
2 \frac{\zeta e^{\eta t} - \eta e^{\zeta t}}{\zeta - \eta}  
- \frac{ \zeta^2 e^{2 \eta t} + \eta^2 e^{2 \zeta t} -2 \eta \zeta e^{(\eta + \zeta) t }} 
{ (\zeta - \eta)^2 } 
\end{eqnarray}

\vspace{-2mm}
\begin{eqnarray}
MTTDL_{mirrored/RAID5}= 
\frac{1}{\zeta-\eta}
[ - 2 \frac{\zeta}{\eta} + \frac{\eta}{\zeta}]          \nonumber
- \frac{1}{(\zeta - \eta)^2}                              
[ - 2 \frac{\zeta^2}{ \eta} + 2 \frac{\eta^2}{\zeta} + 2 \frac{\eta\zeta}{\eta+\zeta} ]
\end{eqnarray}

RAID5 reliability can be expressed as a single component 
whose MTTF equals the MTTDL as was given by Eq. (\ref{eq:appr}).
\vspace{-1mm}
$$R_{RAID15}^{appr} (t) = 2 R_5^{appr} (t) - [ R_5^{appr} (t) ]^2 $$ 
\vspace{-1mm}
$$MTTDL_{RAID15}^{appr}=  \frac{1.5 \mu }{N(N+1) \delta^2} = \frac{1.5 MTTF^2}{N(N+1)MTTR}$$


\subsection{Reliability of RAID5 Consisting of Mirrored Disks - RAID5/1}\label{sec:RAID51}

An approximate expression for mirrored RAID5 MTTDL was given in Xin et al. 2003 \cite{Xin+03}.
The contents of a failed disk are recovered by its mirror,
but otherwise by invoking the RAID5 paradigm
\vspace{-1mm}
$$MTTDL_{RAID5/1} \approx \frac{\mu^3}{4N(N-1)\delta^4} = \frac{MTTF^4}{4N(N-1)MTTR^3}$$

A rigorous method to derive RAID5/1 MTTDL using the shortest path reliability model 
is presented in Iliadis and Venkatesan 2015b \cite{IlVe15b}.
The state tuples $(x, y, z)$ indicate that there are x pairs with both devices in operation, 
$y$ pairs with one device in operation and one device failed, 
and $z$ pairs with both devices failed.
With $D$ devices on each side and failure rate $\delta$ and repair rate $\mu$
we have the following state transitions leading to {\it Data Loss - DL}:

\vspace{-2mm}
\begin{align}\nonumber
(D,0,0) &\xrightarrow{2 D \delta} (D-1,1.0), \hspace{5mm} \xleftarrow{\mu} (D-1,1,0) \\
\nonumber
(D-1,1,0) &\xrightarrow{\delta} (D-1,0,1), \hspace{5mm} (D-1,0,1) \xleftarrow{2\mu} (D-1,1,0) \\
\nonumber
(D-1,1,0) &\xrightarrow{2(D-1)\delta} (D-2,2.0), \hspace{5mm} (D-2,2,01) \xleftarrow{2\mu} (D-1,1,0) \\
\nonumber
(D-1,0,1) &\xrightarrow{2(D-1) \delta} (D-2,1,1), \hspace{5mm} (D-2,1,1) \xleftarrow{\mu} (D-1,0,10) \\
\nonumber 
(D-2,2,0) &\xrightarrow{2\delta}, \hspace{5mm}  (D-2,1,1) \xleftarrow{\mu} (D-2,2,0) \\
\nonumber 
D_2,1,1) &\xrightarrow{\delta} \mbox{DL} 
\end{align}

Taking into account that $\delta \ll \mu$ we get 
the following end-to-end probabilities for the upper and lower paths.
\vspace{-1mm}
$$P_u \approx \frac{\delta}{\mu} \times \frac{2(D-1)\delta}{2 \mu} \times \frac{\delta}{2 \mu} = 
\frac{(D-1) \delta^3 } {2 \mu^3} $$
\vspace{-1mm}
$$P_{ell} \approx \frac{2(D-1)\delta}{\mu} \times \frac{\delta}{\mu} \times \frac{\delta}{2 \mu}=
\frac{(D-1) \delta^3) }{\mu^3}$$
\vspace{-1mm}
$$P_{DL} =  P_u + P_\ell \approx \frac{3 (D-1) \delta^3}{2 \mu^3}$$

It is argued that the MTTDL is a product of two first device failures 
and the expected number of first device failure events:

Given that $MTTDL \approx E[T] / P_{DL} $, where $E[T] = 1/(N\delta)$.
It follows that $ MTTDL \approx 1 / (N \delta P_{DL})$ and noting that $N=2D$ we obtain:

\vspace{-2mm}
\begin{eqnarray}
MTTDL^{(approx)}_{RAID(5/1)} \approx \frac{\mu^3}{3D(D-1)\delta^4} = \frac{MTTF^4}{D(D-1)MTTR^3} 
\end{eqnarray}

\begin{framed}
\subsection{Shortcut Reliability Analysis to compare RAID1/5 and RAID5/1}\label{sec:shortcut2}

That RAID5/1 is more reliable than RAID1/5 as shown 
by the shortcut reliability analysis in Thomasian 2006 \cite{Thom06}.
We consider RAID with $N$ disks and denote disk reliability with $r$.

\vspace{-2mm}
\begin{eqnarray}\label{eq:15}
R_{RAID1/5}= 1 - [1- R_{RAID5}]^2 = 1- [R^N + N r^{N-1} (1-r)]^2
\end{eqnarray}

The reliability of RAID5 with mirrored disks can be expressed as:

\vspace{-2mm}
\begin{eqnarray}\label{eq:16}
R_{RAID5/1} = N R_{RAID1}^{N-1} -(N-1)R_{RAID}^N = N[1 -(1-r)^2]^{N-1} - (N-1) [ 1 -(1-r)^2]^N 
\end{eqnarray}

For all values on $N$ it is easy to show from the above equations
that $RAID_{5/1} > R_{RAID1/5}$, e.g., for $N=3$ we have

\vspace{-1mm}
$$R_{RAID5/1}- R_{RAID1/5}  = 6r^2 (1-r)^4 > 0$$

Setting $r=1-\varepsilon$ in the above equations and 
retaining only the lowest power of $\varepsilon$ we have:

\vspace{-2mm}
\begin{eqnarray}
\nonumber
R_{RAID1/5} \approx 1 -\frac{1}{4} N^2 (N-1)\varepsilon^4 \hspace{5mm}
R_{RAID5/1} \approx 1-\frac{1}{2} N(N-1)\varepsilon^4. 
\end{eqnarray}
It is easy to see that $R_{RAID5/1} > R_{RAID1/5}$. 

\end{framed}

\subsection{AutoRAID Hierarchical Array}\label{sec:autoRAID}

AutoRAID is a two level memory hierarchy with two disks organized as RAID1
holding hot data and cold data with an RAID5/LSA organization Wilkes et al. 1996 \cite{WGSS96}.
Data is is initially written onto RAID1 and as RAID1 disks fill their contents are moved to RID5/LSA.

AutoRAID assumes that part of data that resides on RAID1 
is active at any time and the working set changes slowly.
Whole {\it Relocation Blocks - RBs} are promoted to RAID1 when they are updated.   

\section{Simulation Studies for Reliability Evaluation}\label{sec:simul}

The use of simulation in computer and network performance analysis dates back to late 1950s.
An early application to study network delays is reported in \cite{Klei64},
who in addition to a pseudo-random number generator 
used a builtin radio-active decay random number generator,
which was found not to be random (personal communication from Prof. L. Kleinrock, 6/4/2022). \newline
\begin{scriptsize}
\url{https://en.wikipedia.org/wiki/Radioactive_decay}
\end{scriptsize}

The RAIDframe simulator could also be used for rapid prototyping for disk arrays \cite{CGH+97}.
RAIDframe was developed at CMU to assist in the implementation and  evaluation  of new  RAID architectures.  
DiskSim with its origin at Univ. of Michigan Worthington et al. 1994 \cite{WoGP94} 
was further developed at CMU's PDL, refer to Section \ref{sec:IDRperf}.

The authors of Secction ref{sec:DEH} argue why they did not use HP's Pantheon disk array simulator:       \newline
\begin{scriptsize}
\url{http://shiftleft.com/mirrors/www.hpl.hp.com/research/ssp/papers/PantheonOverview.pdf}     \newline
\end{scriptsize}
or the DiskSim simulator developed at CMU's PDL.                                 \newline
\begin{scriptsize}
\url{https://www.pdl.cmu.edu/DiskSim/index.shtml}                                \newline
\end{scriptsize}
DiskSim relies on Dixtrac automated disk drive characterization,
which provides disk specifications until 2007.                                   \newline
\begin{scriptsize}
\url{https://www.pdl.cmu.edu/Dixtrac/index.shtml}                                \newline
\url{https://www.pdl.cmu.edu/DiskSim/diskspecs.shtml}
\end{scriptsize}

Hsu and Smith 2004 \cite{HsSm04} is a comprehensive study dealing with disk performance,
whose abstract is as follows: \newline
\begin{small}
\begin{quote}
``In this paper, we use real server and personal computer workloads 
to systematically analyze the true performance impact of various I/O optimization techniques, 
including read caching, sequential prefetching, opportunistic prefetching, write buffering, 
request scheduling, striping, and short-stroking. 
We also break down disk technology improvement into four basic effects, faster seeks, 
higher RPM, linear density improvement, and increase in track density and analyze each separately 
to determine its actual benefit. In addition, we examine the historical rates of improvement and 
use the trends to project the effect of disk technology scaling. 
As part of this study, we develop a methodology for replaying real workloads 
that more accurately models I/O arrivals and that allows the I/O rate 
to be more realistically scaled than previously. 
We find that optimization techniques that reduce the number of physical I/Os 
are generally more effective than those that improve the efficiency in performing the I/Os. 
Sequential prefetching and write buffering are particularly effective, 
reducing the average read and write response time by about 50\% and 90\%, respectively. 
Our results suggest that a reliable method for improving performance 
is to use larger caches up to and even beyond 1\% of the storage used. 
For a given workload, our analysis shows that disk technology improvement 
at the historical rate increases performance by about 8\% per year if the disk occupancy rate 
is kept constant, and by about 15\% per year if the same number of disks are used. 
We discover that the actual average seek time and rotational latency are, respectively, 
only about 35\% and 60\% of the specified values. 
We also observe that the disk head positioning time far dominates the data transfer time, 
suggesting that to effectively utilize the available disk bandwidth, 
data should be reorganized such that accesses become more sequential.''
\end{quote}
\end{small}

In what follows several simulation based studies are discussed.

\subsection{Simulation Study of a Digital Archive}\label{sec:uber}

Digital archives require stronger reliability measures than RAID to avoid data loss from device failure.
Multi-level redundancy coding is used to reduce the probability of data loss 
from multiple simultaneous device failures Wildani et al. 2009 \cite{WSML09}.
This approach handles failures of one or two devices efficiently,
while still allowing the system to survive rare-events, 
i.e., larger-scale failures of four or more devices.

The uber/super-parity is calculated from all user data in all disklets belonging to a stripe 
in the uber-group using another erasure code.
The uber-parity can be stored on NVRAM or always powered-on disks to offset write bottlenecks, 
while still keeping the number of active devices low.
The calculations of failure probabilities determined that 
the addition of uber/super-parities allows the system 
to absorb many more disk failures without data loss.
Adding uber-groups negatively impacts performance when these groups need to be used for a rebuild.
Since rebuilds using uber parity occur rarely, they impact system performance minimally.
Robustness against rare events can be achieved for under 5\% of total system cost.

\subsection{Simulation of Hierarchical Reliability}\label{sec:HRAID}

{\it Hierarchical RAID - HRAID} was described in Thomasian 2006 \cite{Thom06a}
and evaluated in Thomasian et al. 2012 \cite{ThTH12}.
There are $N$ {\it Storage Nodes - SNs}, 
where each SN has a DAC - Disk Array Controller with a cache serving $M$ disks.
HRAID$(k/\ell)$ provides recovery against $k$ SN failures and $\ell$ disk failures at each node,
which means $k$ (resp. $\ell$) strips out of $M$ strips per stripe (at each SN) 
are dedicated to inter-SN (resp. intra-SN) check strips.
Data and inter-SN check strips are protected by intra-SN check strips.
It can be said HRAID$k/\ell$ is $k$-Node-Failure Tolerant and $\ell$DFT 
Hierarchical RAID is shown in Figure \ref{fig:hraid}.

\begin{figure}
\begin{tiny}
\begin{tabular}{|c|c|c|c||c|c|c|c||c|c|c|c||c|c|c|c|}\hline
\multicolumn{4}{|c||}{ Node 1} & \multicolumn{4}{c||}{ Node 2}
& \multicolumn{4}{c||}{ Node 3} & \multicolumn{4}{c|}{ Node 4} \\ \hline \hline
$~D_{1,1}^1$ & $~D_{1,2}^1$ & $~P_{1,3}^1$ & $~Q_{1,4}^1$ &
$~D_{1,1}^2$ & $~P_{1,2}^2$ & $~Q_{1,3}^2$ & $~D_{1,4}^2$ &
$~P_{1,1}^3$ & $~Q_{1,2}^3$ & $~D_{1,3}^3$ & $~D_{1,4}^3$ &
$~Q_{1,1}^4$ & $~D_{1,2}^4$ & $~D_{1,3}^4$ & $~P_{1,4}^4$ \\ \hline
\end{tabular}
\end{tiny}
\caption{\label{fig:hraid}HRAID1/1 with $N=4$ nodes, $M=4$ disks per node.
P is the intra-SN and Q the inter-SN parity.
Only the first stripe on all 4 SNs is shown in the figure.} 
\end{figure}

Parities are rotated from row-to-row and SN to SN 
to ensure that Q parities cover all SN strips for inter-SN rebuild.
The P parity is used for intra-SN rebuild due to a local disk failure and 
they cover data and Q parity strips (as stated earlier).

Intra-SN rebuild is carried out by restriping (Rao et al. 2011 \cite{RaHG11}) 
so that we may have the following progression \cite{ThBl09}:
\vspace{-1mm}  
$$ RAID7 \rightarrow RAID6 \rightarrow RAID5 \rightarrow RAID0 \rightarrow Data\_Loss. $$
InterSN rebuild is carried by overwriting Q strips.

Three options for HRAID operation are considered.

\begin{description}

\item[I. Intra-SN but no Inter-SN Redundancy:]
There are $N$ independent SNs which are RAID$(4+\ell)$ disk arrays.
This option serves as a baseline in reliability comparisons comparisons.

\item[Option II: Inter-SN Redundancy and Rebuild Processing:]
Inter-SN rebuild processing is invoked on demand and for rebuild.

\item[Option III: Inter-SN Redundancy, but no Rebuild Processing:] 
Rebuild may not be possible due to limited interconnection network bandwidth.
In the case of SNs considered failed due to the failure of more than $\ell$ failed disks but no controller failures,
operation continues in degraded mode with on-demand reconstruction via inter-SN redundancy, until data loss detected.
This is strictly done to determine the MTTDL,
since the system can continue operation .

\end{description}

Given that $u$ is a uniformly distributed pseudo-random numbers in the range $(0,1)$,
it can converted to (negative) exponential distribution to determine the time to next failures.
Given $u$ exponentially distributed random numbers can be obtained as follows.
\vspace{-1mm} 
$$F(t)=1- e^{-\delta t } = u \mbox{  leads to  }t = (- 1 / \sum_{\forall i} \delta_i) \times ln(1-u)\mbox{  or just  } 
t= - \mbox{MTTF} \times \mbox{ln}(u).$$
$\sum{\forall (i)} \delta_i$ is the sum of the failures rates of all components.

More complex methods for obtaining distributions are available.
The accuracy of simulation results is specified 
by the confidence interval at a given confidence level.
The two topics are discussed in Chapters 5 and 6 in Lavenberg 1983 \cite{Lave83}.

\begin{framed}
{\bf Procedure: Hierarchical RAID Simulator to Determine MTTDL}

\begin{description}
\item[\bf Inputs:]

\noindent
$N$: The number of SNs.                                                             \newline
$M$: The number of disks per SN.                                                    \newline
$D = N \times M$: Total number of disks.                                            \newline
$k$: Internode redundancy level ($k=0$ for Option I).                               \newline
$\ell$: Intranode redundancy level, $\ell=0$ for RAID0, else RAID($4+\ell)$.         \newline
$\delta$: Disk failure rate, $\delta = 10^{-6}$ and MTTF$=1/\delta = 10^6$ hours.   \newline
$\gamma$: Controller failure rate, multiple of $\delta$).

\item
[\bf Outputs] (with initializations):

\noindent
$F_{cd}=0/1$ data loss occurred due to controller/disk failure.                    \newline
$N_c = 0$: number of failed controllers.                                           \newline
Initialize SN/controller state: $C[n]=1, \forall{n}$.                   \newline
Initialize number of failed disks at $n^{th}$ SN: $F[n]=0,\forall{n}$). 

\item
[\bf Simulation Variables:]

\noindent
$Clock=0$: Simulation time initialized.                                    \newline
State 0: Failed. State 1: Operational.                                     \newline
$C[0:N-1]$ the state of controllers. 
$Disks[0:N-1,0:M-1]$ the state of $D= N \times M$ disks.                   \newline
$(Disks[n,m]=1,\forall{n},\forall{m})$. /* initial disk states*/           \newline
$F[0:N-1]$ \# of failed disks at node $n$ 
$\Omega:$ Failure rate for a given configuration.                          \newline
$T_{NF}:$ Time to next failure.                                            \newline
$N_c=0$: Number of failed controllers.                                     \newline
$N_n=0:$ Number of failed nodes for to controller and $>\ell$ disk failures \newline
$N_d=0$: number of failed disks.                                           \newline
$u_i$: $i^{th}$ uniformly distributed pseudo-random variable in $(0,1)$.

\end{description}

{\bf Simulation steps:}

\begin{description}

\item[\bf Step\_1:]
Total failure rate since all failures exponential:       
{\bf $\Omega = (N- N_c) \gamma + (D - N_d) \delta.$}

Compute time to next failure and increment $Clock$:              \newline
$T_{NF} = (-1/\Omega) ln(u_1)$,                                  \newline
$Clock +=  T_{NF}$. /* advance simulation clock */

\item[\bf Step\_2:]
Probability a controller failed: $p=(N-N_c)\gamma/\Omega$.       \newline
If $u_2 \leq p$ then controller failure: goto Step 4.            \newline
Else disk failure: goto Step 4.

\item[\bf Step\_3:]
Determine failed controller: $n= \lfloor N \times u_3 \rfloor $.   \newline
If $C[n]=0$ then regenerate $u_3$ and recompute $n$,              \newline
else \{ $C[n] = 0$, $N_c++$, $N_n++$, $D[n,m]=0, 0 \leq m \leq M-1$, \newline
$t=M-F[n]$ disk inaccessible, so $N_d = N_d + t$. \}               \newline
If $N_n > k$ then \{$F_{cd}=0$, goto Step 6\}, else goto Step\_1.

\item[\bf Step\_4:]
$t=\lfloor D \times u_4 \rfloor$. /* index of failed disk  */      \newline
SN number: $n= \lfloor t / M \rfloor$;                                 \newline
Disk number: $m = $ mod $(t, M)$.                                       \newline
If $Disks[n,m]==0$ (disk already failed) resample $u_4$ and recompute $n$ and $m$, \newline
/*  disks attached to failed controller are considered failed */    \newline
else \{$ Disks[n,m]==0$, $N_d++$, $F[n]++$]\}.                     \newline 
If $F[n] \leq \ell$ goto Step\_1.                                   \newline
else \{ C[n]=0, $t=M-F[n]$, $N_d = N_d + t$, $N_n++$ \}.           \newline
If $N_n \leq k $ then go to Step\_1, else $F_{cd}=1$.

\item[\bf Step\_5:]
Return($Clock$, $F_{cd}$, $N_c$, $N_{df}$).

\end{description}

The simulation is repeated to obtain confidence intervals 
at a sufficiently high confidence level.

\end{framed}

It is shown by the shortcut reliability analysis method Thomasian 2006 \cite{Thom06} 
and results based on the above simulation in Thomasian et al. 2012 \cite{ThTH12}
that a higher MTTDL is attained by associating higher reliability at the intra-SN rather than inter-SN level.
For example, given three check strips P, Q, R, then P and Q can be used for as intra-SN and R for inter-SN recovery. 
This is contradictory to the two preferred configurations 
for IBM's Intelligent Bricks project described in Wilcke et al. 2006 \cite{WGF+06}, \newline
(1) 1DFT (RAID5) at the the node level and 2NFT at the node level provides 40-75\% storage efficiency.\newline
(2) 0DFT in bricks and 3NFT at brick level provides 50-75\% storage efficiency.\newline

\subsection{Proteus Open-Source Simulator}\label{sec:Proteus}

The Proteus open-source simulator developed at
{\it Storage Systems Research Center - SSRC} at {\it U. Calif. at Santa Cruz - UCSC}
can be used to predict the risk of data loss in many disk array configurations: 
mirrored disks, all RAID levels, and two-dimensional arrays Kao et al. 2013 \cite{KPSL13} 
Proteus was used to learn that there is no measurable difference 
between values obtained assuming deterministic versus exponential repair times,
which are used in Markov chain modeling,
which was an issue raised in RAID5 reliability analysis in Gibson 1992 \cite{Gibs92}. 

\subsection{CQSIM\_R Tool Developed at AT\&T}\label{sec:ATT}

CQSIM\_R is a tool suited for predicting the reliability 
of large scale storage systems Hall 2016 \cite{Hall16}. 
It includes direct calculations based on an only-drives-fail failure model
and an event-based simulator for predicting failures.
These are based on a common combinatorial framework for modeling placement strategies. 

CQSIM\_R models common storage systems, 
including replicated and erasure coded designs. 
New results, such as the poor reliability scaling of spread-placed systems and 
a quantification of the impact of data center distribution 
and rack-awareness on reliability, demonstrate the tools' usefulness. 
Analysis and empirical studies show the tool's soundness,  performance, and scalability.

\subsection{SIMedc Simulator for Erasure Coded Data Centers}\label{sec:SIMEDC}

The discrete-event SIMedc simulator for the reliability analysis of erasure-coded data centers 
was developed at {\it Chinese Univ. of Hong-Kong - CUHK} by Zhang et al. 2019 \cite{ZhHL19}.
SIMedc reports reliability metrics of an erasure-coded data center based on data center topology, 
erasure codes, redundancy placement, and failure/repair patterns 
of different subsystems based on statistical models or production traces. 

Simulation is accelerated via importance sampling assisted 
by uniformization which is discussed in Section \ref{sec:2.5}. 
It is shown that placing erasure-coded data in fewer racks generally improves 
reliability by reducing cross-rack repair traffic, 
even though it sacrifices rack-level fault tolerance in the face of correlated failures.

\section{Reliability Modeling Tools}\label{sec:tools}

Analytic and simulation software packages developed to assess
the reliability, availability, and serviceability of computer systems 
are surveyed in Johnson and Malek 1988 \cite{JoMa88}.
Provided are the application of the tool, input, models, and model solution methods.



In what follows we discuss three reliability modeling tools:
ARIES at UCLA in Subsection \ref{sec:ARIES},
SAVE at IBM Research in Subsection \ref{sec:SAVE},
SHARPE at Duke Univ. in Subsection \ref{sec:SHARPE}.


\subsection{Automated Reliability Interactive Estimation System - ARIES Project at UCLA}\label{sec:ARIES}


{\it Triple Modular Redundancy - TMR} with the voter outputting 
the majority outcome until a unit fails was an early proposal to achieve high reliability.
The system continues operation until two units disagree.
Assuming the voter hardware is less complex than its input units
and hence its contribution to system failure is negligible,
so we voter reliability to one.
Given the unit reliabilities: $r(t)= e^{-\delta t}$.

\vspace{-1mm}
$$R_{TMR} (t) = r^3 (t) + 3 r^2 (t) (1-r(t)) = 2e^{-3\delta t} - 3 e^{-2 \delta t},
\mbox{   MTTF}_{TMR}=\int_0^\infty R_{TMR}(t)dt = \frac{5}{6\delta}$$

Note that $MTTF_{TMR} < R_{simplex}=1/\delta$,
but TMR reliability exceeds that of a single unit for $(0,t_m)$, 
where $t_m$ the mission time can be determined as follows:

\vspace{-1mm}
$$3e^{-2 \delta t_m} - 2 e^{-3\delta t_m} > e^{-\delta t_m} \mbox{ so that } t_m < \frac{\mbox{ln}(2)}{\delta}$$ 
TMR/Simplex is a TMR variant that continues its operation after failure with a single unit
according to a hypoexponential distribution with parameters $3 \delta$ and $\delta$ Trivedi 2002 \cite{Triv02}.

\vspace{-2mm}
\begin{eqnarray}\nonumber 
R_{TMR/Simplex} (t) = 1.5 e^{ -\delta t } - 0.5 e^{- 3 \delta t }
\mbox{  sum of MTTFs for 2 phases of operation: } 
MTTF_{TMR/Simplex} = \frac{1}{3\delta} + \frac{1}{\delta} = \frac{4}{3\delta}.
\end{eqnarray}

\begin{framed}
\subsection*{Application of Shortcut Method to Reliability Comparison}

We apply the shortcut method in Thomasian 2006 \cite{Thom06} 
to the problem at hand for a small $t_s$.

\vspace{-1mm}
$$R_{TMR/Simplex} (t_s) = 1.5 e^{ -\delta t_s } - 0.5 e^{- 3 \delta t_s } \approx 
   1.5 (1  -  \delta  t_s + 0.5 (   \delta t_s)^2 - \ldots  )  
-  0.5 (1 - 3 \delta  t_s + 0.5 ( 3 \delta t_s)^2 - \ldots  ) = 
1 - 4 (\delta t_s)^2.$$ 

We also compare TME/Simplex with TMR and its observed that TMR/Simplex is more reliable
\vspace{-1mm}
$$R_{TMR} =  
3 e^{-2 \delta t_s} - 2 e^{-3 \delta t_s} \approx 
3  (1 - 2 \delta t_s + 0.5 (2 \delta t_s)^2 + \ldots ) 
-2 (1 - 3 \delta t_s + 0.5 (3 \delta t_s)^2 + \ldots ) = 
1 - 3 (\delta t_s)^2 $$

\end{framed}


An example of a reliability expression derived for a particular system is that of 
a hybrid $k$-out-of-$n$ system with $n$ components energized (with failure rate $\lambda$) 
and $m$ components deenergized (with failure rate $\mu$) Mathur and Avizienis \cite{MaAv70},
which is given in Example 3.32 in Trivedi 2002 \cite{Triv02}.

Aries is a unifying method for analyzing closed systems Ng and Avizienis 1980 \cite{NgAv80}.
In closed systems no new spares can be introduced as in the case for space-borne computers.
The parameters used in this study are:

\begin{table}[h]
\begin{footnotesize}
\begin{center}
\begin{tabular}{|c|c|} \hline
$N$          &initial number of modules in the active configuration.          \\ \hline
$D$          &number of degradations allowed in the active configuration      \\ \hline
$S$          &number of spare modules.                                        \\ \hline
$Ca$         &coverage for recovery from active module failures.              \\ \hline
$Cd$         &coverage for recovery from spare module failures.               \\ \hline
$\lambda$    &failure rate of active modules.                                 \\ \hline
$\mu$        &failure rate of spare modules                                   \\ \hline
${\bf Y}$    &sequence of allowed degradations of the active configuration.   \\ \hline
${\bf CY}$   &coverage vector for transitions into degraded configuration.    \\ \hline
\end{tabular}       
\end{center}
\end{footnotesize}                              
\caption{\label{tab:NgAv}Parameters used in Ng and Avizienis 1970 \cite{NgAv80}}
\end{table}

The reliability of a closed fault-tolerant system  has a reliability function of the form: 

\vspace{-3mm}
\begin{eqnarray}\label{eq:Ng}
R(t) = \sum_{\forall i} A_i e^{-\sigma_i t}
\end{eqnarray}

The analysis is based on the eigenvalues of the CTMC representing 
the failure of system components and its recovery by using spares.
Repeated roots are not considered.

Let ${\cal S}_i$ indicate a nonfailed state and $\sigma_i$ the failure rate at that state.
$A_i$ can be expressed as a function of state parameters can be quite complex.
The MTFF easily follows from Eq.~\ref{eq:Ng}
\vspace{-1mm}
$$\mbox{MTFF} = \int_{t=0}^\infty R(t) dt = \sum_{\forall i} \frac{A_i}{\sigma_i}.$$
For closed systems $R(t)$ given by Eq. (\ref{eq:Ng}) has a specific form.

\vspace{-3mm}
\begin{eqnarray}
R(t) = X(t) {\cal A} W(t)
\end{eqnarray}

\vspace{-2mm}
\begin{eqnarray}\nonumber
X(t) = ( e^{-Y[0] \lambda t} , \ldots,  e^{-Y[D]\lambda t} ) \mbox{ with }Y[0]=N\hspace{5mm}
\mbox{ and } W(t)=(1,e^{-\mu t}, \ldots, e^{-S \mu t} ),
\end{eqnarray}

\vspace{-2mm}
\begin{eqnarray}
{\cal A} =
\begin{pmatrix}
A_{S,0}^0 & \ldots & A_{S,0}^D   \\
\vdots    & \ddots &\vdots       \\
A_{S,S}^0 & \ldots & A_{S,S}^D   \\
\end{pmatrix}
\end{eqnarray}

Two more parameters used for modeling repairable systems 
are the number of repairman or repair facilities ($M$) 
and the repair rate of one repairman ($\Psi$) 

\subsection{System AVailability Estimator - SAVE Project at IBM Research}\label{sec:SAVE}

An early and influential reliability modeling effort at IBM is Bouricius et al. 1971 \cite{BCJ+71}.
The coverage factor ($c$) which is the probability that the system
can successfully recover from a failure is defined in this study.
An example of such calculations is the probability of encountering 
LSEs in RAID5 with and without IDR or disk scrubbing Iliadis et al. \cite{IHHE11}.

The processors of a duplex systems fail with rate $\delta$ and the coverage factor is $c$
according to Example 8.35 in Trivedi 2002 \cite{Triv02}. 
Given that ${\cal S}_i$ is a state with $i$ failed  disks 
we have the following transitions:

\vspace{-2mm}
\begin{eqnarray}\label{eq:transit}
{\cal S}_0 \xrightarrow{2 \delta c} {\cal S}_1, \hspace{5mm}                                      
{\cal S}_0 \xrightarrow{2 \delta (1-c)} {\cal S}_2, \hspace{5mm}                                   
{\cal S}_0 \xleftarrow{\mu} {\cal S}_1,  \hspace{5mm}                                            
{\cal S}_1 \xrightarrow{\delta} {\cal S}_{2}.
\end{eqnarray}

The SAVE project was started in 
Steve Lavenberg's Systems Modeling and Analysis Dept. at IBM Research in mid-1985. 
The effort first dealt with analytical reliability modeling techniques,
along the lines adopted in ARIES and SHARPE Blum et al. 1994 \cite{BGH+94}. 
 
Adopting simulation as an alternative to analysis for reliability estimation 
led to unified framework for simulating Markovian models 
of highly dependable systems Goyal et al. 1992 \cite{GSH+92},
This is because realistic system models are often not amenable 
to analysis using conventional analytic or numerical methods as discussed in Section \ref{sec:deGa00}.
Straightforward simulations is too costly since failures are rare.
Techniques for fast simulation of models of highly dependable systems 
are reviewed in Nicola et al. 2001 \cite{NiSN01}.
Importance  sampling is one such method and when it works well 
it can reduce simulation run lengths by several orders of magnitude. 

The following example is provided in this study:                    
(1) two sets of processors with four processors per set,                 
(2) two sets of controllers with two controllers per set,
(3) four clusters of disks each consisting of four disks. 
The multiprocessors are connected to all controllers. 
The pair of controllers have two connections to disk clusters.
In a disk cluster, data are replicated so that one disk can fail without affecting the system.
The ID organization for mirrored disks in Section \ref{sec:mirhyb} is utilized.

When a processor fails it has a 0.01 probability of causing another processor to fail.
Each unit in the system has two failure modes which occur with equal probability.
The repair rates for mode 1 and 2 failures are 1 and 0.5 hour, respectively.
The failure rates in hour$^{-1}$ are 1/1000 for processors,  
1/20,000 for controllers, and 1/60,000 for disks.
The repair rates (per hour) are 1 for all mode 1 failures and 1/2 for all mode 2 failures.

Efficient simulation techniques for estimating steady-state quantities
in models of highly dependable computing systems with general component failure
and repair time distributions are developed in Nicola et al. 1993 \cite{NNHP93}.
The regenerative method of simulation for steady state estimation
can be used when the failure time distributions are exponentially distributed.
A splitting technique is used for importance sampling 
to speed up the simulation of rare system failure events during a cycle.
Experimental results show that the method is effective in practice.

\subsection{Symbolic Hierarchical Automated Reliability and Performance Evaluator - SHARPE}\label{sec:SHARPE}

SHARPE is a reliability and performance modeling tool 
was developed at Duke Univ. by Sahner, Trivedi and Puliafito 1996 \cite{SaTP96}.
Several generations of Trivedi's students have contributed to SHARPE.              \newline
\begin{scriptsize}
\url{https://sharpe.pratt.duke.edu/}                                               \newline
\end{scriptsize}

Section 5.4 titled ``Imperfect Fault Coverage and Reliability'' in Trivedi 2002 \cite{Triv02}
provides a good discussion of the topic and some formulas derived in Ng and Avizienis \cite{NgAv80}.
Section 8.5 titled ``Markov Chains with Absorbing States''  
discusses a technique used in Sharpe as discussed in Appendix II.
Hierarchical reliability modeling is discussed in Chapter 16 in Trivedi and Bobbio 2017 \cite{TrBo17}.
The mathematics behind SHARPE is covered in Appendix II.


\section{Reliability Analysis at IBM Zurich Research Lab - ZRL}\label{sec:ZRL}

Earlier work on reliability of storage systems at ZRL dealt with a comparison 
of different {\it Intra-Disk Redundancy - IDR} methods on reducing 
the effect of LSEs on disk failures  in Dholakia et al. \cite{DEH+08}.
Other topics studied at ZRL are discussed below:

\subsection{System Reliability Metrics}\label{sec:metrics} 

Most redundancy schemes have been evaluated via the MTTDL metric, 
which has been proven useful for assessing tradeoffs, 
for comparing reliability schemes 
and for estimating the effect of the various parameters on system dependability. 

That MTTDL derivations based on CTMC models provide unrealistically high reliability estimates 
by Elerath and Schindler 2014 \cite{ElSc14} is refuted by Iliadis and Venkatesan 2015 \cite{IlVe15} 
by showing that MTTDL equations that account for latent sector errors 
and scrubbing operations yield satisfactory results. 

In the context of distributed and cloud storage systems
the magnitude of lost data is as important as the frequency of data loss. 
A general methodology to obtain the EAFDL metric analytically, in conjunction with the MTTDL metric, 
for various redundancy schemes and for the Weiball and Gamma real-world distributions 
Trivedi 2002 \cite{Triv02} is provided in this study. 

It is shown that the declustered placement scheme offers superior reliability in terms of both metrics. 
The parameters used in the study are summarized in Table \ref{tab:IlVe}.

\begin{table}
\begin{footnotesize}
\begin{center}
\begin{tabular}{|c|c|}\hline
Parameter               &Definition               \\ \hline \hline
$n$                     & number of storage devices  \\
$c$                     & amount of data stored on each devise \\
$r$                     & replication factor  \\
$k$                     & spread factor of data placement scheme \\
$b$                     & reserved rebuild bandwidth per device  \\
$1/\lambda$             & mean time to failure of a storage device \\ \hline
$U$                     &amount of user data stored in the system ($U=n c / r$) \\
$1/\mu$                 &time to read data amount $c$ at rate $b$ ($1/\mu=c/b$) \\ \hline  
\end{tabular}
\end{center}
\end{footnotesize}
\caption{\label{tab:IlVe}Systems parameters with derived parameters below the line.}
\end{table}

At any point the system is either in normal or rebuild mode, which starts immediately after a failure.
Deferring rebuild has its advantages, 
since some disk failures are due to transient server failures and are resolved by restarting servers. 
Following a first-device failure, rebuild operations and subsequent device failures may occur, 
which eventually lead the system either to {\it Data Loss - DL} with probability $P_{DL}$ 
or back to the original normal mode by restoring all replicas with probability $1 − P_{DL}$.
Typically, rebuild time is negligible compared to the time between failures: $E[T] = 1 / (n \delta )$.
Given that the expected number of first-device failures until data loss occurs is $1/P_{DL}$. It follows: 
\vspace{-3mm}
\begin{eqnarray}
MTTDL \approx \frac{E(T)}{P_{DL}}
\end{eqnarray}

Let $H$ denote the amount in data loss when data loss occurs 
and $U$ is the amount of stored user data.
EAFDL is then the normalized data lost per MTTDL:
\vspace{-3mm}
\begin{eqnarray}
EAFDL = \frac{E[H]}{MTTDL \cdot U} 
\end{eqnarray}

Define $Q$ as follows: $Q=H$ if DL and $Q=0$  if no DL.
Given $E(Q)$ we can determine EAFDL as follows:
\vspace{-2mm}
\begin{eqnarray}
E(Q) = P_{DL}  \cdot E(H),\mbox{  hence: }EAFDL = \frac{E(Q)}{E(T) U}
\end{eqnarray}

\subsection{Clustered versus Declustered Data Placements}\label{sec:placement}

To justify the EAFDL measure consider clustered and declustered data placements
In {\it Clustered Placement - CP} the $k$ blocks on a device are replicated in another device,
while in {\it Declustered Placement - DP} the $k$ blocks 
are distributed over $k$ blocks as shown in Figure \ref{fig:dataplacement}.

\begin{figure}
\begin{footnotesize}
\begin{minipage}[b]{0.4\linewidth}
$$
\begin{array}{|c|c|c|c|c|c|}\hline
D_1  &D_2    &D_3  &D_4  &D_5    &D_6   \\ \hline
A    &A      &K    &K    &\cdot      &\cdot     \\
B    &B      &L    &L    &\cdot      &\cdot     \\
C    &C      &\cdot    &\cdot    &\cdot      &\cdot     \\
D    &D      &\cdot    &\cdot    &\cdot      &\cdot     \\
E    &E      &\cdot    &\cdot    &\cdot      &\cdot     \\
F    &F      &\cdot    &\cdot    &\cdot      &\cdot     \\
G    &G      &\cdot    &\cdot    &\cdot      &\cdot     \\
I    &I      &\cdot    &\cdot    &\cdot      &\cdot     \\
J    &J      &\cdot    &\cdot    &\cdot      &\cdot     \\ \hline
\end{array}
$$
\end{minipage}
\hspace{5mm}
\begin{minipage}[b]{0.4\linewidth}
$$
\begin{array}{|c|c|c|c|c|c|}               \hline
D_1  &D_2    &D_3  &D_4  &D_5    &D_6   \\ \hline
A    &A      &B    &C    &D      &E     \\
B    &F      &G    &H    &I      &J     \\
C    &K      &K    &L    &\cdot  &\cdot \\
D    &D      &\cdot    &\cdot    &\cdot &\cdot   \\
E    &E      &\cdot    &\cdot    &\cdot &\cdot   \\
F    &F      &\cdot    &\cdot    &\cdot &\cdot   \\
G    &G      &\cdot    &\cdot    &\cdot &\cdot   \\
I    &I      &\cdot    &\cdot    &\cdot &\cdot   \\
J    &J      &\cdot    &\cdot    &\cdot &\cdot   \\ \hline
\end{array}
$$
\end{minipage}
\end{footnotesize}
\caption{\label{fig:dataplacement}Clustered and declustered data placement 
with six devices with degree of replication $r=2$.}
\end{figure}

In both cases two node failures will lead to data loss,
but two node failures with DP result in a lower data loss than CP.
Expressions for MTTDL and EAFDL derived in Iliadis and Venkatesan 2014 \cite{IlVe14} are as follows:

\vspace{-1mm}
$$
MTTDL \approx
\begin{cases}
( \frac{b}{\delta c} )^{r-1} \frac{1}{n \delta}  &\mbox{  for CP}\\
( \frac{b}{2 \delta c} )^{r-1} \frac{(r-1)!}{n \delta}
\prod_{e=1}^{r-2} ( \frac{n-e}{r-e} )^{r-e-1} &\mbox{  for DP}
\end{cases}
$$
\vspace{-2mm}
$$
EAFDL \approx
\begin{cases}
( \frac{\delta c}{b} )^{r-1} \lambda &\mbox{  for CP}\\
( \frac{2 \delta c}{b})^{r-1} \frac{\delta}{(r-1)!} \prod_{e=1}^{r-1}
(\frac{r-e}{n-e})^{r-e} &\mbox{  for DP}
\end{cases}
$$

Analytical reliability expressions for the symmetric,
clustered, and declustered data placement schemes are derived in Iliadis 2022 \cite{Ilia22},
where it is demonstrated that the employment of lazy rebuild results 
in a reliability degradation of orders of magnitude. 
The degradation of MTTDL due to sector errors is observed in all cases, 
those that apply lazy rebuild, and those that do not.
By contrast there is no degradation for the EAFDL.
It is also shown that the declustered data placement scheme offers superior reliability.

The adverse effect of LSEs on the MTTDL and the EAFDL reliability metrics is evaluated in Iliadis 2023 \cite{Ilia23}. 
A theoretical model is developed, and closed-form expressions for the metrics of are derived. 
Highly reliable storage devices with a very small ratio for mean time 
to repair ($1/\mu)$ to mean time failure ($1/lambda$),  expressed as:
\vspace{-1mm}
$$ \mu \int_0^\infty F_\lambda *t) [ 1 - F_X (t)[] dt \ll \mbox { with } \frac{ \lambda} {\mu} \ll 1 $$ 
The MTTDL and EAFDL are obtained analytically for 
(i) the entire range of bit error rates; 
(ii) the symmetric, clustered, and declustered data placement schemes; and 
(iii) arbitrary device failure and rebuild time distributions under network rebuild bandwidth constraints. 
For realistic values of sector error rates the MTTDL degrades, 
while EAFDL remains practically unaffected. 
In the range of typical sector error rates and for very powerful erasure codes, EAFDL degrades as well. 
It is also shown that the declustered data placement scheme offers superior reliability.
The actual probability of data loss in the case of correlated symbol errors is smaller 
than that obtained assuming independent symbol errors and of the same order of magnitude.

\subsection{Performance Analysis of a Tape Library System at ZRL}\label{sec:tapelibrary}

An analytical model to evaluate the performance of a tape library system 
is presented in Iliadis et al. 2016 \cite{IKSV16}. 
The analysis takes into account the number of cartridges and tape drives 
as well as different mount/unmount policies to determine the mean waiting time for tapes. 
The accuracy of the model is confirmed by validation against measurements. 
Earlier work on this topic is also reviewed.

A tape library consists of tape drives, robot arms,
a storage rack for the tape cartridges, and a cartridge control unit.
To serve a request, a robot arm fetches the appropriate tape cartridge
from the storage rack and delivers it to a free tape drive.
The tape drive control unit mounts the tape, 
positions the head to the desired file and then transfers its data.
To free a tape drive, a robot arm unmounts the tape cartridge and returns it to the storage rack.
Tape read/write requests contain the cartridge id, 
the position of the data block in the cartridge, and its size.
Requests submitted for cartridges are queued and served according to a scheduling policy.
The hierarchical scheduling algorithm ensuring fairness and avoiding starvation is as follows:

\begin{description}
\item[Upper level:]
Cyclic or round-robin scheduling among the queues (mounting cartridges).
\item[Lower level:]
A FCFS policy for serving requests within a queue (reading from a mounted cartridge).
\end{description}


When all requests for a cartridge are served (exhaustive service),
and there are still pending requests at some other, non-mounted cartridge,
an unused cartridge is unmounted and another cartridge with pending requests is mounted.
If, however, there are no other pending requests to any other non-mounted cartridge,
the cartridge can either remain mounted in anticipation of future requests
or be unmounted so as to save time when future requests arrive for other non-mounted cartridges.
Two mount/unmount policies deployed in this context are:

\begin{description}
\item[\bf Always-Unmount (AU):]
A tape cartridge is immediately unmounted upon completion of all pending requests for it
in anticipation of the next request arriving for another non-mounted cartridge.
\item[Not-Unmount (NU):]
A tape cartridge remains mounted upon completion of all pending requests for it
in anticipation of the next request arriving for this same currently mounted, but idle cartridge.
\end{description}

The performance analysis of the resulting polling system 
relies on state-dependent queues Takagi 1986 \cite{Taka86}  
Parameters considered in this study are given in Table \ref{tab:tapelibrary}
based on IBM 2021 \cite{IBMC21}.

\begin{table}
\begin{footnotesize}
\begin{center}
\begin{tabular}{|c|c|c|}    \hline
Parameter        &Values        & Definition            \\ \hline \hline
c               &3200           & number of cartridges  \\ 
d               &32             & number of tape drives \\ 
a               &1,2            & number of arms        \\ 
R               &5 s (fixed)    & robot transfer time   \\ 
$t_L$           &24 s (fixed)   & load ready time        \\
$\bar{t}_R$     &59 s           & mean rewind time      \\  
$\bar{t}_U$     &24 s           & unload ready time     \\ 
$s_{max}$       &118 s          & maximum seek time     \\ 
$\bar{Q}$       &843 MB         & mean request size      \\ 
$\overline{Q^2}$ &2.8 GB         & standard deviation     \\ 
$b_w$           &360 MB/s       & bandwidth             \\ \hline
$n$             &100            & \# cartridges per tape drive \\ 
$M$             & 20 s (fixed)  & mount time $M=R+t_L$                   \\ 
$\bar{U}$       & 88 s          & mean unmount time ($\bar{U}=\bar{t}_R+t+U +R$) \\
$\bar{t}_T$     &2.34           & mean transfer time ($t_T = \bar{Q}/b_w$   \\ \hline
\end{tabular}
\end{center}
\end{footnotesize}
\caption{\label{tab:tapelibrary}
Parameter values according to Table II in Iliadis et al. 2021 \cite{IJLS21}.}
\end{table}

Predicting the performance of a tape library system is key to efficiently dimensioning it,
possibly in the context of multi-tiered storage systems that include tape libraries.
The effect of two mount/unmount policies was analytically assessed.
For light loads, the AU - Always-Unmount policy yields a mean delay
that is lower than that of the NU - Not-Unmount policy,
with the difference becoming negligible as the load increases.
The effect of the number of robot arms was assessed
by means of simulation shows that given fast robot arms
multiple arms do not provide a significant performance improvement,
but multiple arms may be necessary to attain higher availability.
The analysis has been extended in Iliadis et al. 2021 \cite{IJLS21}

\section{Flash Solid State Drives - SSDs}\label{sec:flash}

Flash SSDs which are fast, nonvolatile, consume less power than HDDs
are replacing them for high performance applications as their prices are dropping.
\begin{scriptsize}
\url{https://en.wikipedia.org/wiki/Hard_disk_drive}   \newline
\url{https://en.wikipedia.org/wiki/Flash_memory}
\end{scriptsize}

Flash SSDs are classified according to the cell type, i.e., number of bits per cell: 
SLC (single), MLC (multiple-2), TLC (triple), QLC (quad), PLC (penta). 
There is a rapid drop in {\it Program/Erase - P/E} cycles as the cell size 
is decreased from 5 nanometer-nm to 1nm (Fig. 1 in Jaffer et al. 2022 \cite{JaMS22}
and given in Table \ref{tab:penta}.               \newline
\begin{scriptsize}
\url{https://blocksandfiles.com/2019/08/07/penta-level-cell-flash/}
\end{scriptsize}

\begin{table}[h]
\begin{center}
\begin{footnotesize}
\begin{tabular}{|c|c|c|c|c|c|}\hline
Cell type   &SLC (1 bit) &MLC (2 bits) &TLC (3 bits) & QLC (4 bits) &PLC (5 bits) \\ \hline \hline 
5Xnm        &11,00       &10,000        &2,500        & 800          & 400         \\ \hline
3Xnm        &10,000      &5,000         &1,250        & 350          & 175         \\ \hline
2Xnm        & 7,500      &3,000         & 750         & 150          &  75         \\ \hline
1Xbn        &5,000       &1,500         & 500         & 70           &  35         \\ \hline
\end{tabular}
\end{footnotesize}
\caption{P/E cycles for varying number of bits-per-cell as cell size is varied.
The number for PLC are estimated.\label{tab:penta}}
\end{center}
\end{table}
To summarize (10 QLC WOM-v(2,4) can store 2 logical bits per cell. 
(20 MLC drives also store 2 logical bits per cell. 
(3) Without WOM-v coding, MLC drive has better endurance than QLC drive.

A comparison of flash SSDs and HDDs from several viewpoints is presented in Table \ref{tab:comp}.
Flash storage is discussed in more detail in Section 2.3.20 in Thomasian 2021 \cite{Thom21}.

\begin{table}
\begin{footnotesize}
\begin{center}
\begin{tabular}{|c|c|c|c|c|c|c|}\hline
Device    &Capacity  &Cost   &Cost/MB &Random Access      &Seq'l Access     &Device     \\
\& Year   &(GB)      &(\$)   &        &Latency/Bandwidth  &Bandwidth        &Scan (s)   \\ \hline \hline
HDD'08    &0.1       &2K     &200     &28/0.28            &1.2              &83         \\ \hline
HDD'10    &1000      &300    &0.0003  &8/0.98             &80               &12,500     \\ \hline
SSD'10    &00        &2000   &0.02    &0.026/300          &700              &143        \\ \hline
\end{tabular}
\end{center}
\end{footnotesize}
\caption{\label{tab:comp}
Comparison of HDDs and SSDs (latency in msec, bandwidth in MB/s) Athanassoulis et al. 2010 \cite{AAC+10}}
\end{table}


Performance-wise SSDs outperform HDDs.
For example., Huawei OceanStor Dorado 18000 V6, 
set a new peak for SPC-1 Benchmark performance in Oct. 2020, 
achieving 21,002,561 SPC-1 IOPS according to {\it Storage Performance Council - SPC}: \newline
\begin{scriptsize}
\url{https://e.huawei.com/en/products/storage/all-flash-storage/dorado-8000-18000-v6}  \newline
\end{scriptsize}
HDDs led in price-performance with the SPC-2 benchmark in March 2017.  \newline
\begin{scriptsize}
\url{https//www.spcresults.org}.
\end{scriptsize}

Rebuild times for HDDs and SSDs are given in Table \ref{tab:SSD}.                      \newline
\begin{scriptsize}
\url{https://www.theregister.com/2016/05/13/disak_versus_ssd_raid_rebuild_times/}      \newline
\end{scriptsize}
The one and four TB drives just differ in the number of platters, 
but are based on the same technology, 
Seagate 18 TB drives with 564 MB/s transfer rate takes about nine hours to read,       
which is due to the fact that disk transfer rates have not kept up with disk capacities.
The Seagate dual actuator HDDs area is a partial solution doubling transfer rate from 6 to 12 Gb/s. \newline
\begin{scriptsize}
\url{https://www.tomshardware.com/news/seagate-launches-2nd-gen-dual-actuator-hdds-18-tb}  \newline
\url{https://www.storagereview.com/news/seagate-exos-2x18}                                 \newline                          
\end{scriptsize}
The much higher bandwidth provided by Flash SSDs is a better solution.

\begin{table}
\begin{footnotesize}
\begin{center}
\begin{tabular}{|c|c|c|c|} \hline
RAID             & Capacity      & Seq'l Write        & Rebuild time      \\
                 & TB            & MB/sec             & Minutes           \\  \hline \hline
Disk-1/2/3       & 0.72/1/4      & 80/115/460         & 15/145/580        \\ \hline
FlashMax III     & 2.2           &1,400               & 26                \\ \hline
Intel D3600      & 2             &1,500               & 22                \\ \hline
Micron 9100      & 3.2           &1,500               & 27                \\ \hline
Intel DC 3608    & 4             &3,000               & 22                \\ \hline
\end{tabular}
\end{center}
\end{footnotesize}
\caption{\label{tab:SSD}Minimum rebuild times for disks and flash SSDs.
Capacity in TB times 1000 divided by Seq'l write in MB/s yields rebuild time in seconds}
\end{table}

In spite of problems such as ``flash wear'' Flash SSDs are much more reliable than HDDs.
Counting the AFRs - Annual Failure Rates  of Backblaze's drives;      
HDDs had a 10.56\% AFR, while SSDs had just 0.58\% AFR.   \newline
\begin{scriptsize}
\url{https://www.backblaze.com/blog/backblaze-hard-drive-stats-q1-2021/}          \newline
\end{scriptsize}
This makes higher levels of redundancy less of a consideration for SSDs.  
Given the superior performance of SSDs with respect to HDDs hybrid arrays are of no interest,
unless SDD capacities are exceeded in which SSDs may cache HDD data Yu et al. 2012 \cite{Yu++12}.
Conversely increasing SSDs sizes allow them to hold large databases,
but a portion of the database can be supplemented with disks.
Indexing structures support on flash SSDs is discussed in Fevgas et al. 2020 \cite{Fev+20}.

Flash wear can be reduced by first writing into {\it Storage Class Memory - SCM},
before being downloaded in large chunks to Flash SSD to reduce wear. \newline
Evolving SCMs are discussed in Freitas and Wilcke 2008 \cite{FrWi08}


HDDs still provide less costly storage per GB,
but NAND Flash technology is progressing rapidly yielding higher capacity chips at lower cost. \newline
\begin{scriptsize}
\url{https://www.intel.com/content/www/us/en/products/docs/memory-storage/solid-state-drives/ssd-vs-hdd.html} \newline
\end{scriptsize}     
The disk/flash crossover is discussed in this work.                        \newline
\begin{scriptsize}
\url{https://blocksandfiles.com/2023/02/15/pure-the-disk-flash-crossover-event-is-closer-than-you-think/} \newline
\end{scriptsize}
and the Purestorage CEO projected in 2023 that no disk drives will be sold after 2028.   \newline
\begin{scriptsize}
\url{https://blocksandfiles.com/2023/05/09/pure-no-more-hard-drives-2028/} 
\end{scriptsize}

Data updates in SSDs are carried out by writing a new copy, as in LFS, 
rather than overwriting old data, marking prior copies of data invalidated.
Writes are performed in units of pages, similarlty to disks, even if data to be written is smaller.
Space is reclaimed in units of multipage erasing,
which necessitates copying of any remaining valid pages in the block before reclamation.

The efficiency of the cleaning process greatly affects performance under random workloads;
{\it Write Amplification factor - WAF} defined as the ratio of 
the amount of data an SSD controller writes in relation to the amount of data 
that the host's flash controller writes reduces application throughput.
A nearly-exact closed-form solution for write amplification
under greedy cleaning for uniformly-distributed random traffic is presented in 
Desnoyers 2014 \cite{Desn14} and validated using simulation.

Log-Structured File System for Infinite Partition is a scheme
eliminates garbage collection in flash SSDs Kim et al 2022 \cite{Kim+22}.
The scheme separates {\it Logical Partition Size - LPS} from the physical storage size 
and the LPS is large enough so that there is no lack of free segments during SSD's lifespan, 
allowing the filesystem to write the updates in append-only fashion without reclaiming the invalid filesystem blocks. 
The metadata structure of the baseline filesystem, 
so that it can efficiently handle the storage partition with $2^{64}$ sectors. 
Interval mapping minimizes the memory requirement for the {\it Logical Block Address - LBA}-to-PBA 
translation in {\it Flash Translation Layer - FTL}.

FTL is a hardware and software layer is part of flash memory controllers.   \newline
\begin{scriptsize}
\url{https://en.wikipedia.org/wiki/Flash_memory_controller}                 \newline
\end{scriptsize}
FTL's role is to emulate a block-type peripheral such as HDDs.

The effect of Flash memories on transaction performance
has been investigated using simulation  and measurement studies.
NoFTL allows for native Flash  access  and integrates parts  of  the FTL  functionality
into the DBMS yielding significant performance increase and simplified I/O stack Hardock et al. 2013 \cite{Har+13}.
A Flash emulator integrated with a DBMS (Shore-MT)
demonstrate  a  performance improvement of $\approx$ 3.7-fold under various TPC workloads.

Performance evaluation of the write operation in flash-based SSDs is reported in Bux 2009 \cite{BuxW09}.
Performance of greedy garbage collection in flash-based SSDs by Bux and Iliadis 2010 \cite{BuIl10}.
Scheduling and performance modeling and optimization by Bux et al. 2012 \cite{BHIH12}.
The complicating factor in flash SSDs is write performance which is analyzed in Desnoyers 2014 \cite{Desn14}.

Flash SSD storage is available from PureStorage, Dell/EMC, NetApp, Fungible, Oracle (Exadata), etc.
Flash SSDs outperform HDDs in power consumption 3-fold (2-5 watts) versus (6-15 watts). 
From a TCO viewpoint SSDs will be preferred when they are less than 5-fold expensive than HDDs per GB. \newline
\begin{scriptsize}
\url{https://www.intelice.com/ssd-or-hdd-hard-drives/}  
\end{scriptsize}

Vastdata is a Flash SSD-based storage system which adopts LSA,
but does not provide block storage, which takes any data, like a file or database entry,
and divides it into blocks of equal sizes.                                \newline
\begin{scriptsize}
\url{https://vastdata.com/whitepaper}                                     \newline
\url{https://vastdata.com/blog/providing-resilience-efficiently-part-ii}  
\end{scriptsize}

Since SSDs play an important role in data centers their failures affect 
the stability of storage systems and cause additional maintenance overhead. 
The {\it Multi-View and Multi-Task Random Forest - MVTRF} scheme described in \cite{ZHN+23} 
is a scheme to predict SSD failures based on multi-view features extracted from 
both long and short-term monitoring of SSD data.  
MVTRF can simultaneously predict the type of failure it is and when it will occur. 
These are are useful for handling SSD failures. 
MVTRF has been evaluated on the large-scale Tencent data centers showing its high failure prediction accuracy 
and improves precision by 46.1\% and recall by 57.4\% on average compared with the existing schemes. \newline
\begin{scriptsize}
\url{https://en.wikipedia.org/wiki/Precision_and_recall}
\end{scriptsize}

\subsection{Write-Once Memory - WOM Codes to Enhance SSD Lifetimes}\label{sec:WOM}

Increased storage density is achieved with higher bits per cell, 
but is accompanied by order of magnitude lower P/E cycles,
which decreases the number of times the SSD can be rewritten and their lifetime. 

{\it Write-Once Memory - WOM} codes are one way of improving drive lifetime.  
i.e., rewrite on top of pre-existing data without erasing previous data. 
WOM codes alter the logical data before it is physically written, 
thus allowing the reuse of cells for multiple writes. 
On every consecutive write, zeroes may be overwritten with ones, but not vice versa 
Yaakobi et al. 2012 \cite{Yaa+12}.
A WOM(x,y) code encodes a code word of x bits into a code word of y bits

WOM increases the total logical data that can be written on the physical medium 
before requiring an erase operation.  
Traditional WOM codes are not scalable and only offer up to 50\% increase 
in total writable logical data between any two erase operations. 
Simple and highly efficient family of generic WOM codes that could be applied to any N-Level cell drive, 
The focus of Jaffer et al. \cite{JaMS22} is QLC drives.
The application of coding at various levels is given in Table \ref{tab:WOM}.

Microbenchmarks and trace-driven simulation shows that WOM-v codes can reduce erase cycles
4.4-11.1-fold with minimal performance overheads.
The increase in the total logical writable data before an erase is 50-375\%. 
The total logical writable data between two erase operations may be increased
by up to 500\% by the choice of internal ECC.

\begin{table}
\begin{center}
\begin{scriptsize}
\begin{tabular}{|c|c|c|}                                 \hline
v(3,4)                  &v(2,4)            &v(1,4)      \\  \hline \hline
D111,V15,G2             &D11,V15,G5        &D1,V15,-    \\ \hline
D110,V14,G2             &D10,V14,G5        &D0,V14,G14  \\ \hline
D101,V13,G2             &D01,V13,G5        &D1,V13,G13  \\ \hline          
D100,V12,G2             &D00,V12,G5,G4     &D0,V12,G12  \\ \hline
D011,V11,G2             &D11,V11,G4        &D1,V11,G11  \\ \hline
D010,V10,G2             &D10,V10,G4        &D0,V10,G10  \\ \hline
D001,V09,G2             &D01,V09,G4,G3     &D1,V09.G09  \\ \hline
D000,V08,G2             &D00,V08,G3        &D0,V08,G08  \\ \hline
D111,V07,G1,G1          &D11.V07,G3        &D1,V07,G07  \\ \hline 
D110,V06,G1             &D10,V06,G3,G2     &D0,V06,G06  \\ \hline
D101,V05,G1             &D01,V05,G2        &D1,V05,G05  \\ \hline
D100,V04,G1             &D00,V04,G2        &D0,V04,G4  \\ \hline
D011,V03,G1             &D11,V03,G2,G1     &D1,V03,G3  \\ \hline
D010,V02,G1             &D10,V02,G1        &D0,V03,G2  \\ \hline
D001,V01,G1             &D01,V01,G1        &D1,V02,G1   \\ \hline
D000,V00,G0             &D00,V00,G0        &D0,V01.-     \\ \hline
\end{tabular}
\end{scriptsize}
\caption{Voltage-based codes for encoding 3,2,1 bits into 16 voltage levels.
D=data,V-voltage level,G-generation, shared sates have two generations.\label{tab:WOM}}
\end{center}
\end{table}

\subsection{Predictable Microsecond Level Support for Flash}\label{sec:predictable}

Barroso et al. 2017 \cite{BMPR17} discuss the issue of lack of support for microsecond ($\mu$s) scale events. 
It is argued that disk accesses are in milliseconds (ms)
and the CPU accesses its registers, cache and even main memory in nanoseconds (ns) (refer to Table~\ref{tab:times}).
The emerging Flash memory access times 
which are in $\mu$s are not adequately supported by the interface (refer to Table \ref{tab:micro}).

\begin{table}[h]
\begin{footnotesize}
\begin{center}
\begin{tabular}{|c|c|c|}\hline
Nanosecond events       &Microsecond events                    &Millisecond events    \\ \hline \hline             
register file: 1-5 ns   & datacenter networking O(1 $\mu$s)    &disk O(10) ms \\ \hline
cache accesses:4-30 ns  &new NVM memory O(1 $\mu$s)            &low-end-flash O(1) ms   \\ \hline
memory access 100 ns    &high end flash O(10 $\mu$s)           &wide area networking O(10) ms \\ 
                        &GPU/accelerator O(10 $\mu$s)          &                              \\ \hline
\end{tabular}
\end{center}
\end{footnotesize}
\caption{\label{tab:times}Events and their latencies based on \cite{BMPR17}.}
\end{table}

\begin{table}[h]
\begin{footnotesize}
\begin{center}
\begin{tabular}{|c|c|}\hline
Flash                       & 225K instructions=O(100$\mu$s                \\ \hline
{\bf Fast Flash}            & {\bf 20K instruction=O(10$\mu$s}             \\ \hline 
{bf New NVM memory}         & {\bf 2K instructions =O(1$\mu$s}             \\ \hline
DRAM                        & 500 instructions = O(100ns-1$\mu$s)          \\ \hline
\end{tabular}
\end{center}
\end{footnotesize}
\caption{\# instructions between I/O events based on \cite{BMPR17}.
Data in bold is for unavailable new memories based on trace analysis. \label{tab:micro}} 
\end{table}

SSD performance is inherently non-deterministic due to the 
internal management activities such as the garbage collection, wear leveling, and internal buffer flush. \newline
\begin{scriptsize}
\url{https://en.wikipedia.org/wiki/Wear_leveling}
\end{scriptsize}

I/O Determinism interface is a host/SSD co-designed flash array with
predictable latency which does not sacrifice aggregate bandwidth Li et al. 2023 \cite{LPS+23}.
Minimal changes to the {\it NonVolatile Memory Express - NVMe} interface and 
flash firmware was required to achieve near ideal latencies. \newline
\begin{scriptsize}
\url{https://en.wikipedia.org/wiki/NVM_Express}
\end{scriptsize}

\subsection{Differential RAID for SSDs}\label{sec:diffRAID}

Diff-RAID design takes into account the fact that flash memories
have very different failure characteristics from HDDs,
i.e., the {\it Bit Error Rate - BER} of an SSD increases 
with more writes Balakrishnan et al. 2010 \cite{BKPM10}.

By balancing writes evenly across the array,
RAID schemes wear together at similar rates, 
making all devices susceptible to data loss at the same time.
Diff-RAID reshuffles the parity distribution on each drive replacement
to maintain an age differential when old devices are replaced by new ones,

Diff-RAID distributes parity blocks unevenly masking higher BERs on aging SSDs while:   
(1) Retaining the low overhead of RAID5.                               
(2) Extending the lifetime of commodity SSDs.   
(3) Alleviating the need for expensive error correction hardware.


For a  workload consisting  only  of  random  writes,  
the relative aging rates of devices for a given parity assignment. 
Let $a_{i,j}$ represent the ratio of the aging rate of the $i^{th}$ device to that of the $j^{th}$ device, 
and $p_i$ and $p_j$ the percentages of parity allotted to the respective devices, then:

\vspace{-2mm}
\begin{eqnarray} 
a_{i,j}= \frac{p_i (n-1) + (100-p_i)}{p_j(n-1) + (100-p_j)}.
\end{eqnarray}

Consider $n=4$ SSDs and 50\% of the parity at the first device
and the rest evenly distributed over three disks, so that $(70,10,10,10)$ 
Using the above formula the aging rate of the first device is twice that of the others.
\vspace{-1mm}
$$(70 \times  3 + 100-70) / (10 \times 3 + 100 - 10) = 240/120=2   $$                                    
After numerous replacements the ages of remaining devices 
at replacement time converge to $(5750, 4312.5, 2875, 1437.5)$.
The implementation of device replacement can be accomplished via rebuild processing,
while maintaining the same parity layout.

A simulator to evaluate Diff-RAID's reliability by using BERs from 12 flash chips
showed that it is more reliable than RAID5 by orders of magnitude.

\subsection{Fast Array of Wimpy Nodes - FAWN}\label{sec:FAWN}

FAWN was a project at CMU's PDL
combined low-power nodes with flash cluster storage providing fast
and  energy  efficient  processing  of  random queries in a {\it Key Value - KV} store. \newline
\begin{scriptsize}
\url{https://en.wikipedia.org/wiki/Key%E2%80%93value_database}                          \newline
\end{scriptsize}
Andersen et al. 2022 \cite{AFK+11}.
FAWN-KV begins with a log-structured per-node datastore to serialize writes and make them fast on flash.
KV-stores store retrieves and manages associative arrays using a dictionary or hash table.
FAWN uses replication between cluster nodes to provide reliability and strong consistency.
The FAWN prototype with 1000 queries/Joule demonstrated a significant potential for I/O-intensive workloads
In 2011 4-year-old FAWN nodes delivered over an order of magnitude 
more queries per joule (watt$\times$second) than disk systems.                \newline
\begin{scriptsize}
https://en.wikipedia.org/wiki/Joule
\end{scriptsize}

\subsection{Distributed DRAM-based Storage - RAMCloud}\label{sec:RAMCloud}

The crossover point for at which the more costly DRAM 
will replace disks is explored in Gibson 1992 \cite{Gibs92}.
Plotted in 1992 Graph 2.1(a) and 2.1(b) give 2010 and 2020 as optimistic and pessimistic crossover points.

RAMCloud is a distributed storage system keeping all data in DRAM 
offering exceptionally low latency for remote accesses,
e.g., small reads completed in less than 10 $\mu$ s in a 100,000 node datacenter Ousterhout et al. 2015 \cite{Oust15}.
RAMCloud's 1 PB or more storage supports Web applications, via a single coherent key-value store.
an associative array where each key is associated with one and only one value in a collection.
RAMCloud maintains backup copies of data on HDDs to ensure high durability and availability.
Memory prices per GB=$2^{30} \approx 10^9$ bytes which are dropping are given here
\begin{scriptsize}
\url{https://thememoryguy.com/dram-prices-hit-historic-low/}  \newline
\end{scriptsize}
NAND prices are lower than DRAM, but Flash is slower than DRAM.

\section{Conclusions}\label{sec:conclusions}

We discussed the reliability analysis of various RAID configurations, 
most notably RAID5 w/o and with IDR,
analyzing reliability models via numerical methods and discrete-event simulation.
We have mentioned but not delved into fast simulation methods 
such as importance sampling for reliability modeling.

We presented the performance analysis of RAID5 disk arrays implemented as HDDs 
in normal, degraded and rebuild modes is presented using the M/G/1 queueing model.
Performance analysis of F/J queues arising in degraded mode operation and 
VSM for rebuild processing are discussed.

There has been significant increase in bandwidths used 
at hyperscalar {\it Data Storage Centers DSCs}, i.e., 100 Gb/s. \newline 
\begin{scriptsize}
\url{https://www.nextplatform.com/2019/08/18/the-future-of-networks-depends-on-hyperscalers-and-big-clouds/}\newline
\end{scriptsize}
Two important issues are Interconnection network reliability Goyal and Rajkumar 2020 \cite{GoRa20} 
and performance Rojas-Cessa 2017 \cite{Roja17} (Part III - Data Center Networks).
``The design space for large, multipath datacenter networks is large and complex, 
and no one design fits all purposes. Network architects must trade off many criteria to design cost-effective, 
reliable, and maintainable networks, and typically cannot explore much of the design space.'' 
Condor enables a rapid, efficient design cycle Schlinker et al. 2015 \cite{Sch+15}. 

Interested readers should follow advances in the field by following relevant trade magazines
and conferences mentioned in this article.

In conclusion there is a need to increase the knowledge of computer scientists and engineers 
in stochastic processes and operations research techniques by expanding the curriculum. 
Steps in this direction are proposed in Hardin 2102 \cite{Hard12}, SEFI 2013 \cite{SEFI13}.

 

\section*{Appendix I: Transient Analysis of State Probabilities in Markov Chains}\label{sec:deGa00}

This section summarizes the review paper by de Souza and Gail 2000 \cite{deGa00}.
Considered are time-homogeneous, finite-state CTMCs.
The state-space is ${\cal \bf S}=\{ s_i, i=1, \ldots, M \}$ and the infinitesimal generator: 
\vspace{-2mm}
$$
{\bf Q}=
\begin{pmatrix}
-q_{1,1}  &q_{1,2} &\ldots &q_{1,M} \\
q_{2,1}  &-q_{2,2}   &\ldots  &q_{2,M} \\ 
\vdots & \vdots  & \ddots  &\vdots \\
q_{M,1} & q_{M,2}  &\ldots   &-q_{M,M}     \\
\end{pmatrix}
$$

Due to time homogeneity: 
\vspace{-1mm}
$$ {\Pi}_{i,j} (t) = P \{ X (t'+t)= s_j | X (t') =s_i \}$$ 

The backward and forward Kolmogorov equations according to Kleinrock 1975 \cite{Klei75}:

\vspace{-3mm}
\begin{eqnarray}\label{eq:30}
{\Pi}'(t) = {\bf Q} {\Pi}(t) \mbox{  and  }
{\Pi}'(t) = {\Pi}(t) {\bf Q}.
\end{eqnarray}
where the state probabilities for CTMC are 
$$ \underline{\pi}(t) = [ \pi_1 (t), \ldots , \pi_M (t) ], 
\mbox{ where } \pi_i (t) =  P \{ X (t) = s_i \} $$ 



Methods to obtain transient solutions to finite state CTMCs are as follows:

\subsection{Solutions Based on ODEs}

A simple method for solving ODEs is dividing 
the time interval $(0,t($ to $n$ subintervals of length $h$ and 
then approximating the derivative at the point $(y(nh),nh)$
with the difference $[y(n+1)h) - y(nh)]/h$ which can be applied to Eq. (\ref{eq:30}).
A better approximation to the derivative using Euler's single step method is the two-step method

\vspace{-3mm}
\begin{eqnarray}\label{eq:24}
y'(t_n) = \frac {y (t_{n+1}) - y(t_{n-1})} {2h} + O(h^2)
\end{eqnarray}

Applying Eq. \ref{eq:24} to Eq. (\ref{eq:30}) we have:

\vspace{-1mm}
$${\Pi}(t) = {\Pi} (t-2h) +2h {\Pi} (t-h)$$

Generally, there is the Taylor series expansion 
\vspace{-3mm}
\begin{eqnarray}\label{eq:Taylor}
y(t+h) =  y(h) + h y^{(1)}(t) + \frac {h^2}{2} y^{(2)}(t) + \frac{h^3}{3!} y^{(3) }(t) + \ldots  
\end{eqnarray}

The ODE of interest is ${\Pi}' (t) = {\bf Q} {\Pi} (t)$: 
\vspace{-2mm}
$${\Pi}^{(n)} (t) = {\bf Q}^n {\Pi} (t)$$  

Similarly to Eq. (\ref{eq:Taylor})
\vspace{-3mm}
\begin{eqnarray}
{\Pi}(t+h) = \left[ {\bf I} + h{\bf Q} +\frac{h^2}{2!} 
{\bf Q}^2 + \frac{h^3}{3!} {\bf Q}^3 + \ldots \right] {\Pi}(t) 
\end{eqnarray}

\subsection{Solutions Based on the Exponentials of a Matrix}

Eq. (\ref{eq:30}) has the solution

\vspace{-3mm}
\begin{eqnarray}\label{eq:31}
{\Pi}(t) = e^{ {\bf Q} t} \mbox{   where   }e^{{\bf Q} t} =  \sum_{n=0}^\infty \frac{ ( {\bf Q} t)^n }{n!}
\end{eqnarray}

One method to compute ${\Pi}(t)$ is as follows:

\vspace{-3mm}
\begin{eqnarray}
{\Pi}(t) = \sum_{n=0}^N {\bf F}_n(t) +  {\bf E}(N)
\end{eqnarray}
where ${\bf E}(N)$ is a matrix which represent the error introduced by truncation at $N$.
${\bf F}_n (t)$ is defined by the recursion:

\vspace{-3mm}
\begin{eqnarray}\label{eq:34}
{\bf F}_n (t) = {\bf F}_{n-1} (t) {\bf Q} \frac{t}{n}.
\end{eqnarray}

Instead of calculating ${\Pi}(t)$ it is more efficient to calculate:

\vspace{-2mm}
$${\Pi} (t) = \sum_{n=0}^N {\bf F}_n(t) + {\bf E}(N) 
\mbox{  where  }{\bf F}_n (t) = {\bf F}_{n-1} (t) {\bf Q} \frac{t}{n}$$
A related approach is to quantize time, such that $t=nh$ and apply Eq. (\ref{eq:31}).

Another simple approach is to perform a similarity transformation of ${\bf Q}$. 
Assuming that all the eigenvalues of Q are distinct, 
or more generally that {\bf Q }is diagonizable, ${\bf Q}$ can be written as:

\vspace{-3mm}
\begin{eqnarray}\label{eq:36}
{\bf Q} = {\bf V} {\Lambda} {\bf V}^{-1}
\end{eqnarray}
where ${\Lambda}$ is a diagonal matrix with the $M$ eigenvalues 
$\lambda_i$ of ${\bf Q}$ and ${\bf V}$ holds the eigenvectors.

\vspace{-3mm}
\begin{eqnarray}\label{eq:37}
{\Pi} (t) = \sum_{n=0}^\infty {\bf Q}^n \frac{t^n}{n!} = 
{\bf V}  \left[ \sum_{n=0}^\infty {\Lambda}^n \frac{t^n}{n!} \right] {\bf V}^{-1}  \\ \nonumber
{\bf V} e^{{\Lambda} t} {\bf V}^{-1}= 
{\bf V} \mbox{diag} \{ e^{\lambda^1 t}, \ldots, e^{\lambda_M t} \}{\bf V}^{-1}
\end{eqnarray}

The problem of finding ${\Pi}(t)$ reduces to the problem of finding 
the eigenvalues and eigenvectors of ${\bf Q}$. 
Note that the matrix $Q$ has an eigenvalue equal to $0$, 
and so the corresponding eigenvector gives the steady state solution ($t \rightarrow \infty$).
Furthermore, the convergence to the steady state solution is determined 
by the subdominant eigenvalue of ${\bf Q}$.

\subsection{Laplace Transform Methods}

Applying the LT to the backward Kolmogorov equation in Eq (\ref{eq:30}) we obtain:

\vspace{-2mm}
\begin{eqnarray}
[s {\bf I} - {\bf Q}] {\Pi}^* (s) = {\bf I}
\end{eqnarray}
the LT of ${\Pi}(t)$ is ${\Pi}^*(s)$ and therefore 
\vspace{-1mm}
$$ s {\Pi}^* (s) - {\Pi}(0 )= s {\Pi}^*(s)- {\bf I}\mbox{ is the LT of }{Pi}'(t)$$

\vspace{-2mm}
\begin{eqnarray}\label{eq:39}
{\Pi}^* (s) =\frac{1}{s} [ I - \frac{Q}{s} ]^{-1}
= \sum_{n=0}^\infty \frac{ {\bf Q}^n } {s^{n+1}} \mbox{ inverting } 
{\Pi}(t) = \sum_{n=0}^\infty \frac{Q^n}{n!} =  e^{{\bf Q} t}
\end{eqnarray}

\subsection{Krylov Subspace Method} 

This method is not discussed.

\subsection{Uniformization or Randomization}\label{sec:2.5}

This method is described in Grassman 1991 \cite{Gras91} and de Souza and Gail 1996 \cite{deGa96}. 
For the CTMC ${\cal X}$ for the finite space ${\cal S}$ and generator ${\bf Q}$ 
let ${\Lambda} \geq \mbox{max}\{q_{i,i} \}$.
Define the matrix ${\bf P}={\bf I}+{\bf Q}/{\Lambda}$, which is stochastic by choice of ${\Lambda}$.
From Eq. (\ref{eq:31}) we have:

\vspace{-2mm}
\begin{eqnarray}\label{eq:49}
\pi (t) = \pi (0) \sum_{n=0}^\infty e^{-{\lambda} t} \frac { ( {\Lambda} t )^n}{n!} {\bf P}^n
\end{eqnarray}

This equation can be evaluated recursively

\vspace{-2mm}
\begin{eqnarray}\label{eq:50}
v(n) =  \pi (0) {\bf P}^n \mbox{   and noting that  }{\bf v}(n)={\bf v}(n-1) {\bf P}.
\end{eqnarray}

The reader is also referred to Section 8.6 ``Solution Techniques'' in Trivedi 2002 \cite{Triv02}. 

\section*{Appendix II: The Mathematics Behind SHARPE Reliability Modeling Package}

Section 4.7.2 in \cite{SaTP96} considers a CTMC with $m=2$ non-absorbing and $n=2$ absorbing states,
states with no transitions to other states Trivedi 2002 \cite{Triv02}.
In the case of RAID5 failures there are two failure states during rebuild:
failure due a second disk failure or unreadable sector as discussed in Section \ref{sec:DEH}. 
The infinitesimal generator matrix with the states indexed as $(0:m-1)$ and $(m:m+n-1)$ is given as:

\vspace{-2mm}
$$
{\bf Q}= 
\begin{pmatrix}
-(\gamma+\lambda) &\gamma &\lambda &0 \\
\beta & -(\beta+\delta + \mu)    &\mu   &\delta \\
0 &0 &0 &0 \\ 
0 &0 &0 &0 \\ 
\end{pmatrix}
$$

Differential equations and respective LSTs are as follows where
${\bf \alpha} = {\bf \pi}(0) = \{ \alpha_0,\alpha_1,\alpha_2,\alpha_3 \}$ denotes the initial state vector.

\vspace{-2mm}
\begin{align}\nonumber
\frac{d\pi_0 (t)}{dt} &= - (\gamma + \lambda) \pi_0(t) +\beta \pi_1(t), \hspace{5mm}
s L^*_0 (s) - \alpha_0 =  -(\gamma +\lambda) L^*_0 (s) + \beta L^*_1 (s) \\
\nonumber
\frac{d\pi_1 (t)}{dt} &= -(\beta +\delta +\mu) \pi_1 (t_ +\gamma \pi_0 (t),  \hspace{5mm} 
s L^*_1 (s) - \alpha_1 = -(\beta +\delta + \mu) L^*_1(s) +\gamma L^*_0 (s) \\
\nonumber
\frac{d\pi_2 (t)}{dt} &= \lambda \pi_0 (t) +\mu \pi_1 (t),   \hspace{5mm}    
s L^*_2 (s) - \alpha_2 = \gamma L^*_0 (s_ +\mu L^*_1 (s) \\
\nonumber
\frac{d\pi_3 (t)}{dt} &= \delta \pi_1 (t), \hspace{5mm}   
s L^*_3 (s) - \alpha_3 = \delta L^*_1 (s)
\end{align}

Let ${\bf L^*} = \left( L^*_0, \ldots , L^*_{m-1}\right)$ 
and ${\bf T}$ the upper $m \times m$ upper left hand corner of generator matrix ${\bf Q}$, we have:

\vspace{-3mm}
\begin{eqnarray}\nonumber
s {\bf L}^*  - {\bf \alpha} = {\bf L}^* {\bf T}  \Rightarrow
{\bf  L}^* ={\bf \alpha}  \left( s  {\bf I} - {\bf T} \right)^{-1}
\end{eqnarray}

We now have the LST to the transient distribution functions for nonabsorbing states
We define ${\bf T}_i$ to be the $m\times 1 $ vector that is transpose of 
$(q_{1i}, q _{2i}, \ldots, q_{mi}) $ so

\vspace{-1mm}
$$s L^2 (s) = \alpha_2 + {\bf L} T_2 = \alpha_2 +{\bf \alpha} (s{\bf I} - {\bf T} )^{-1} T_2$$

For the absorbing state $L^*_i$ we have:
\vspace{-1mm}
$${\bf L}^*_i (s) = \frac{1}{s} \left( \alpha_i + {\bf \alpha} 
\left( s {\bf I} - {\bf T} \right)^{-1} \right) {\bf T}_i ,$$ 

The real work is inverting the LST using partial fraction expansion:
\vspace{-1mm}
$${\bf P} (s) = \frac{1}{s} \left( {\bf \alpha} \left( s {\bf I} - {\bf T} \right)^{-1} {\bf T}_i \right) $$

Vector ${\bf X}(s)$ is defined as 
\vspace{-1mm}
$${\bf X}(s) = \left( s {\bf I} - {\bf T} \right)^{-1} {\bf T}_i \mbox{  then  }
\left( s {\bf I}- {\bf T} \right) {\bf X}(s) = {\bf T}_i$$

Applying Cramer's rule 
\footnote{\url{https://en.wikipedia.org/wiki/Cramer's_rule}}
the $i^{th}$ element of ${\bf X}$ is
\vspace{-1mm}
$$X_i (s) = \frac{N_i(s)}{D(s)}\mbox{  where  }D(s) = det({\bf I} - {\bf T})$$

$N_i (s)$ is the determinant of the matrix obtained by replacing the $i^{th}$ column by the vector ${\bf T}_i$. 
Given the column vector ${\bf X}$ 
\vspace{-3mm}
\begin{eqnarray}\nonumber 
P(s) = \frac{1}{s} \left( {\bf \alpha} {\bf X}(s) \right) = \frac {\sum \alpha_i N_i(s) }{s D(s)}
\end{eqnarray}

To invert $P(s)$ we need to carry out a partial fraction expansion. 
The zero root of the denominator yields the steady state solution.
Given $D(s) = det(s {\bf I} - {\bf T})$ eigenvalues of ${\bf T}$ are given as sorted list of the roots"
$( s_1, s_2, \ldots , s_{m+1})$, where root $s_k$ may occur $d_k$ times. 
Then after determining $a_k$ we have the inversion:

\begin{eqnarray}\nonumber 
P^*(s) = \sum_{k=1}^{m+1} \frac{a_k}{ (s-s_k)^{d_k}} \Rightarrow
F(t) = \sum_{k=1}^{m+1} \frac{a_k} {(d_k - 1)!} t^{d_k -1} e^{s_k t}
\end{eqnarray}

This solution yields transient state probabilities. 
However, the number and types of matrix manipulations make the algorithm unstable for some CTMCs 
Purely numerical algorithms for computing the transient probabilities are more stable.
Use is made of Eq. (\ref{eq:31}) discussed in Section \ref{sec:deGa00}. 

\vspace{-2mm}
$${\bf \pi} (t) ={\bf \pi}(0) e^{{\bf Q} t}\mbox{ given }q \geq \mbox{max}_i |q_{ii}| 
e^{ {\bf Q} t} = \sum_{i=0}^\infty \frac{({\bf Q}t)^i}{i!}$$

\vspace{-2mm}
$${\bf \pi}(t) =\sum_{k=1}^\infty {\bf \theta} (k) e^{-q t} \frac{(q t)^k}{k!}$$

where 
${\bf \theta}(0) = {\bf \pi}(0)$, ${\bf Q}^* = \frac{ {\bf Q}} {q} + {\bf I}$ and 
${\bf \theta} (k) = {\bf \theta} (k-1) {\bf Q}^*$

Incidentally hierarchical performance modeling for queueing networks 
in Thomasian and Bay 1986 \cite{ThBa86} is incorporated into SHARPE. 

\subsection*{Abbreviations} 

\begin{small}
{\bf BIBD} - Balanced Incomplete Block Design,
{\bf CTMC} - Continuous Time Markov Chain,
{\bf DAC} - Disk Array Controller,
{\bf DAR} - Disk Adaptive Redundancy,
{\bf DRC} - Degraded Read Cost,
{\bf EAFDL} - Expected Annual Fraction of Data Loss,
{\bf EB} - exabyte=$10^{18}$,
{\bf FBA} - Fixed Block Architecture.
{\bf F/J} - Fork/Join,
{\bf FSQ} - Fair Share Queueing,  
{\bf GB} - gigabyte=$10^9$,
{\bf HDFS} - Hadoop Distributed File System,
{\bf IOPS} -Input/Output Per Second,
{\bf IDR} - Intra-Disk Redundancy,
{\bf IPC} - Interleaved Parity Check,
{\bf KB} - kilobyte=$10^3=2^{10}$,
{\bf kDFT} - k Disk-Failure Tolerant,
{\bf LRC} - Local Redundancy Code.
{\bf LSE} - Latent Sector Error,
{\bf LST} - Laplace Stieltjes Transform,
{\bf MDS} - Maximum Distance Separable,
{\bf MB} - megabyte=$10^6$,
{\bf MTTD} - Mean Time to Detection,
{\bf MTTF} - Mean Time To Failure,
{\bf MTTR} - Mean Tine To Repair, 
{\bf MTTDL} - Mean Time To Data Loss,
{\bf NRC} - Normalized Repair Cost,
{\bf NRP} - Nearly Random Permutations,
{\bf NVMeoF} - NVMe\textregistered over Fabrics (NVMe-oF\texttrademark ),
{\bf NVRAM} - Non-Volatile Random Access Memory,
{\bf ODE} - Ordinary Differential Equation,
{\bf OLTP} - OnLine Transaction Processing,
{\bf PB} - petabyte=$10^{15}$, 
{\bf PCM} - Permanent Customer Model,
{\bf PDF} - Probability Distribution Function,
{\bf PDL} - Parallel Data Laboratory,
{\bf PUE} - Power Usage Effectiveness,
{\bf RAID} - Redundant Array of Inexpensive/Independent Disks, 
{\bf RDP} - Rotated Diagonal Parity,
{\bf RMW} - Read-Modify Write,
{\bf RPM} - Rotations Per Minute,
{\bf RS} - Reed-Solomon (code).
{\bf RU} - Rebuild Unit,
{\bf SATF} - Shortest Access Time First,
{\bf SDC} - Silent Data Corruption,
{\bf SPC} - Single Parity Check, 
{\bf SPTF} - Shortest Processing Time First,
{\bf SSD} - Solid State Disk,
{\bf SU} - Stripe Unit,
{\bf SWP} - Small Write Penalty,
{\it TCO} - Total Cost of Ownership,
{\bf UDE} - Undetected Disk Errors,
{\bf UPS} - Uninterruptible Power Supply,
{\bf VSM} - Vacationing Server Model, 
{\bf TB} - terabyte=$10^{12}$,
{\bf XOR} - eXclusive OR,
{\bf ZB} - zettabyte=$10^{21}$,
{\bf ZBR} - Zoned Bit Recording,
{\bf ZLA} - Zero Latency Access.
{\bf ZRL} - Zurich Research Lab (IBM)
\end{small}


\end{document}